\documentclass[10pt,aps,pre,amsmath,amsfonts,amssymb,floatfix,superscriptaddress,notitlepage,nofootinbib]{revtex4-2}
\usepackage{mathrsfs} \usepackage{graphicx,color} 
\usepackage{multirow}
\usepackage{bm}
\usepackage[normalem]{ulem}
\let\a=\alpha \let\b=\beta \let\g=\gamma \let\d=\delta
\let\e=\epsilon \let\z=\zeta \let\h=\eta \let\k=\kappa
\let\l=\lambda \let\m=\mu \let\n=\nu \let\x=\xi 
\let\s=\sigma \let\t=\tau \let\f=\varphi \let\c=\chi
 \let\y=\upsilon  
\let\D=\Delta \let\Th=\Theta\let\L=\Lambda \let\X=\Xi 
   
\let\ee=\varepsilon \let\r=\rho \let\th=\theta \let\io=\infty

\def\ie{{\textit{i.e.} }}\def\eg{{\textit{e.g.} }}

\def\PP{{\cal P}}\def\MM{{\cal M}} 
\def\CC{{\cal C}}\def\FF{{\cal F}} \def\HH{{\cal H}}
 
\def\cR{{\cal R}}  \def\OO{{\cal O}}
 \def\SS{{\cal S}}

 \def\XXX{{\bf X}}  
\def\xx{{\bf x}} \def\yy{{\bf y}}

\def\ul{\underline}

\def\redv{\bar v}

\def\sl{s_{\rm liq}}

\def\rr{\mathbf{r}}

\def\xx{\mathbf{x}}
\def\de{\mathrm d}

\def\to{\rightarrow} \def\la{\left\langle} \def\ra{\right\rangle}

\newcommand{\beq}{\begin{equation}} \newcommand{\eeq}{\end{equation}}
\newcommand{\wh}{\widehat} 
\newcommand{\Tr}{\text{Tr}}

\def\bX{\boldsymbol{\X}}

\newcommand{\moy}[1]{\left\langle  #1 \right\rangle }

\begin{document}
\title{
Gradient descent dynamics and the jamming transition in infinite dimensions
}

\author{Alessandro Manacorda}
\affiliation{Laboratoire de Physique de l’Ecole Normale Sup\'erieure, ENS, Universit\'e PSL, CNRS, Sorbonne Universit\'e, Universit\'e Paris Cit\'e, F-75005 Paris, France}
\affiliation{Department of Physics and Materials Science, University of Luxembourg, L-1511 Luxembourg}

\author{Francesco Zamponi}
\affiliation{Laboratoire de Physique de l’Ecole Normale Sup\'erieure, ENS, Universit\'e PSL, CNRS, Sorbonne Universit\'e, Universit\'e Paris Cit\'e, F-75005 Paris, France}

\begin{abstract}
Gradient descent dynamics in complex energy landscapes, \ie featuring multiple minima, finds application in many different problems, from soft matter to machine learning.
Here, we analyze one of the simplest examples, namely that of soft repulsive particles in the limit of infinite spatial dimension~$d$. The gradient descent dynamics
then displays a {\it jamming} transition: at low density, it reaches zero-energy states in which particles' overlaps are fully eliminated, while at high density
the energy remains finite and overlaps persist. At the transition, the dynamics becomes critical. In the $d\to \io$ limit, a set of 
self-consistent dynamical equations can be derived via mean field theory. 
We analyze these equations and we present some partial progress towards their solution. We also study the Random Lorentz Gas in a range of $d=2\ldots 22$, and
obtain a robust estimate for the jamming transition in $d\to\io$. The jamming transition is analogous to the capacity transition in supervised
learning, and in the appendix we discuss this analogy in the case of a simple one-layer fully-connected perceptron.
\end{abstract}

\maketitle

%\tableofcontents

%\clearpage

\section{Introduction}

Gradient descent (GD) dynamics is one of the simplest dynamics one can imagine. It consists in following the gradient of an energy function in search of a local minimum.
Yet, if the energy function is sufficiently complex, \ie it features multiple minima and saddle points, GD dynamics can lead to an unexpectedly complex phenomenology. For example, in simple models
for spin glasses, it features persistent {\it aging} dynamics, as first demonstrated in the pioneering work of Cugliandolo and Kurchan~\cite{CK93,CK94,Cu02}, and subsequently developed in, among others, Refs.~\cite{BFP97,MR04,Ri13,FFR20,altieri2020dynamical,kurchan2021time}: the energy decays to its asymptotic value as a power law, and the correlation functions continue
to evolve at all times. The system keeps descending into the landscape, going through saddle points of lower and lower order, 
getting closer and closer to local minima but never reaching them~\cite{CK93,FFR20}.

GD dynamics finds application in many contexts, especially related
to theoretical computer science and optimization problems, \ie when one 
needs to minimize a {\it cost} or {\it loss} function.
A renowned example 
is given by machine learning, especially considering
{\it supervised learning}. In this case, one has to learn an 
unknown function to relate an input data $x$ to an output label 
$y=f(x)$, based on a training set where the relation between the
input data (\eg the pixels of an image) and the output label (\eg 
``cat'' or ``dog'') is known. The unknown function $f(x)$ is then parametrized by a guess function $g(x;\th)$, whose
 parameters $\th$ can be learnt by 
minimizing a loss function accounting for the error made in labelling
data from the training set. Also in this case
the GD dynamics can be surprisingly complex~\cite{Baity_Jesi_2019,mannelli2019passed,sarao2019afraid,mannelli2020complex,sclocchi2021high,biroli2020iron,mannelli2020marvels,mignacco2021stochasticity}.

A special class of such loss minimization problems is obtained whenever 
the loss function associated to a single data point is continuous and perfectly vanishing when the data point
is correctly classified, and positive otherwise, \eg the so-called squared hinge loss~\cite{FP16,FPSUZ17,franz2019jamming,spigler2019jamming,franz2021surfing}.
The question then becomes whether a choice of parameters $\th$ exists, such that all data points in the training set are perfectly classified,
leading to a zero loss.
This question defines a {\it satisfiability} transition~\cite{transition_exp,FSS_KiSe,monasson1999determining,AMSZ09,FMZ22arxiv}: 
if such a configuration exists, then all the constraints are satisfied (SAT phase), whereas the existence of 
some violated constraints for all possible configurations defines the UNSAT 
phase. In the UNSAT case, one can then 
shift the problem to an optimization problem, \ie finding a configuration that minimizes the number of unsatisfied constraints.
The SAT/UNSAT transition becomes sharp in the 
thermodynamic limit $N\to\infty$, and search algorithms critically
slow down when looking for solutions near the transition~\cite{AMSZ09,FSS_KiSe,monasson1999determining,transition_exp,HI20}.
In neural networks, the transition corresponds to a {\it capacity} transition~\cite{GD88,krauth1989storage,brunel1992information,monasson1995learning};
in the SAT phase, the network is able to perfectly learn the input-output association $y=f(x)$ for all training examples, while in the UNSAT phase this relation
is not correctly learned for some of the training examples.
In a typical setting~\cite{GD88}, $P=\a N$ random inputs with corresponding labels are given as training data
to a neural network of $N$ units, and there is a
critical value $\a_c$ above which the system is unable to perfectly learn all the training examples, \ie the available information exceed the storage {\it capacity} of the network.
GD dynamics and constraint satisfaction problems (CSP) also find application in many other contexts, \eg in the study 
of complex ecological~\cite{tikhonov2017collective,landmann2018systems,altieri2019constraint} and economical systems~\cite{de2004statistical,moran2019may,sharma2021good}.

We will be particularly interested here in the application to soft matter problems, in which one can consider an idealized model of emulsions or soft athermal colloids, 
\ie an assembly of finite-range interacting soft repulsive particles~\cite{durian1995foam}.
Such a system displays a sharp {\it jamming} transition~\cite{OLLN02,OSLN03,LN10,LNSW10}, which is fully analogous to the capacity or satisfiability transition introduced above~\cite{FP16,FPSUZ17}.
At low density, the GD dynamics converges exponentially to a {\it floppy}, zero-energy state in which particle overlaps are fully eliminated (unjammed phase, SAT)~\cite{IKBSH20,NIB21}. 
At high density instead, the GD dynamics closely resembles
that of mean field spin glasses~\cite{chacko2019slow,nishikawa2021relaxation}, \ie it displays power-law relaxation to the final energy with persistent aging, while particle overlaps persist at long times, 
leading to a finite asymptotic energy (jammed phase, UNSAT). The {\it jamming} phase transition that separates the two regimes displays
diverging length and time scales~\cite{OLLN02,OSLN03,LN10,LNSW10}.
The aim of this work is to investigate to some extent the dynamical mean field theory (DMFT) equations~\cite{CK93, CK94,Cu02,sompolinsky1981dynamic,sompolinsky1982relaxational,MKZ15,Sz17,ABUZ18,AMZ18,mannelli2020complex,mignacco2021stochasticity,sclocchi2021high,liu2021dynamics,altieri2020dynamical,kurchan2021time} that describe GD dynamics in mean-field complex systems, which display a jamming transition.
We will focus in particular on the infinite-dimensional limit of soft repulsive particles~\cite{AMZ18}, and we will thus use the jamming terminology in the main text.

Our results are the following. First of all, in section~\ref{sec:DMFT_dinf},
we present some analytical results on the long-time behavior of the DMFT equations for GD dynamics in the unjammed (SAT) phase, 
and we obtain simple analytical expressions for the response kernel and
the response function in terms of the long-time limit of a single one-time quantity, namely the contact number.
From these results we derive analytically the vibrational spectrum of the Hessian in the final unjammed state, which is of the Marcenko-Pastur form, 
as suggested numerically for infinite-dimensional particles in~\cite{IS21}, see also~\cite{SMBI20} (the same result had been previously obtained in the perceptron model~\cite{FPUZ15}).
Our calculation, however, misses the isolated eigenvalue that is responsible for critical slowing
down around jamming~\cite{LDW13,HI20,IKBSH20,Ik20,NIB21}. 
Moreover, our analysis of the DMFT equations does not provide the location of the jamming transition, because we are unable to evaluate the long-time limit of the contact number.

We thus turn in section~\ref{sec:num_dinf} to the numerical study of a finite-dimensional model to get additional insight on the convergence to the DMFT limit when $d\to\io$. 
It has been shown~\cite{biroli2021interplay,biroli2021mean} that the Random Lorentz Gas (RLG) is a very convenient model for this kind of investigation.
In the present context, the RLG model consists in a single tracer that interacts via a soft repulsive interaction with a set of randomly drawn obstacles. Despite the striking difference between a many-body dynamics
as in particle systems and a single-particle dynamics as in the RLG,
it has been shown that these problems converge to the same DMFT equations in the $d\to\io$ limit~\cite{biroli2021interplay,biroli2021mean}. The simplicity of the
RLG allows one to investigate it numerically over a wide range of $d$. We then confirm that the dynamics converge to the DMFT limit, and obtain some
additional insight on the range of validity of the analytical results. We also obtain an estimate of the location of the jamming transition in $d\to\io$.

In the conclusion section~\ref{sec:conclusions}, we briefly discuss the state-of-the-art of the analytical calculations of the asymptotic energy for GD dynamics in complex landscapes. We show that
none of these methods is able to provide the correct asymptotic energy in the UNSAT phase, and we discuss some possible routes towards the solution of the problem.

In the appendix, in order to show the generality of the approach, we discuss the single-layer fully-connected perceptron~\cite{FP16,FPSUZ17,ABUZ18,FPUZ15,HI20}, 
which is one of the simplest data classifiers in supervised machine learning.
We show how the results of the main text can be translated to that case.

\section{Dynamical mean field theory}
\label{sec:DMFT_dinf}

In this section, we recapitulate the DMFT equations for infinite-dimensional particle systems,
as derived in~\cite{MKZ15,Sz17,ABUZ18,AMZ18,liu2021dynamics}. 
We focus on the specific case of GD dynamics of soft repulsive spheres, see \eg the discussion in~\cite[chapter 9]{parisi2020theory}.
Unfortunately, these equations are particularly difficult
to solve numerically~\cite{manacorda2020numerical}. We discuss here some analytical results, 
while the numerical solution will be discussed in section~\ref{sec:num_dinf}.

\subsection{Definitions}
\label{sec:DMFT_def}

For a system of $N$ particles with positions $\xx_i(t)$ confined in a periodic volume $V \subset \mathbb{R}^d$, 
starting in equilibrium at infinite temperature $\b_0=0$ (that corresponds to a uniformly random initial condition),
consider the gradient descent equations:
\beq
\label{eqC3:GENLang-active}
 \z \dot \xx_i(t)
 	=  \bm F_i(t)  \ , \qquad
 \bm F_i(t)  = -\frac{\partial V(\ul X(t))}{\partial \xx_i(t)} \ , \qquad
 	V(\ul X) = \sum_{i<j} v(\xx_i - \xx_j) \ .
\eeq
Important observables that characterize the dynamics are the correlation and response function
\beq
\CC(t,t') = \frac{d}{\ell^2 N}\sum_i \la [\xx_i(t)-\xx_i(0)]\cdot [\xx_i(t')-\xx_i(0)] \ra \ ,
\qquad
 \cR(t,t') = \frac{d}{\ell^2 N}\sum_{i\m} \frac{\delta \la x_{i\m}(t) \ra}{\delta \l_{i\m}(t')} \ ,
\eeq
where $\m=1,\cdots,d$ is a coordinate index and ${\bm\lambda}_i(t)$ is an external field added
to the force $\bm F_i(t)$ in order to compute the response, and the mean square displacement (MSD)
\beq
\D(t,t') = \frac{d}{\ell^2 N}\sum_i \la [\xx_i(t)-\xx_i(t')]^2 \ra \ , \qquad \D_r(t) = \D(t,0) \ .
\eeq

We take first the thermodynamic limit $N \to \io$ and
$V\to\io$ at constant number density $\rho=N/V$.
In~\cite{AMZ18} it is shown that if the limit $d\to\io$ is taken next, with
\begin{itemize}
\item  time $t$ remaining finite;
\item  potential $v(r) = \redv(h)$, with $h = d(r/\ell-1)$ the scaled inter-particle gap and $\ell$ the particle diameter; 
\item packing fraction scaled as $\wh\f = 2^d \f/d$, with $2^d \f = \r V_d \ell^d$ and $V_d$ the volume of a $d$-dimensional unit sphere;
\item friction coefficient scaled as
\beq\label{eqdio:AMZ19}
\wh\z = \frac{\ell^2}{2 d^2} \z \ ;
\eeq
\end{itemize}
then the system dynamics is described by a set of one-dimensional stochastic equations with a Gaussian colored noise $\X(t)$, given by
\beq\label{eqdio:y}
\begin{split}
 &  \wh\z \dot y(t)
 	=	- \k(t) y(t)
 		+ \int_0^t \!\! \de t' \, \MM_R(t,t') \, y(t')
 		- \redv'(h_0+y(t)+\D_r(t)) +  \X(t)
 \ , \qquad y(0)=0 \ ,\\
 & \moy{ \X(t)}=0
  	\ , \quad
 	\moy{ \X(t) \X(t')}
 		=    \MM_C(t,t') \ ,\\
\end{split}
\eeq
with ${h(t) = h_0+y(t)+\D_r(t)}$ being the time-dependent inter-particle gap, and with memory kernels
\beq\label{eqdio:ker}
\begin{split}
 \k(t)
 	&= \frac{\wh \f}2 \int^{\infty}_{-\infty} \, \de h_0 \, e^{h_0 }   \la \redv''(h(t)) + \redv'(h(t)) \ra_{h_0}
 \ , \\
 \MM_C(t,t')
 	&=  \frac{\wh\f}2 \int^{\infty}_{-\infty}  \de h_0 \, e^{h_0}  
 		\langle \redv'(h(t)) \redv'(h(t')) \rangle_{h_0} 
 \ , \\
 \MM_R(t,t')
 	&=  \frac{\wh\f}2 \int^{\infty}_{-\infty}  \de h_0 \, e^{h_0 }  
		\left. \frac{\d \langle \redv'(h(t))  \rangle_{h_0,\PP}}{\d \PP(t')}\right\vert_{\PP=0} 
		= \frac{\wh\f}2 \int^{\infty}_{-\infty}  \de h_0 \, e^{h_0 }  
		\langle \redv''(h(t)) H(t,t') \rangle_{h_0}
 \ .
\end{split}
\eeq
Here, the averages $\la \bullet \ra_{h_0}$ are over the noise $\X(t)$ at fixed $h_0=h(0)$, the perturbation ${\PP(t)}$ acts via the replacement ${\redv'(h_0+y(t)+\D_r(t)) \to \redv'(h_0+y(t)+\D_r(t) - \PP(t))}$ in Eq.~\eqref{eqdio:y}, and $H(t,t') = \d h(t) /\d \PP(t') \vert_{\PP=0}$,
which, differentiating Eq.~\eqref{eqdio:y}, satisfies
 \beq\label{eqdio:H}
 \wh\z \frac{\partial}{\partial t}  H(t,t') = - \k(t) H(t,t') - \redv''(h(t)) \left[ H(t,t') - \d(t-t') \right] + \int^t_{t'} \de u \, 
 \MM_R(t,u) H(u,t') \ .
 \eeq 
Note that $H(t,t')$ is a functional of $h(t)$, but we omit this dependence in order to simplify the notation.
The correlation and response functions and the MSD
 are given by
\begin{equation}\label{eqdio:corr}
\begin{split}
\wh \z \frac{\partial}{\partial t} \CC(t,t')
  	=& -\k(t)\CC(t,t')+\int_0^{t}\de u\,\MM_R(t,u)\CC(u,t')
 		 +\int_0^{t'}\de u\,\MM_C(t,u)\cR(t',u)
 \ ,\\
\wh \z \frac{\partial}{\partial t}
 \cR(t,t')
 	=& \frac{\d(t-t')}{2}-\k(t)\cR(t,t')+\int_{t'}^{t}\de u\,\MM_R(t,u)\cR(u,t')
 \ , \\
  \frac{\wh \z}2  \frac{\partial}{\partial t} \D(t,t')
  	=&-\frac{\k(t)}{2}\left[\D(t,t')+\D_r(t)-\D_r(t')\right]+\frac12\int_0^{t} \de u\,\MM_R(t,u)\left[\D_r(t)-\D_r(t')+\D(u,t')-\D(u,t)\right]\\
  	& 
  	+\int_0^{\max(t,t')}\de u\,\MM_C(t,u)\left[\cR(t,u)-\cR(t',u)\right]
 \ . \\	
 \end{split}
\end{equation}
Here $\D(t,t') = \CC(t,t) + \CC(t',t') - 2 \CC(t,t')$ with $\CC(t,0)=\CC(0,t)=0$, hence
$\D_r(t) \equiv \D(t,0) = \CC(t,t)$.

It is convenient to define the integrated responses,
\beq \label{eqdio:interesp}
\c(t,t') = \int^t_{t'} \de u \, \cR(t,u) \ ,
\qquad
\wh\c(t,t') = \k (t) - \int^t_{t'} \de u \, \MM_R(t,u) \ ,
\eeq
and $\wh\c(t,t')$ encodes the response kernels as
\beq
\k(t) = \wh\c(t,t) \ , 
\qquad \MM_R(t,t') = \th(t-t') \, \partial_{t'} \wh\c(t,t') \ .
\eeq

Eqs.~\eqref{eqdio:y}, \eqref{eqdio:ker}, \eqref{eqdio:H} and \eqref{eqdio:corr} form a closed set that can be in principle solved numerically.
Using the integrated response $\wh\c(t,t')$ instead of $\MM_R(t,t')$, one can write them in an alternative form
(appendix~\ref{app:B}):
\begin{equation}\label{eq:DMFT-set}
\begin{split}
 \wh\z \dot y(t) &= - \int_0^t \!\! \de t' \, \wh\c (t,t') \, \dot y(t') - \redv'(h(t)) +  \X(t) \ , \qquad y(0)=0 \ ,\\
\moy{ \X(t)} &=0 \ , \quad \moy{ \X(t) \X(t')} = \MM_C(t,t') \ ,\\
h(t) &= h_0 + y(t) + \D_r(t) \ , \\
\wh\z \frac{\partial}{\partial t}  H(t,t') &= - \frac1{\wh\z} \redv''(h(t')) \wh\c(t,t') - \int^t_{t'} \de u \, \wh\c (t,u) \frac{\partial}{\partial u} H(u,t') - \redv''(h(t)) \left[ H(t,t') - \d(t-t') \right]  \ , \\
\wh \z \frac{\partial}{\partial t} \CC(t,t') &= - \int_0^{t}\de u\,\wh\c(t,u)\frac{\partial}{\partial u}\CC(u,t') +\int_0^{t'}\de u\,\MM_C(t,u)\cR(t',u) \ ,\\
\wh \z \frac{\partial}{\partial t} \cR(t,t') &= \frac{\d(t-t')}{2}-\frac1{2\wh\z} \wh\c(t,t')- \int_{t'}^{t}\de u \, \wh\c (t,u) \frac{\partial}{\partial u} \cR(u,t') \ , \\
  \frac{\wh \z}2  \frac{\partial}{\partial t} \D(t,t') &= -\frac12\int_0^{t} \de u\,\wh\c(t,u)\frac{\partial}{\partial u} 
  \left[\D(u,t')-\D(u,t)\right]+\int_0^{\max(t,t')}\de u\,\MM_C(t,u)\left[\cR(t,u)-\cR(t',u)\right] \ , \\
  \D_r(t) &= \D(t,0) \ , \\
 \wh\c(t,t') &= \frac{\wh \f}2 \int^{\infty}_{-\infty} \, \de h_0 \, e^{h_0} \moy{ \redv''(h(t)) \left[ 1 - \int^t_{t'} \de u \, H(t,u) \right]  +
\redv'(h(t)) }_{h_0} \ , \\
 \MM_C(t,t') &=  \frac{\wh\f}2 \int^{\infty}_{-\infty}  \de h_0 \, e^{h_0} \moy{ \redv'(h(t)) \redv'(h(t')) }_{h_0}  \ ,
\end{split}
\end{equation}
having exploited the initial conditions $y(0)=\CC(0,t)=0$ and 
$\cR(t^+,t) = 1/(2\wh\z)$, $H(t^+,t) = \redv''(h(t))/\wh\z$. The time
integrals are taken over $[t'+\e,t-\e]$ for arbitrarily small $\e$, excluding the singular contributions that have been accounted for separately.

We also recall the definition of the scaled energy, pressure and isostaticity index at infinite 
dimensions~\cite{AMZ18}, which read, respectively,
\beq\label{eq:one-time-q}
\begin{split}
e(t) &= \frac{\wh\f}2 \int^{\infty}_{-\infty} \, \de h_0 \, e^{h_0}  \moy{\redv(h(t)}_{h_0} \ , \\
p(t) &= -\frac{\wh\f}2 \int^{\infty}_{-\infty} \, \de h_0 \, e^{h_0}  \moy{\redv'(h(t)}_{h_0} \ , \\
c(t) &= \frac{\wh\f}2 \int^{\infty}_{-\infty} \, \de h_0 \, e^{h_0}  \moy{\th(-h(t))}_{h_0} \ , \\
\end{split}
\eeq
where the isostaticity index $c(t) = z(t)/(2d)$ is the number of contacts per particle $z(t)$ divided by the 
isostatic number $z=2d$, which is the minimal number of contacts required for mechanical stability~\cite{parisi2020theory}.

\subsection{Long-time limit of the dynamics in the unjammed phase}
\label{sec:TTI}

In the following we consider the harmonic soft sphere potential~\cite{durian1995foam}, 
which in infinite dimensions corresponds to $\redv(h) = \ee h^2\th(-h)/2$~\cite{parisi2020theory}, 
and we fix $\ee=1$ and $\wh \z=1$ for simplicity, and without loss of generality.

For infinite-dimensional particle systems, the GD dynamics in the jammed phase is expected to display a non-trivial aging dynamics, belonging
to the same universality class of $p$-spin models~\cite{CK93,Ri13,FFR20}. The energy should then decay as a power-law and correlation functions should age indefinitely;
numerical evidence for this has been given in~\cite{chacko2019slow,nishikawa2021relaxation}. 
It is not clear if the memory of the initial condition is lost or not~\cite{Ri13,FFR20}:
in finite dimensions, the MSD with respect to the initial condition,
\ie $\D_r(t)$, reaches a finite plateau, thus suggesting a persistent memory of the initial condition~\cite{nishikawa2021relaxation}. However, this pleateau seems to increase upon increasing $d$. Our numerical results (see section~\ref{sec:num_dinf})
also suggest that, in the $d\to\io$ limit, $\D_r(t)$ keeps growing with $t$, but they are not conclusive. 
Because the asymptotic dynamics in this regime is extremely difficult to describe~\cite{CK93,Ri13,FFR20,altieri2020dynamical}, we do not consider the jammed phase here; some speculations
will be presented in the conclusion section~\ref{sec:conclusions}.

We focus instead on the
unjammed phase, where the gradient descent dynamics converges exponentially  to a unique final configuration~\cite{HI20,NIB21},
and we follow similar steps as in Ref.~\cite{sclocchi2021high}.
Because at long times motion is arrested, we have
\beq\label{eq:MSD-TTI}
\lim_{t\to\io}\D(t+\tau,t)=\D^\io(\tau) =0 \ ,\qquad  \forall \t \ ,
\qquad
\text{and}
\qquad
\lim_{t\to\io}\D_r(t)=\D_r^\io \ ,
\eeq
and a constant effective gap 
\beq
\lim_{t\to\io} h(t) = h_\infty = h_0 + \D_r^\io + y_\io \geq 0 \ ,
\eeq
which is a positive random variable because by definition of the unjammed phase, all
overlaps between particles are removed in the final state. Therefore
\beq\label{eqdio:TTIlong}
\begin{split}
 \k_\io	&=  \lim_{t\to\io} \k(t) = \frac{\wh \f}2 \int^{\infty}_{-\infty} \, \de h_0 \, e^{h_0 }   \la \redv''(h_\io) + \redv'(h_\io) \ra_{h_0} 
 = \frac{\wh \f }2 \int^{\infty}_{-\infty} \, \de h_0 \, e^{h_0 }   \la \th(-h_\io)  \ra_{h_0} =  c_\io \ , \\ 
\MM_C^\io(\tau) &= \lim_{t\to\io} \MM_C(t+\t,t) =\frac{\wh\f}2 \int^{\infty}_{-\infty}  \de h_0 \, e^{h_0}  
 		\langle \redv'(h_\io)^2 \rangle_{h_0} =0  \ ,\qquad  \forall \t\ .
\end{split}\eeq
%where
%\beq
%c_\io = \frac{z_\io}{2d} = \frac{\wh \f}2 \int^{\infty}_{-\infty} \, \de h_0 \, e^{h_0 }   \la \th(-h_\io)  \ra_{h_0}
%\eeq
%is the number of contacts divided by the isostatic number in the final configuration~\cite{parisi2020theory}, \ie it is the {\it isostaticity index}.

Following Ref.~\cite{sclocchi2021high},
we make an additional assumption, \ie that the decay of the response kernel is fast enough to ensure that,
for any function $f(t)$ that has a finite long-time limit, $f(t\to\io) = f_\io$,
\beq\label{eq:fastcond}
\int_0^t \de u \MM_R(t,u)f(u) \underset{t\to\io}{\sim} \int_0^t \de u \MM_R(t,u) f_\io = [\k(t) - \wh\c(t,0)] f_\io \ .
\eeq
Under this assumption, the explicit time-dependence in Eq.~\eqref{eqdio:H} disappears when $t,t'\to\io$, and as a result response
functions become time-translationally invariant (TTI):
\beq
\lim_{t\to\io}H(t+\tau,t) = H^\io(\t) \qquad \Rightarrow \qquad \lim_{t\to\io}\MM_R(t+\tau,t) = \MM_R^\io(\t)
\qquad \Rightarrow \qquad \lim_{t\to\io}\cR(t+\tau,t) = \cR^\io(\t) \ .
\eeq
We also define the long-time limit of the integrated response kernels as
\beq
\c=\lim_{t\to\io} \chi(t,0) = \int_0^\io \de \t \cR^\io(\t) \ ,\qquad  \wh\c=\lim_{t\to\io} \wh\c(t,0) =\k_\io -  \int_0^\io \de \t \MM_R^\io(\t) \ .
\eeq
Combining the previous results, the equation for $H^\io(\t)$ is
\beq
\partial_\t H^\io(\t) = -[ c_\io+  \th(-h_\io)] H^\io(\t) + \th(-h_\io)\d(\t) +\int_0^\t \de u \MM^\io_R(\t-u) H^\io(u)  \ .
\eeq
Note that $h_\io > 0$ then implies $H^\io(\t)=0$ because the source term vanishes.
Multiplying by $\redv''(h_\io)= \th(-h_\io)$ and averaging over $h_0$ as in Eq.~\eqref{eqdio:ker},
we obtain a closed equation for $\MM^\io_R(\t)$:
\beq\begin{split}
\partial_\t \MM_R^\io(\t) &= -( c_\io+  1)  \MM_R^\io(\t) 
+  c_\io \d(\t) +\int_0^\t \de u \MM^\io_R(\t-u) \MM_R^\io(u)  \ .
\end{split}\eeq
In Laplace space,
$\MM_R^\io(s) = \int_0^\io \de \t \MM_R^\io(\t) e^{-s \t}$,
this gives
\beq\label{eqdio:MRLap}
s \MM^\io_R(s) = -  (1+c_\io) \MM_R^\io(s) +  c_\io + [\MM^\io_R(s)]^2 \quad \Rightarrow \quad
\MM^\io_R(s) = \frac{(1+ c_\io+s) - \sqrt{(1+ c_\io+s)^2 - 4 c_\io}}2 \ .
\eeq
Note that at short times we must have
\beq
\MM^\io_R(\t\to 0) \sim  c_\io + \OO(t) \qquad\Leftrightarrow \qquad
\MM^\io_R(s\to\io) = \frac{c_\io}{s} + \OO(1/s^2) \ ,
\eeq
which fixes the choice of sign in Eq.~\eqref{eqdio:MRLap}, because the other solution has an unphysical behavior
$\MM^\io_R(s) \sim s$ at large~$s$.
Note also that in order for Eq.~\eqref{eqdio:MRLap} to be well defined we need the condition
\beq
(1+c_\io)^2 \geq 4 c_\io  \qquad\Leftrightarrow \qquad (1 -c_\io)^2 \geq 0 \ ,
\eeq
which is always satisfied. We also obtain
\beq\label{eqdio:kMconsist}
\wh \chi =\k_\io- \MM^\io_R(s=0) =\k_\io-  c_\io = 0 \ ,
\eeq
which guarantees a proper cancellation of the response kernels in the long time limit.
Indeed,
recalling that in the long time limit the noise vanishes and $h(t\to\io) = h_\io$,
and applying Eq.~\eqref{eq:fastcond} to the equation for $y(t)$ in Eq.~\eqref{eqdio:y}, Eq.~\eqref{eqdio:kMconsist} implies:
\beq
0 = -\k_\infty y_\infty+  \MM^\io_R(s=0) y_\infty - \redv'(h_\infty) = - \redv'(h_\infty) \qquad \Rightarrow \qquad h_\io \geq 0 \ ,
\eeq
which leaves $h_\io$ indeterminate but ensures its positivity. 
Eq.~\eqref{eqdio:kMconsist} is therefore crucial for the consistency of the initial assumptions.

The Laplace transform in Eq.~\eqref{eqdio:MRLap} can be inverted as
\beq\label{eqdio:MRio_gen}
\MM^\io_R(\t) = \frac{ \sqrt{c_\io}}{\t} e^{- (1+c_\io)\t} I_1(2\sqrt{ c_\io} \t) \ ,
\eeq
where $I_1(x)$ is the modified Bessel function of first kind.
Because $I_1(x)\sim e^x/\sqrt{2 \pi x}$ at large $x$, we obtain that $\MM^\io_R(\t) \sim e^{- (1-\sqrt{c_\io})^2 \t}$ at large $\t$, provided $c_\io < 1$,
but when $c_\io=1$ (the isostatic point) we obtain $\MM^\io_R(\t) \sim \t^{-3/2}$. Note that in both cases, the decay is fast enough to ensure
the validity of Eq.~\eqref{eq:fastcond}.

Under the assumption of Eq.~\eqref{eqdio:TTIlong} we can inject the expression of $\MM^\io_R(\t)$ in the equation for $\cR(t,t')$ and
in Laplace space we get
\beq\label{eqdio:RioLap}
\cR^\io(s) =\frac12 \frac1{s+ c_\io-\MM_R^\io(s)}
= \frac{1}{c_\io-1+s + \sqrt{(1+ c_\io+s)^2 - 4 c_\io}} 
= \frac{1-c_\io-s + \sqrt{(1+ c_\io+s)^2 - 4  c_\io}}{4 s } \ ,
\eeq
hence
\beq
\chi  = \cR^\io(s=0) =\io \ ,
\eeq
\ie $\chi$ diverges in the whole unjammed phase, which indicates that the response function reaches a plateau at long times.
When $s\to 0$ we have, from Eq.~\eqref{eqdio:RioLap},
\beq
\cR^\io(s) \sim \frac{1-c_\io}{2s} \qquad \Rightarrow \qquad \cR^\io(\t\to\io) \to \frac{1-c_\io}{2} \ .
\eeq

\subsection{Density of vibrational states}
\label{sec:DOS}

From the results of section~\ref{sec:TTI} we can derive the shape of the density of vibrational states in the unjammed phase, and get some physical insight
on the origin of the TTI regime and of the plateau in the response function.
At long times, the system reaches a unique configuration $\ul X^* = \{\xx_i^*\}$. We can then linearize the dynamics around this configuration, 
with $\yy_i(t) = \xx_i(t) - \xx_i^*$,
and
Eq.~\eqref{eqC3:GENLang-active}, with the inclusion of the external field $\bm \l_i(t)$, becomes
\beq
 \z \dot \yy_i(t)
 	=  -\sum_j \frac{\partial V(\ul X^*)}{\partial \xx_i \partial \xx_j}\cdot  \yy_j(t) + \bm \l_i(t) 
	= -\sum_j \HH_{ij} \cdot  \yy_j(t) + \bm \l_i(t) \ ,
\eeq		
which is solved by
\beq
\ul Y(t) = \frac1\z\int_0^t \de u \, e^{- \HH (t-u) /\z} \ul \L(u) \ ,
\eeq
where $\HH$ is the Hessian in the minimum and $\ul \L(t) = \{ \bm \l_i(t)\} $ is the external field. The response function in this approximation can then 
be expressed in terms of the density of scaled vibrational states,
\beq\label{eq:rhola}
\r(\l) = \frac{1}{N d} \sum_{\a=1}^{Nd} \d\left(\l - \frac{\ell^2 \l_\a}{2\ee d^2} \right) \ ,
\eeq
as
\beq\begin{split}
\cR^\io(t) &=\left. \frac{d}{\ell^2 N}\sum_{i\m} \frac{\d y_{i\mu}(t)}{\d \l_{i\m}(0)} \right|_{\bm \l=0} = 
\frac{d}{\ell^2 \z N} \Tr \, e^{- \HH t /\z}
=\frac{1}{2 \wh\z N d} \Tr \, e^{- \HH t /\z} \\
&=\frac{1}{2 \wh\z N d} \sum_{\a=1}^{Nd} e^{- \l_\a t /\z}
=\frac{1}{2 \wh\z} \int \de \l \r(\l) e^{- \ee \l t  /\wh\z} \ ,
\end{split}\eeq
which gives in Laplace space
\beq\label{eqdio:Rs2}
\cR^\io(s) = \frac{1}{2 \ee}\int \de\l \frac{\r(\l)}{ \l + \wh \z s/\ee} \ .
\eeq
Note that the potential $V(\ul X)$ has an energy scale $\ee d^2$, hence the Hessian
has a natural scale $\HH \propto \ee d^2 /\ell^2$, which explains the scaling of $\l_\a$ in Eq.~\eqref{eq:rhola}.

Combining Eqs.~\eqref{eqdio:Rs2} and \eqref{eqdio:RioLap} (now with $\wh\z=1$ and $\ee=1$)
we obtain the Cauchy transform (see \eg~\cite{bun2017cleaning}) of $\rho(\l)$ in the form:
\beq
g(z) = \int \de\l \frac{\r(\l)}{z - \l} = - 2 \cR^\io(s=- z)
= \frac{1-c_\io + z + \sqrt{(1+ c_\io- z)^2 - 4 c_\io}}{2z } \ ,
\eeq
which shows that $\r(\l)$ is the Marcenko-Pastur distribution with parameter $c_\io<1$, \ie
\beq
\r(\l) = (1-c_\io) \d(\l) + \frac{\sqrt{(\l_+-\l)(\l-\l_-)}}{2 \pi \l} \ , \qquad \l_{\pm} = (1 \pm \sqrt{c_\io})^2 \ .
\eeq
This result provides a mathematical derivation of the conjecture
proposed in Ref.~\cite{IS21}. The density of states displays a finite density of zero modes in the unjammed phase with
$c_\io<1$, which explains the finite plateau in the response function: if the system is perturbed away from the final state
of the GD dynamics, it can be displaced along the zero modes, so that it never returns to the state it had before the perturbation.
Finally,
note that the present calculation gives the density of states in the thermodynamic limit, and it is thus unable to detect the isolated eigenvalue that is responsible for the
critical slowing down upon approaching the jamming transition~\cite{LDW13,Ik20,NIB21}.

\subsection{Dilute limit}
\label{sec:dilute}

Because the numerical solution of the DMFT equations is difficult, it is useful to consider here a low-density, dilute limit,
to have an idea of what to expect at higher densities.
In the very low-density limit, we can neglect all the kernels, because they have a factor $\wh\f$ in front~\cite{manacorda2020numerical}. Following the same calculation line exposed in~\cite{AdPMvWZ21}, the equations for 
correlation and response reduce to
\beq
\begin{split}
\wh\z \frac{\partial}{\partial t} \cR (t,t') &= \frac{\d(t-t')}2  \quad \Rightarrow \quad \cR(t,t') = \frac1{2\wh\z} \th(t-t') \ , \\
\wh\z \frac{\partial}{\partial t} \CC (t,t') &= 0  \quad \Rightarrow \quad \CC(t,t')= \D(t,t') = 0 \ .
\end{split}
\eeq
The evolution equation for $h(t)$ then becomes
\beq
\wh\z \dot h(t) = - \redv'(h(t)) \quad  \Rightarrow \quad h(t) = 
\begin{cases}
h_0 \, e^{-\ee t /\wh\z} & h_0 <0 \\
h_0 & h_0 \geq 0 
\end{cases}
\ ,
\eeq
and the fluctuating response $H(t,t')$ then satisfies
\beq
\wh\z \frac{\partial}{\partial t} H (t,t') = - \redv''(h(t)) \left[ H(t,t') - \d(t-t') \right] \quad \Rightarrow \quad H(t,t') = \frac{\ee}{\wh\z} \,
e^{-\ee (t-t') /\wh\z} \, \th(t-t') \, \th(-h_0) \ .
\eeq
With these results we can compute the kernels. The instantaneous response is
\beq
\k (t) = \frac{\wh\f}2 \int^0_{-\io} \de h_0 \, e^{h_0} \, \ee \, \left( 1 + h_0 \, e^{-\ee t /\wh\z} \right) = \frac{\wh\f}2 \ee \left( 1 - 
e^{-\ee t /\wh\z} \right) \ ,
\eeq
the retarded response is
\beq
\MM_R(t,t') = \frac{\wh\f}2 \int^0_{-\io} \de h_0 \, e^{h_0} \, \frac{\ee^2}{\wh\z} e^{-\ee (t-t') /\wh\z} \, \th(t-t') = \frac{\wh\f}2 
\frac{\ee^2}{\wh\z} \, e^{-\ee (t-t') /\wh\z} \, \th(t-t') \ ,
\eeq
and the integrated response is
\beq
\wh \c (t,0) = \int_0^t \de s \, \MM_R(t,s) = \frac{\wh\f}2 \ee \left( 1 - e^{-\ee t /\wh\z} \right) = \k(t) \ ,
\eeq
which shows that $\wh \c(t,0) = \k(t)$ at all times in this limit. 
The noise correlation is
\beq
\MM_C(t,t') = \frac{\wh\f}2 \int^0_{-\io} \de h_0 \, e^{h_0} \, \ee^2 \,h_0^2 e^{-\ee (t+t') /\wh\z} = \wh\f \, \ee^2 \, e^{-\ee (t+t') /\wh\z} \ .
\eeq
Note that this correlation matrix is a projector, and as a consequence in the dilute limit the only randomness in the noise comes from its initial value, \ie $\Xi(t) = \Xi(0) e^{-\ee t /\wh\z}$.

\subsection{Methods for the numerical solution of the DMFT equations}
\label{sec:DMFT-inte}

 The DMFT equations can be solved by means of two 
alternatives methods~\cite{roy2019numerical,manacorda2020numerical,mignacco21jsm,folena2021marginal,MU21}:
\begin{itemize}
    \item An \textit{iterative} method, where we fix a maximum time $t_{\rm max}$ and we discretize the time with a step $\D t$, leading to
    a grid of $N_t = t_{\rm max}/\D t$ points. We then solve the self-consistent DMFT equations in the following way:
    \begin{enumerate}
        \item Start with an initial guess for the memory kernels in Eq.~\eqref{eqdio:ker}. 
        We typically set them to zero, as this corresponds to the dilute limit.
        \item Integrate the equations for the correlation and response functions in Eq.~\eqref{eqdio:corr}.
        \item Simulate the stochastic dynamics of Eqs.~\eqref{eqdio:y} to generate trajectories $y(t)$, and for each of them,
        integrate Eq.~\eqref{eqdio:H} to obtain  $H(t,t')$.
        \item Compute the new kernels in Eq.~\eqref{eqdio:ker} averaging over the stochastic dynamics.
        \item Repeat until convergence.
    \end{enumerate}
    This scheme relies on the structure of the self-consistent equations for the memory kernels, which can be written as a kind of low-density expansion, \ie
    \beq
    \begin{split}
    \left\{ \k_{i+1}(t), \MM_{R,i+1}(t,t'), \MM_{C,i+1}(t,t') \right\} &= \wh\f \, \FF[\left\{ \k_i(t), \MM_{R,i}(t,t'), \MM_{C,i}(t,t') \right\}] \\
    &\Downarrow \\ 
    \k(t) &= \wh\f \, \k^{(1)}(t) + \wh\f^2 \, \k^{(2)}(t) + \ldots \ ,
    \end{split}
    \eeq
    and so on for all the kernels, being $i$ the iteration step. The outcome of the first iteration 
    is then given by the dilute solution at $\wh\f\ll 1$ derived in section~\ref{sec:dilute}, as we explicitly checked in 
    the numerical implementation.
    This algorithm is the extension to the out-of-equilibrium case of the equilibrium algorithm 
    implemented in~\cite{manacorda2020numerical}.
    \item A \textit{step-by-step} method, where we exploit the causal structure of the DMFT equations as follows. First of all, 
    note that we only need to know the kernels for $t\geq t'$, either because they are symmetric under exchange of times (correlation functions),
    or because they vanish for $t < t'$ (response functions). Then, we can proceed as follows:
    \begin{enumerate}
        \item Set the initial value of the kernels using their analytical expression in 
        Eq.~\eqref{eqdio:ker} at $t=t'=0$.
        \item Integrate the first time step $\D t$ of the dynamical 
        Eqs.~\eqref{eqdio:y}, \eqref{eqdio:H} and~\eqref{eqdio:corr} with the memory kernels 
        computed above; in particular we obtain a sample of values of $h(\D t)$.
        \item From these, we can numerically compute the kernels at $(t=\D t, t'=0)$ and $(t=\D t, t' = \D t)$ by averaging over the 
        stochastic process.
        \item Repeat the procedure by extending at each step the stochastic trajectories $\{h(0), h(\D t), \ldots, h(t)\}$ to add a new point $h(t+\D t)$; from this,
        compute the kernels at subsequent times $t+\D t$ and $0 \leq t' \leq t+\D t$, given the 
        kernels at previous times.
    \end{enumerate}
\end{itemize}

The two methods are theoretically equivalent and so are the expected outcomes. Using them, we could achieve the numerical integration
of the DMFT equations for short trajectories of total duration $t \sim 10$ in natural 
units $\wh\z=\ee=1$. However, both methods become unreliable at long times because of 
a numerical instability in the generation of the correlated noise $\X(t)$ in Eq.~\eqref{eqdio:y}.
More precisely, the noise is generated in the following ways:
\begin{itemize}
    \item Iterative algorithm: solving the eigenproblem for the correlation matrix 
    $M_{ij}=\MM_C(i\D t,j\D t)$ at any iteration and generating the correlated noise $\X(t)$ independently before constructing each trajectory
    as
    \beq
    \X(i\D t) \equiv \X_i = \sum_{j=0}^{N_t-1} \sqrt{\l_j}\, v_i^j \, \x_j \ ,
    \eeq
    being $\l_j$ the $j$-th eigenvalue of the matrix $M$ and $v_i^j$ the $i$-th component of the 
    corresponding eigenvector, and $\x_j$ a Gaussian white noise of zero mean and covariance $\moy{\x_i \x_j} = \d_{ij}$.
    \item Step-by-step algorithm: in this case the matrix $M$ is constructed step-by-step by extending existing trajectories,
    so we need to
    generate the noise $\X_i$ at time $t=i\D t$ for each trajectory, from the knowledge of the 
    correlation matrix elements $M_{nm}$ at $n,m=0,\ldots,i$ and conditioned to the previous noise realization 
    $(\X_0,\ldots,\X_{i-1})$. This is done through the relation
    \beq
    P(\X_i \vert \X_0,\ldots,\X_{i-1}) \propto \exp \left[ -\frac12 \left(M^{-1}\right)_{ii} \X_i^2
    - \sum_{j=0}^{i-1} \left(M^{-1}\right)_{ij} \X_j \, \X_i \right] \ .
    \eeq
    Therefore $\X_i$ is Gaussian with mean 
    $\moy{\X_i} = -\dfrac{1}{(M^{-1})_{ii}} \sum_{j=0}^{i-1} (M^{-1})_{ij} \X_j$ and variance $1/(M^{-1})_{ii}$.
\end{itemize}

The numerical instability in the noise generation is related to the physical behavior of the 
trajectories; when the GD dynamics approaches its final state, 
the force due to the self-consistent bath converges to a constant value, either vanishing (unjammed phase) 
or finite (jammed phase). In both cases, the correlation matrix
develops almost-zero modes that lead to numerical instabilities either in the solution of the eigenproblem (iterative algorithm)
or in the inversion of the correlation matrix (step-by-step algorithm).
A second source of instability comes from the computation of memory kernels from Eq.~\eqref{eqdio:ker}: the 
trajectories flowing towards negative $h(t)$ starting from large positive $h_0$ have an exponential weight $e^{h_0}$ in the kernels, and are at the same time exponentially rare 
(as it should be to ensure convergence of the integration over $h_0$), thus leading to strong fluctuations in the numerical computation of the kernels. We 
were unfortunately not able to solve this problem. As a consequence of that, the numerical solution of the DMFT equations 
becomes unreliable at high density, and it is unable to describe the jamming transition properly, as we discuss in more details 
in sections~\ref{sec:num_dinf}.

\section{Numerical results}
\label{sec:num_dinf}

We now compare the  numerical solution
of the DMFT equations discussed in section~\ref{sec:DMFT_dinf}
with finite-dimensional numerical simulation results.
For a study of the GD dynamics in many-body (MB) soft harmonic particle systems in varying $d$ we refer to Refs.~\cite{OLLN02,OSLN03,chacko2019slow,IKBSH20,NIB21,nishikawa2021relaxation}.
Because finite-dimensional MB systems converge quite slowly to the asymptotic $d\to\io$ limit~\cite{CCPZ15,MZ16,charbonneau2021dimensional,sartor2021mean},
here we focus on the simpler Random Lorentz Gas (RLG)~\cite{biroli2021interplay,biroli2021mean}, 
which is a single particle tracer with $d$ degrees of freedom embedded in a sea of random obstacles. 
In the limit $d\to\infty$, the MB problem can be mapped onto the RLG~\cite{biroli2021interplay,biroli2021mean}, via a simple rescaling
described in appendix~\ref{app:equivalence}.
In short, a given value of density $\wh\f$ in the RLG corresponds to twice that value in the MB problem. At corresponding densities,
one-time observables such as the energy and isostaticity index in Eq.~\eqref{eq:one-time-q} have equal values, the memory function of the MB problem is twice that of the RLG, and the MSD of the MB problem is half that of the RLG.
Finally, the friction coefficient of the RLG has to be set as half that of the MB problem.
All the results presented in this section will be expressed in RLG units, \ie the DMFT results
have been rescaled by appropriate factors as described above.

\begin{figure}[b]
%\includegraphics[width=.45\textwidth]{traj_f1_d2_A7_lowmsd}
%~
\includegraphics[width=.3\textwidth]{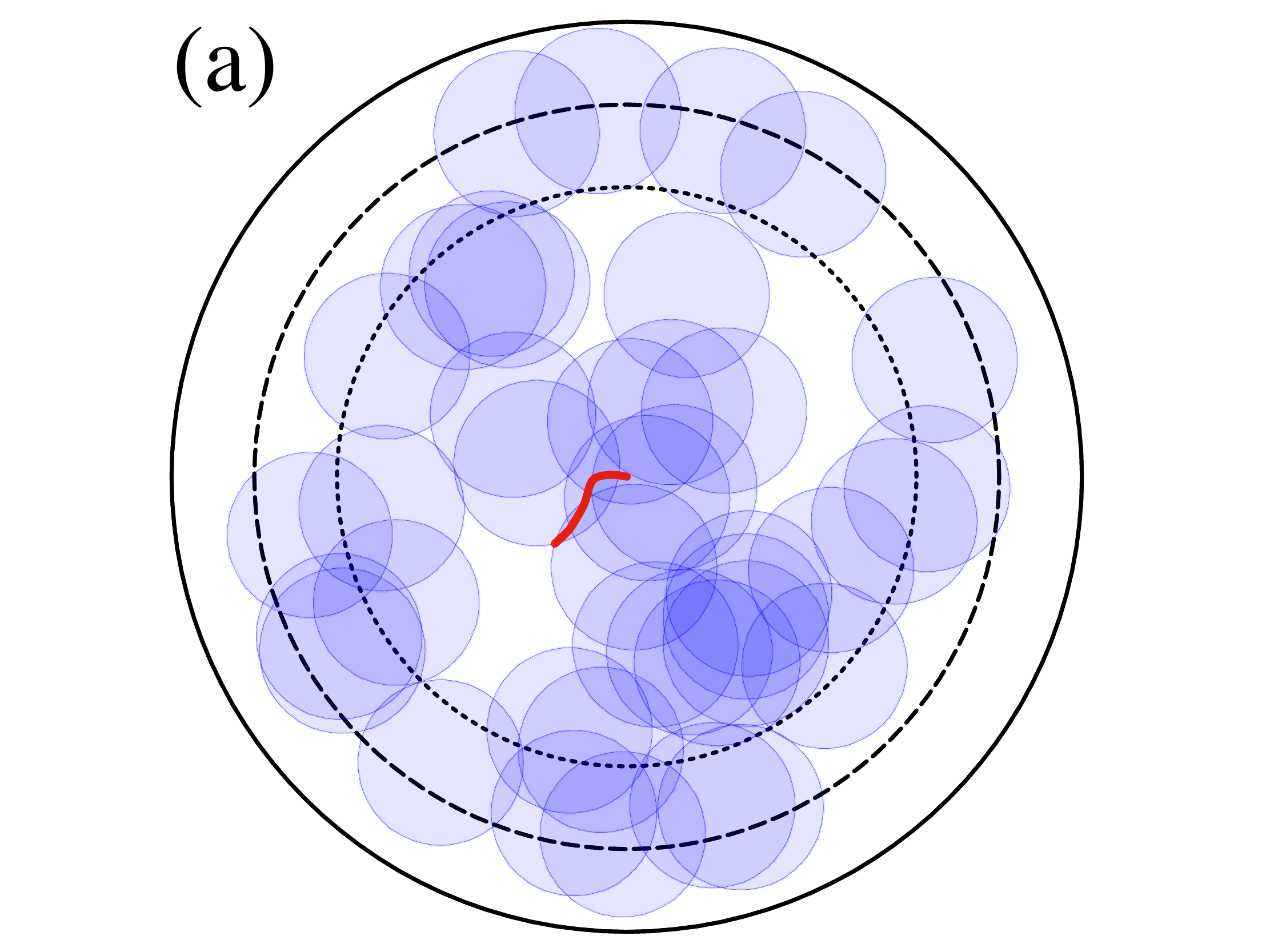}
\includegraphics[width=.3\textwidth]{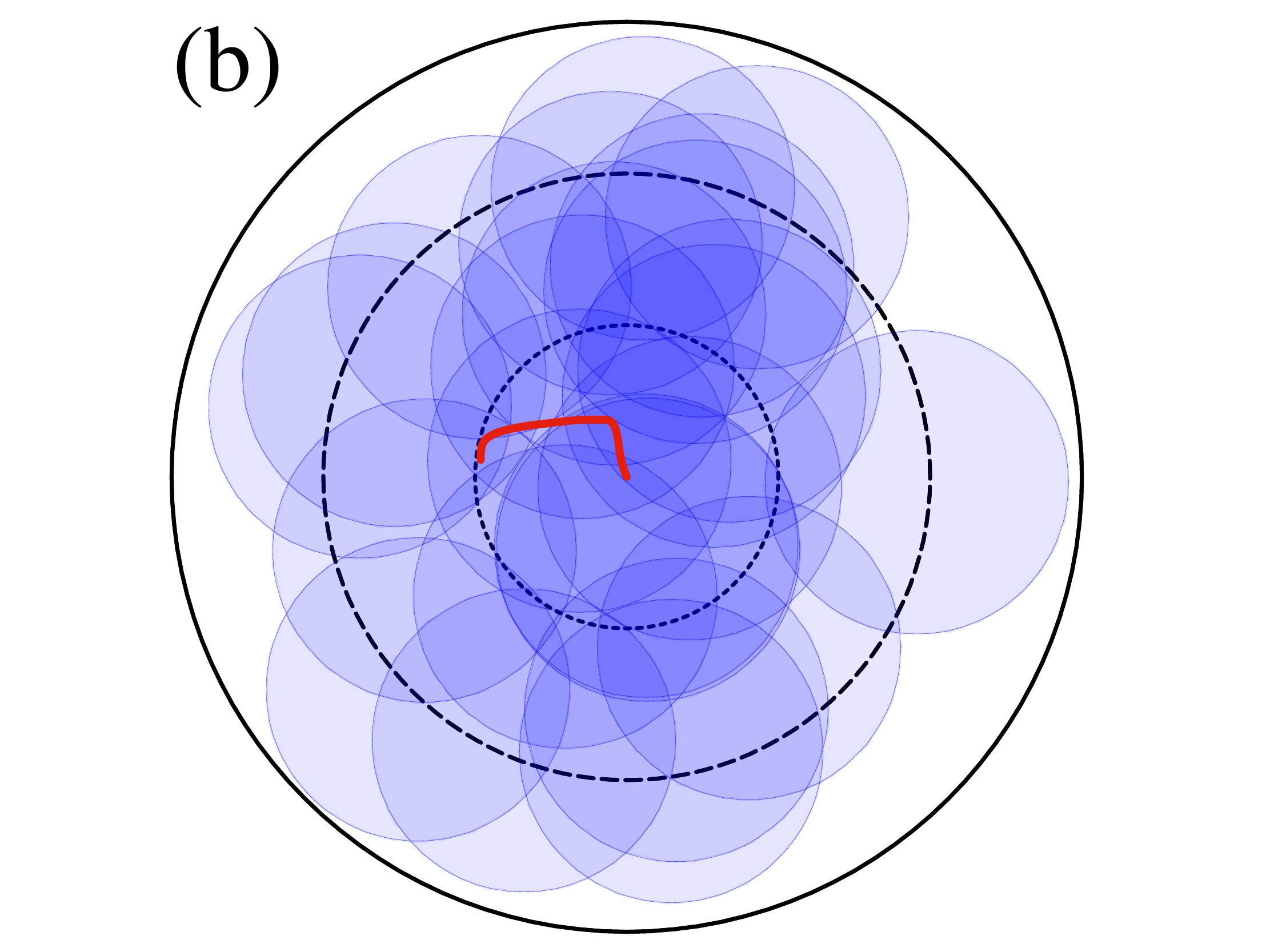}
\includegraphics[width=.3\textwidth]{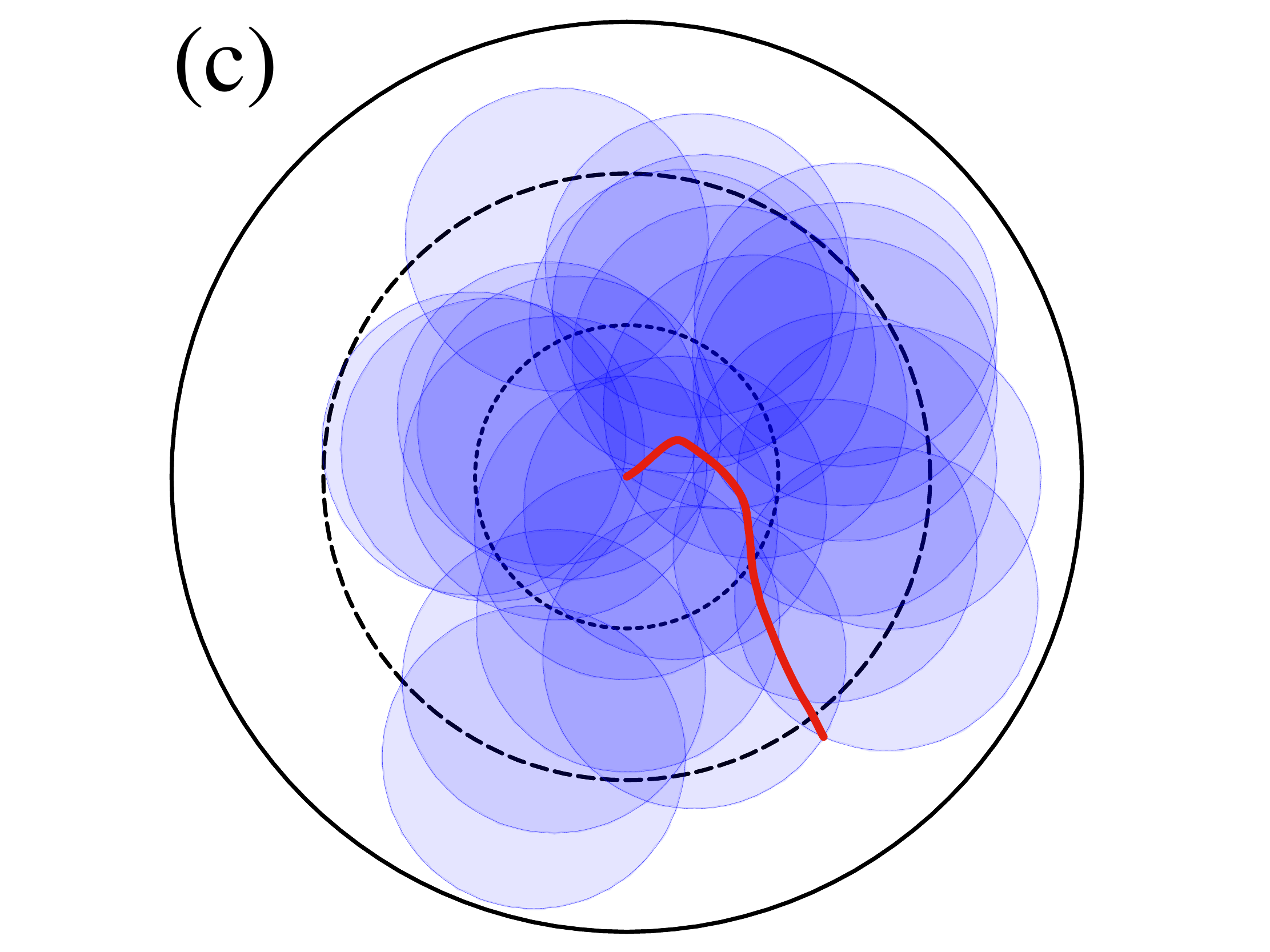}
\caption{Illustration of three typical trajectories in the 2$d$ RLG, 
with $\wh\f=1, A=7$ (a), $\wh\f=3, A=2$ (b,c). In panels a (unjammed phase) and b (jammed phase), the tracer descends the energy landscape until 
it reaches a local minimum where it gets arrested, and the trajectory never escapes the 2$d$ 
sphere of radius $A/d$. In panel c, the tracer reaches the zero-energy endpoint in a region 
outside the cutoff radius, therefore leading to an unphysical trajectory for this problem.
The obstacle radius is $\ell=1$, the dotted, dashed and full circle have radius $A/d$, $1+A/d$ and $2+A/d$ respectively.
}
\label{fig:traj}
\end{figure}

\subsection{Random Lorentz Gas}
\label{sec:RLG-num}

\begin{figure}[t]
\includegraphics[width=.4\textwidth]{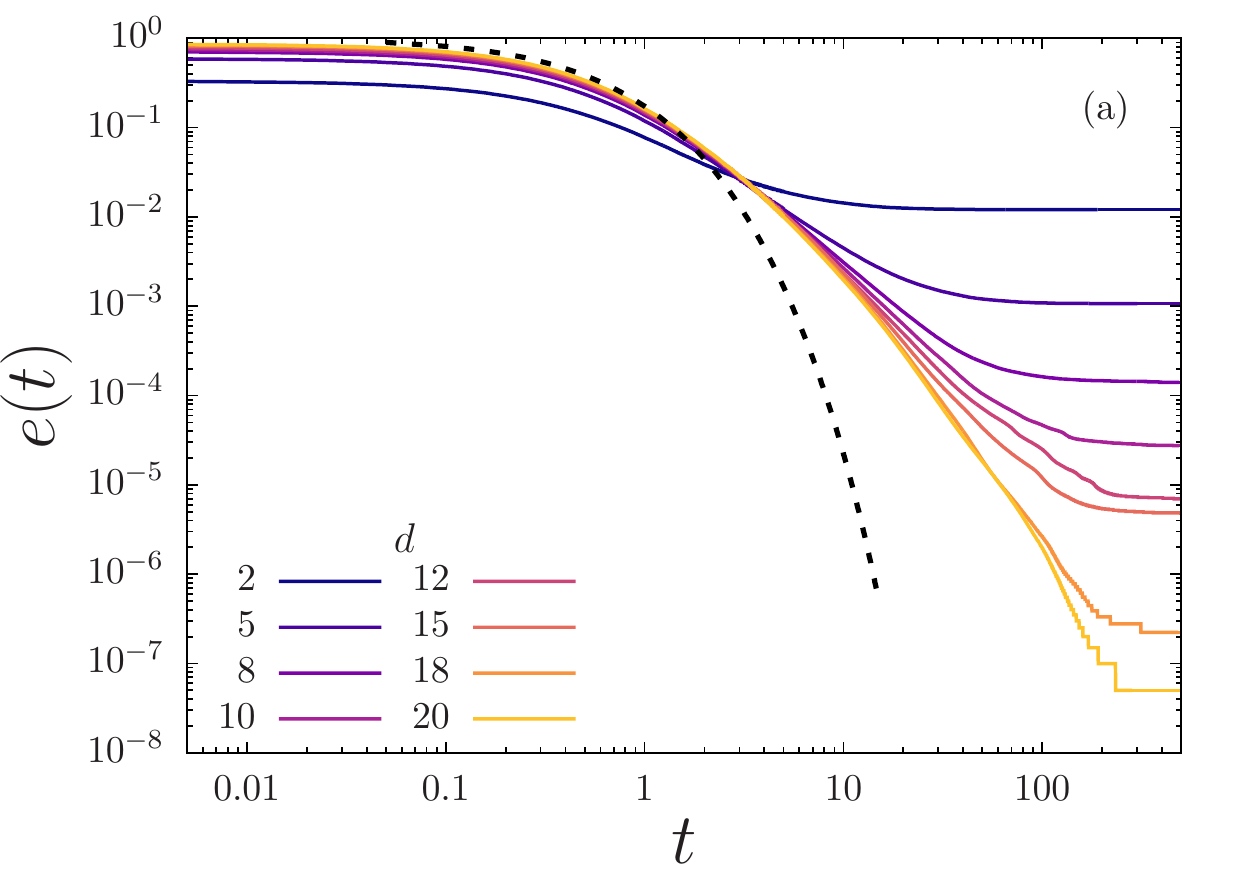}
~
\includegraphics[width=.4\textwidth]{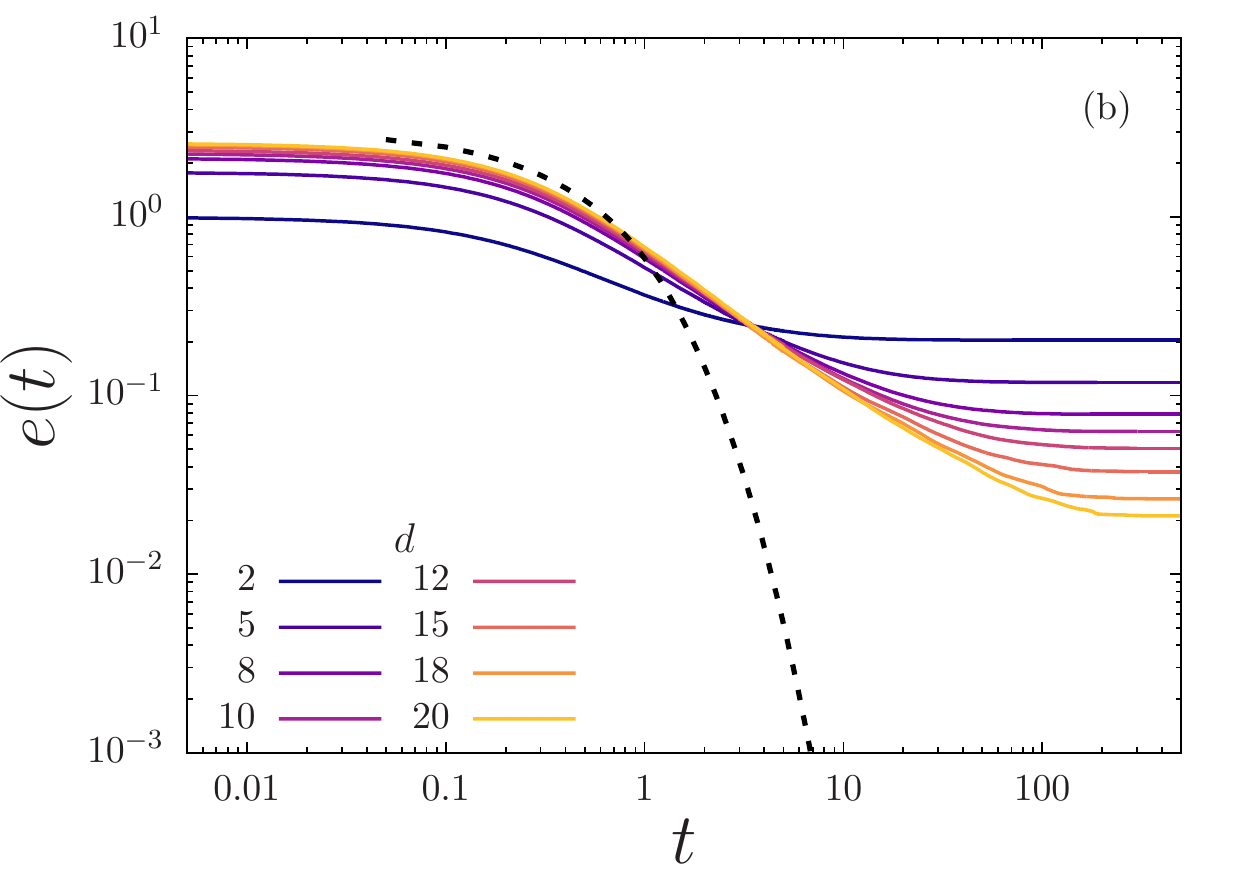}
~
\includegraphics[width=.4\textwidth]{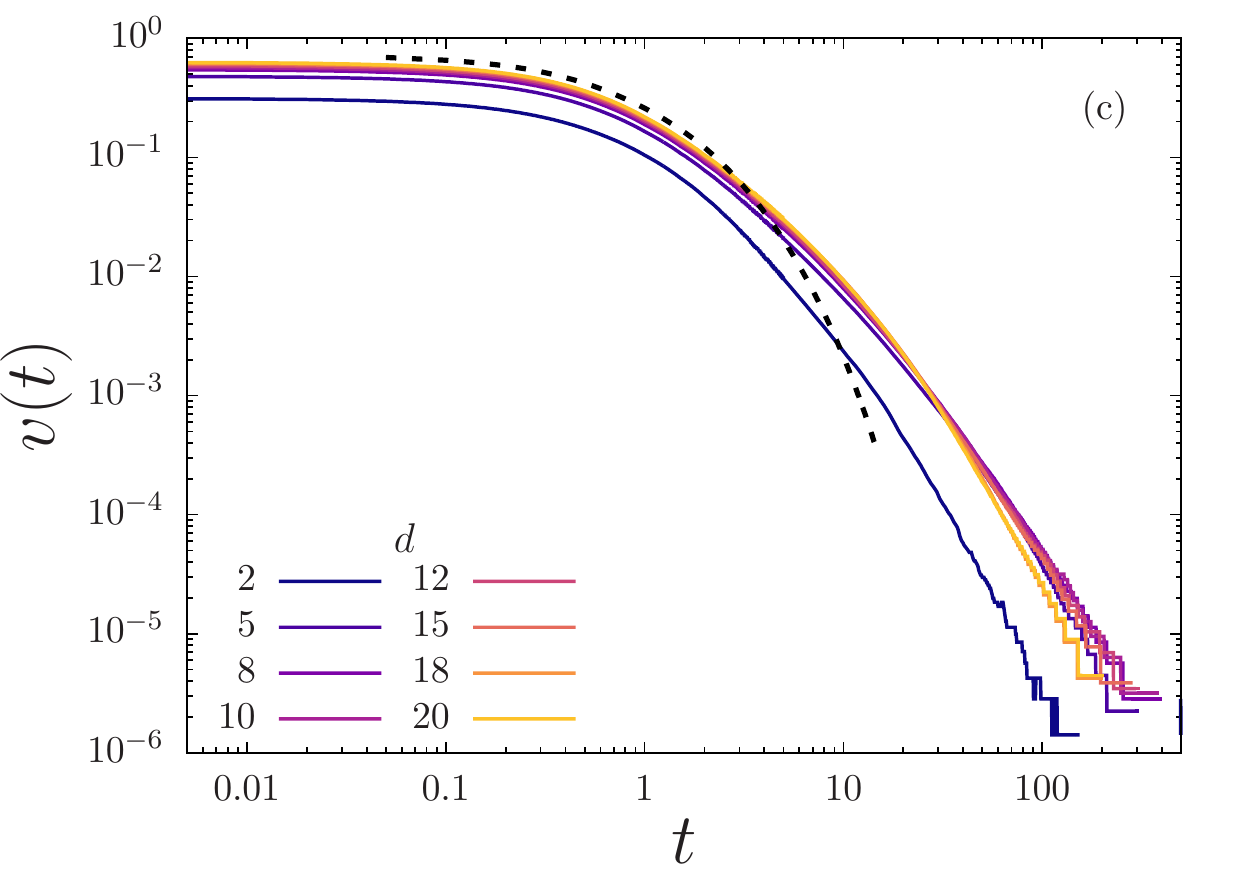}
~
\includegraphics[width=.4\textwidth]{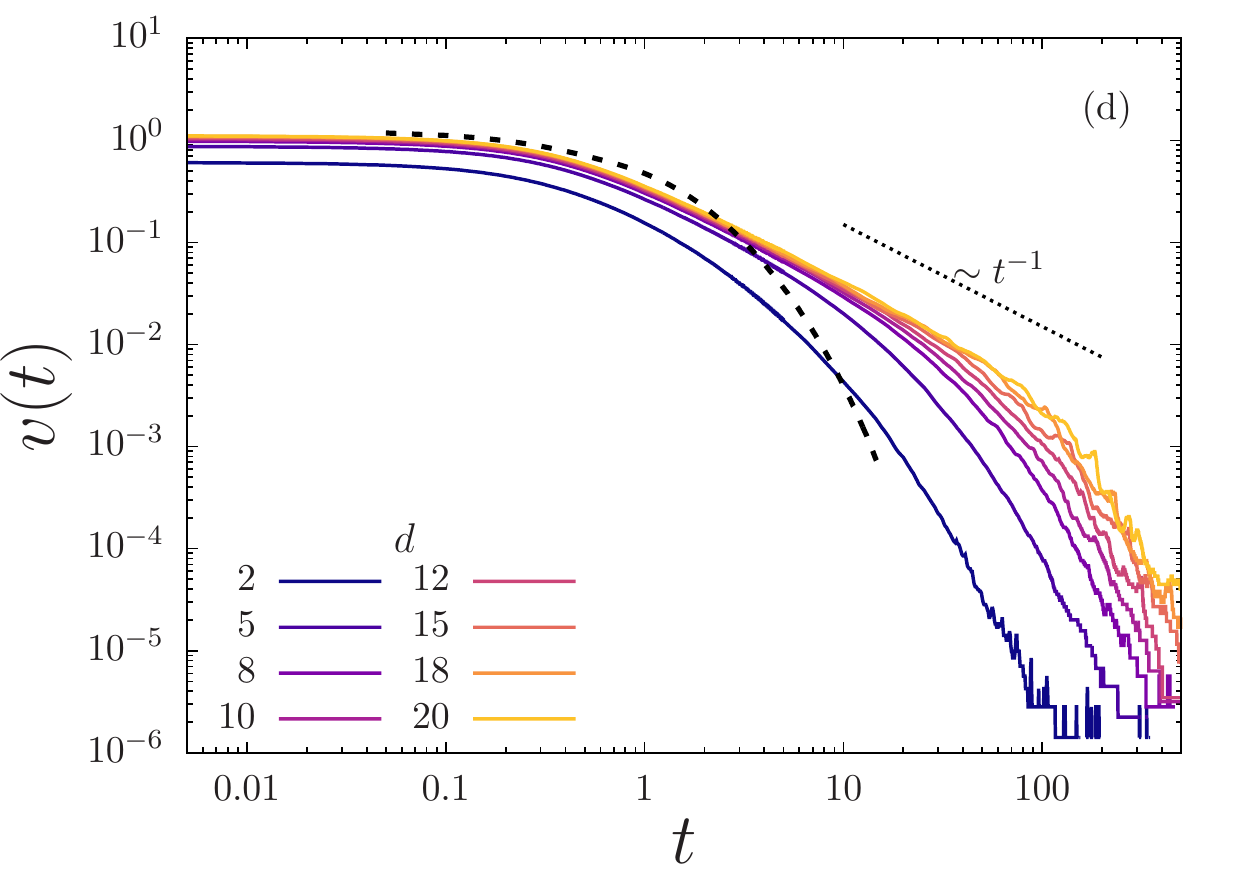}
~
\includegraphics[width=.4\textwidth]{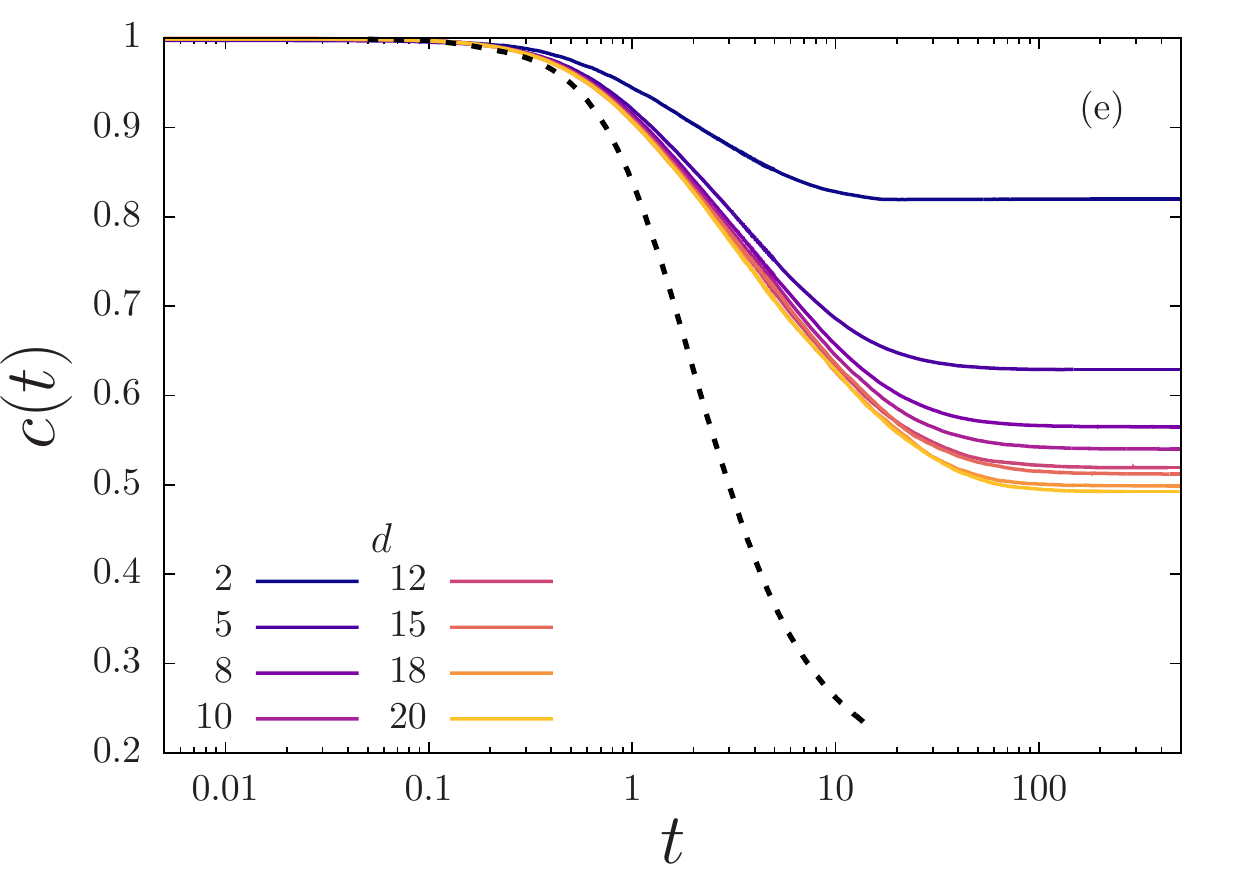}
~
\includegraphics[width=.4\textwidth]{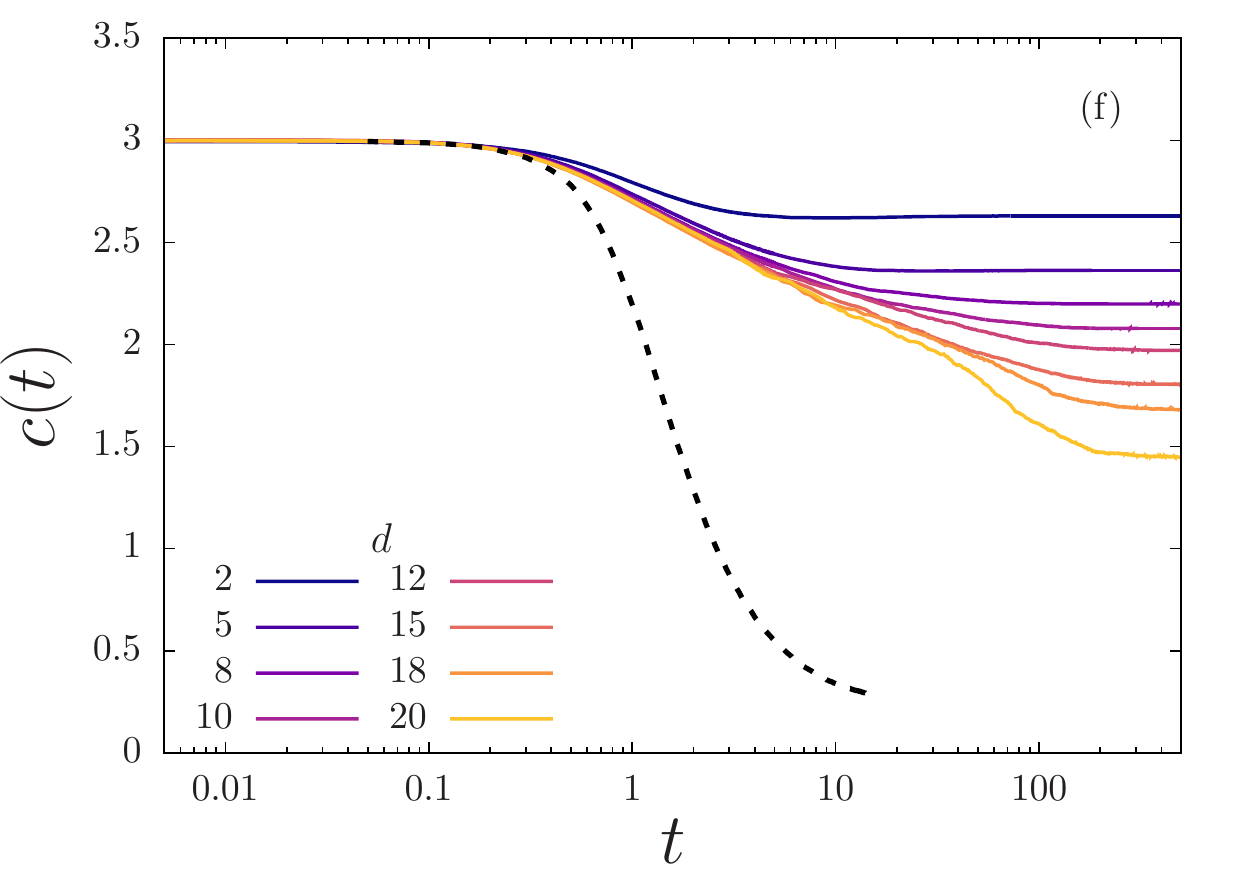}
~
\includegraphics[width=.4\textwidth]{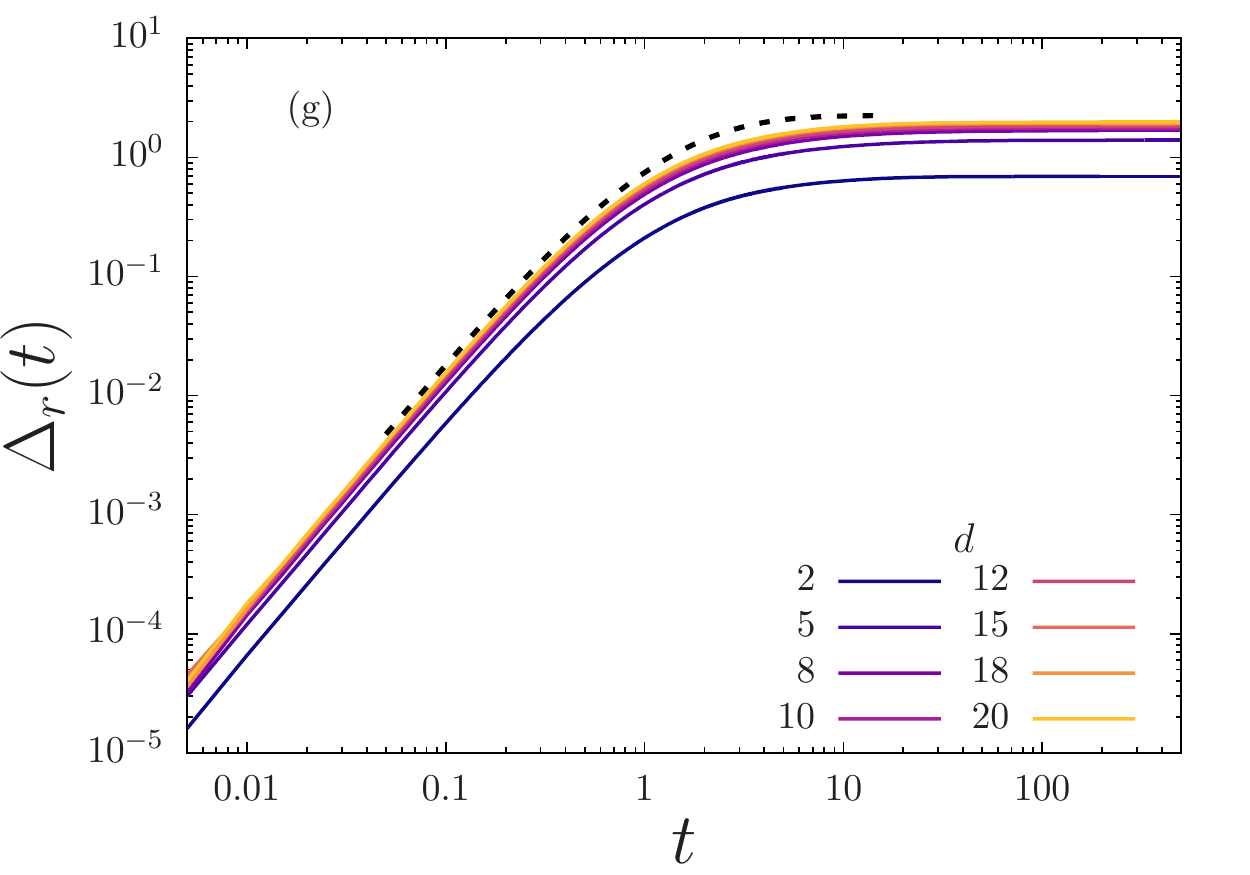}
~
\includegraphics[width=.4\textwidth]{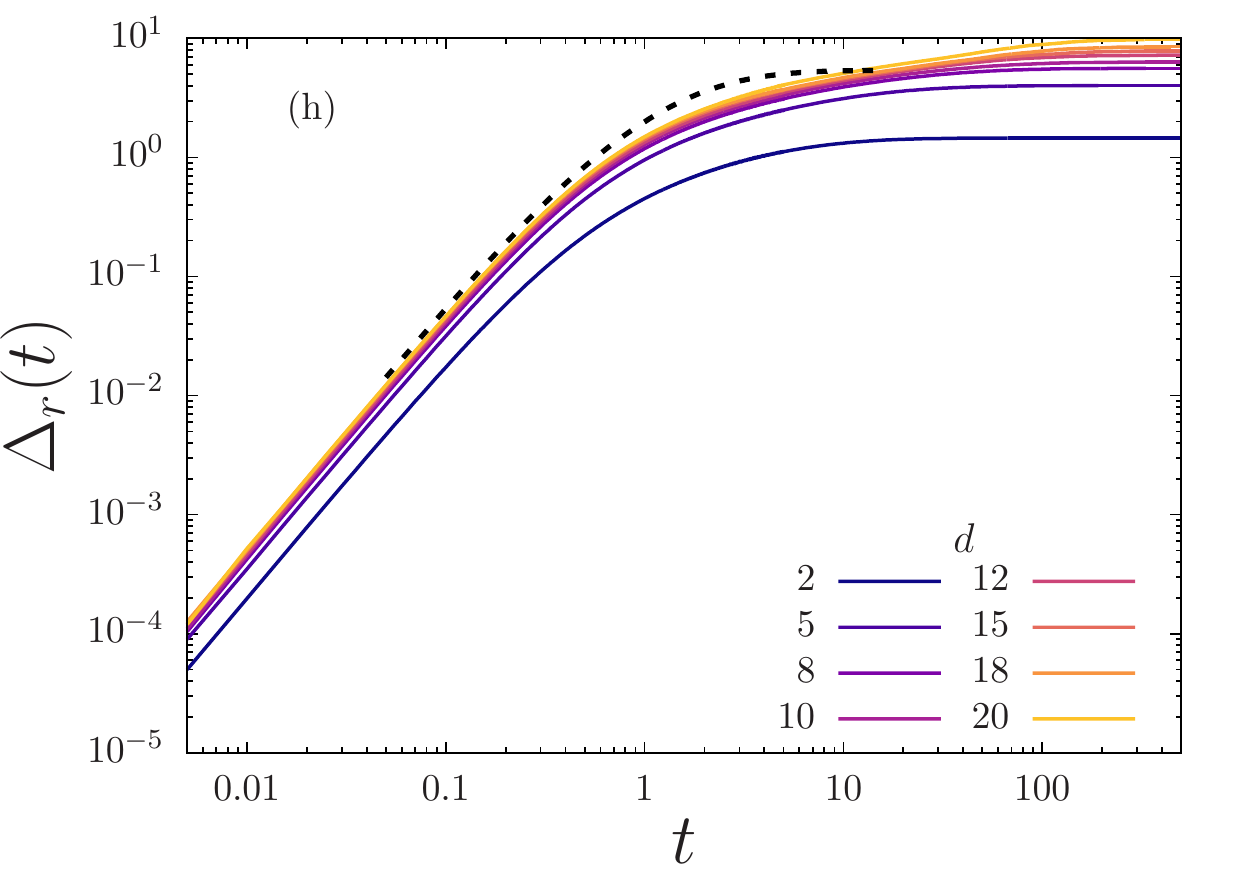}
\caption{Time dependence of $e(t)$, $v(t)$, $c(t)$ and $\D_r(t)$ for the RLG, as obtained from the numerical solution of the DMFT equations in $d\to\io$ (black dashed line: iterative method; 
the step-by-step method coincides at short times but becomes unreliable around $t\sim1$) 
and from the direct numerical simulation at finite $d$ (colored lines). The finite-$d$ data converge to the DMFT predictions at short times, and converge upon increasing $d$ at all times, but the DMFT predictions are not reliable at long times, as also evidenced by the discrepancy between the two DMFT solution methods (not shown). At low density $\wh\f=1$, the energy (a) and velocity (c) both decay exponentially, the isostaticity index (e) converges to a value $c_\io<1$, and the MSD (g) goes to a finite plateau (unjammed phase). At large density $\wh\f=3$, the energy (b) goes to a finite plateau, the velocity (d) decays as a power law over a range of times that increases with $d$, the long-time isostaticity index (f) is $c_\io >1$, and the MSD (h) keeps increasing slowly with time when $d$ increases (jammed phase).
}
\label{fig:RLG-DMFT}
\end{figure}

We simulate the dynamics of a point-like tracer moving in a $d$-dimensional space occupied 
by $N$ obstacles of radius $\ell$; the positions $\mathbf{X}_i$ ($i=1,\ldots,N$) of the latter are drawn
independently and uniformly in space. Because of translational invariance, we can assume that the tracer starts at the origin without loss of generality.
The microscopic dynamics of the tracer position $\xx(t)$ then reads
\beq\label{eq:RLG_dyn}
\z \dot \xx(t) = - \nabla V (\xx(t)) \ , \qquad V(\xx) = \sum_{i=1}^N v(\vert \xx - \mathbf{X}_i \vert) \ , \qquad \xx(0)=0 \ .
\eeq
This dynamics corresponds to the zero-noise limit of Ref.~\cite[Eq.~(30)]{biroli2021mean}.
We analyze the case of a soft sphere interaction potential $v(r) = (d^2 \ee/2) (r/\ell-1)^2 
\th(\ell-r)$, whose infinite-dimensional limit is equivalent to the potential $\redv(h) = 
(\ee/2) h^2 \th(-h) $ defined in section~\ref{sec:DMFT_dinf}. The friction coefficient must also 
be scaled with dimension through the relation $ \z = (2 d^2/\ell^2) \wh\z$, with $\wh\z$ remaining finite in 
the infinite-dimensional limit. We fix $\ee=1$, $\wh\z=1/2$ 
(in such a way that the RLG value is half that of the MB problem, see appendix~\ref{app:equivalence})
and $\ell=1$ without loss of generality, because these choices define a time and length unit. Under these assumptions, the dynamics read
\beq\label{eq:RLG_dyn_dinf}
\dot \xx(t) = -\frac1{d} \sum^N_{i=1} \redv'(h_i) \, \hat{\rr}_i = -\frac1{d} \sum^N_{i=1} h_i \th(-h_i) \, \hat{\rr}_i  \ , \qquad \hat{\rr}_i = \dfrac{\xx-\mathbf{X}_i}{\vert \xx-\mathbf{X}_i \vert} \ , \qquad h_i = d \left(\frac{\vert \xx - \mathbf{X}_i \vert}\ell -1 \right) \ .
\eeq
The dynamics is 
therefore deterministic once the obstacles have been drawn. The number of obstacles represents the 
main computational challenge to the simulation. We draw obstacles in a spherical region of radius $R$, which then provides a cutoff 
on the maximal distance of an obstacle from the origin where the tracer starts its dynamics. This 
cutoff is justified if the displacement of the tracer's trajectory from the origin never exceeds $R-\ell$, meaning that the 
tracer always explores a portion of space with a uniform obstacle density. The number of obstacles 
is then $N = \r V= d \wh\f (R/\ell)^d$, being $\rho = d \wh\f / (V_d \ell^d)$ and $V = V_d R^d$ the 
volume of the $d$-dimensional sphere of radius $R$. We therefore define $R=\ell(1+A/d)$, introducing 
the cutoff parameter $A$, so that at fixed $A$ the number of obstacles grows linearly rather than exponentially with $d$, namely $N \approx d \wh\f e^A$. This choice will be justified 
\textit{a posteriori} through the distribution of the tracer's MSD.

The simulations are run by numerical integration of Eq.~\eqref{eq:RLG_dyn_dinf} using the Euler scheme with a 
fixed time step ${\D t=10^{-3}}$. The trajectories continue until the velocity becomes smaller than a 
threshold value, typically $\vert \dot \xx(t) \vert < 10^{-8}$, where we assume that a 
local minimum has been reached.
Three examples of 2$d$ trajectories are shown in Fig.~\ref{fig:traj}, illustrating the descent 
of the tracer in the obstacles' energy landscape (both in the unjammed and jammed phases),
and the escape from the cutoff radius in a high 
density trajectory. The dependence on the cutoff will be discussed in section~\ref{sec:cutoff}.

\subsection{Time-dependent results}
\label{sec:timedep}

In Fig.~\ref{fig:RLG-DMFT} we plot the time 
evolution of the average (over the realizations of the obstacles) MSD $\D_r(t) = d \la \vert \xx(t) \vert^2 \ra$, energy $e(t) = \frac1{2d} \la \sum_i h_i^2 \th(-h_i)\ra$, velocity 
$v(t)=\sqrt{- \dot e(t)}$ and 
isostaticity index $c(t) = \frac1d \la \sum_i  \th(-h_i)\ra$, at two representative density values $\wh\f=1.0$ and $3.0$, which are expected to be 
respectively below and above the jamming transition~\cite{parisi2020theory}. In all cases, we observe that at very short times, the two DMFT solution algorithms described in section~\ref{sec:DMFT-inte} coincide, and
correctly describe the large-$d$ limit of the numerical data. Unfortunately, however, the agreement between the two DMFT solution schemes quickly worsen upon increasing time, 
and the DMFT numerical 
solution is thus not reliable at intermediate and long times. While we were unable to overcome this numerical limitation, we can still compare some exact asymptotic long-time
predictions of DMFT with numerical data, see section~\ref{sec:mem-corr}.

From the finite-$d$ simulation we can obtain insight on the behavior of the system (both MB, RLG and DMFT) in the $d\to\io$ limit.
Let us discuss first the case $\wh\f=1$. We observe that the numerical data converge quite well upon increasing $d$, such that $d\approx 20$ with a cutoff $A\approx 7$ seems
representative of the asymptotic limit. We observe that the energy and velocity both decay exponentially, 
and the isostaticity index converges to a value $c_\io<1$. This is consistent with the expectation in the unjammed phase~\cite{IKBSH20,NIB21}.
Furthermore, the MSD goes to a finite plateau, indicating that the tracer relaxes over a few valleys before settling on the boundary of a final ``lake'' of zero energy, and that a lake can always be
found at finite distance from any initial condition.

Conversely, for $\wh\f=3$, the energy approaches a finite plateau at long times, while the velocity seems to decay as a power law, and the isostaticity index converges to a value
$c_\io >1$, consistently with the expectation for a jammed phase~\cite{chacko2019slow,nishikawa2021relaxation}. In this case, while for any finite $d$ the MSD seems to converge
to a finite plateau, the value of the plateau keeps increasing with dimension~\cite{nishikawa2021relaxation}. Furthermore, a slow time-dependence of the MSD seems to appear upon
increasing $d$. This suggests the existence of persistent aging with weak ergodicity breaking (\ie loss of the memory of the initial condition)~\cite{bouchaud1992weak} in the jammed phase~\cite{chacko2019slow,nishikawa2021relaxation}, 
akin to that of $p$-spin models~\cite{CK93,FFR20}. Because in the infinite-dimensional 
RLG the only non-trivial correlation is the MSD, weak ergodicity breaking implies 
that the MSD must keep growing at all times. Our data are, however, not conclusive on this issue, because the growth is modest and it is not clear whether it persists at longer times.

\begin{figure}[b]
\includegraphics[width=.3\textwidth]{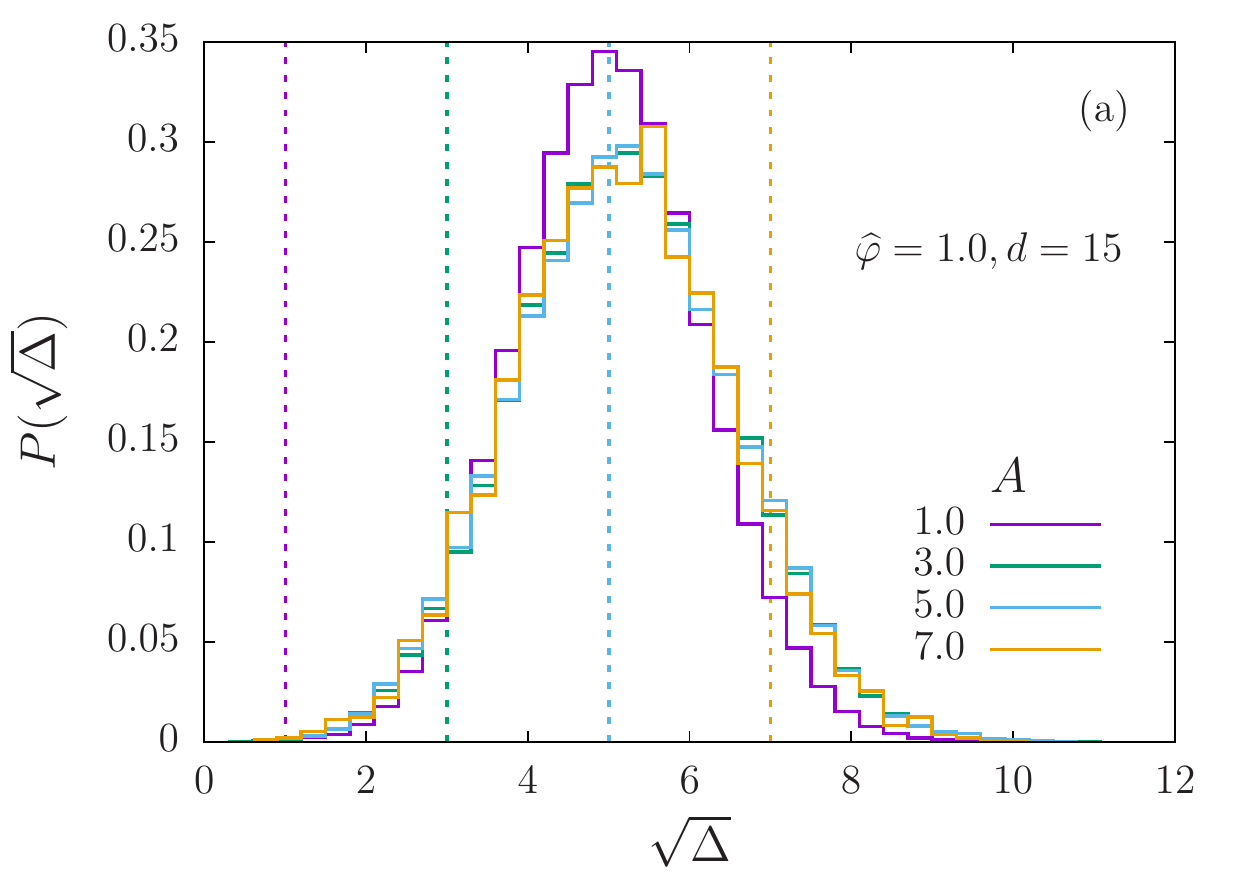}
\includegraphics[width=.3\textwidth]{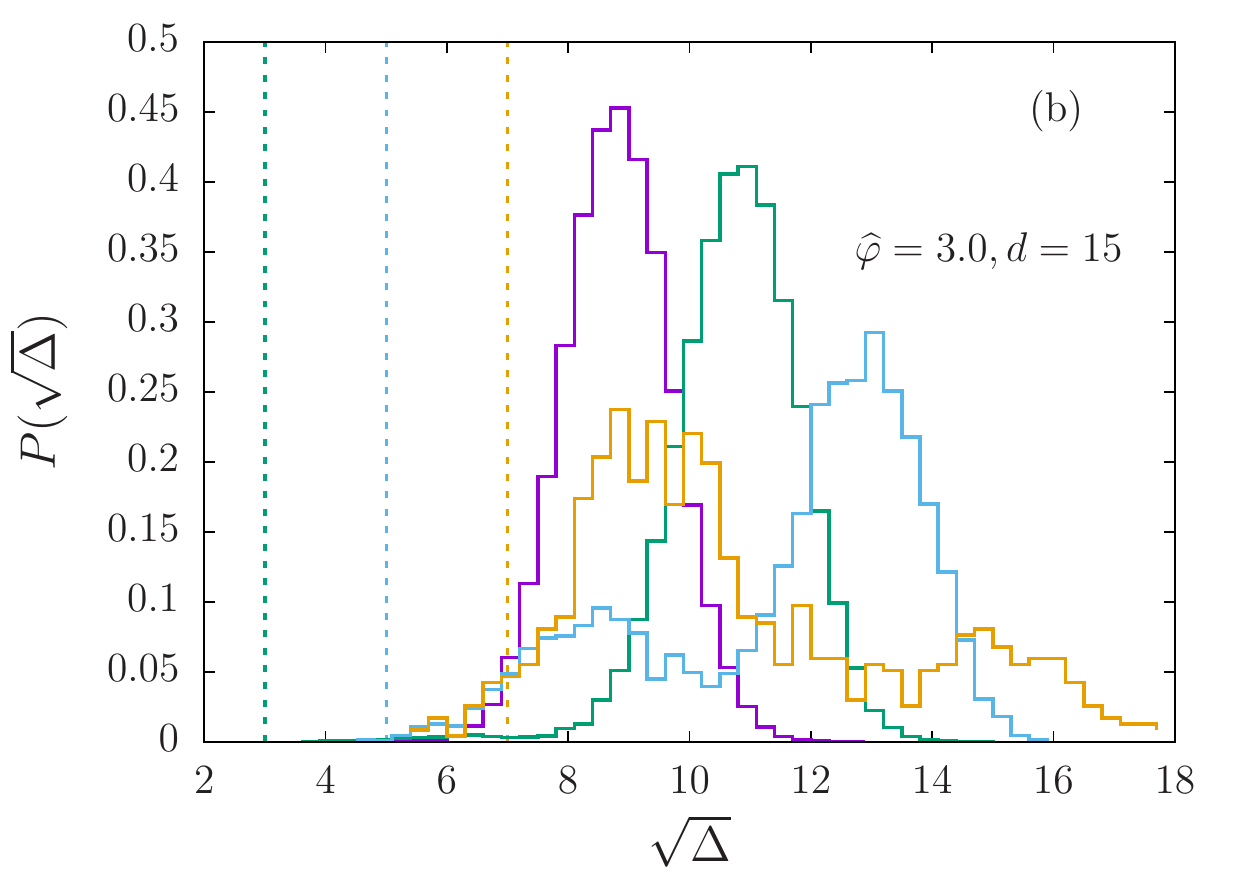}
\includegraphics[width=.3\textwidth]{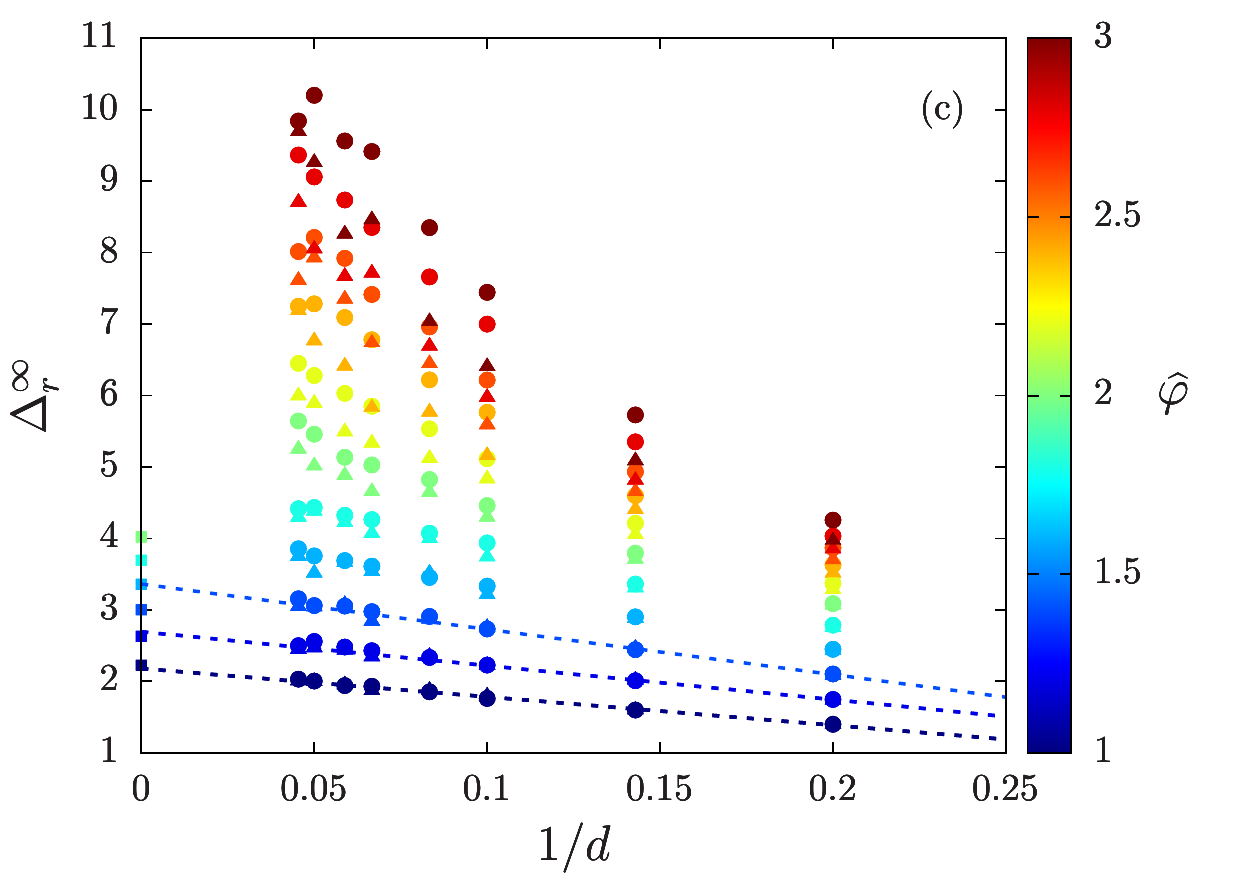}
\caption{Probability distribution of the displacement $\sqrt{\D} = \sqrt{d} \,  \vert \xx_\io \vert$ in the RLG at $\wh\f=1$ (a) and $\wh\f=3$ (b), 
$d=15$ and several values of $A$. At low densities the distributions converge upon increasing the cutoff when $A\geq3.0$, 
while at high densities the distribution has not 
converged at $A=9.0$ and two peaks emerge. The vertical lines are placed at $x = A/\sqrt{d}$ with corresponding color codes. 
Panel c shows the average MSD versus $1/d$ for several values of $\wh\f$ with $A=5$ (squares) and $A=7$ (circles). The 
points at $1/d=0$ are estimated from DMFT and the dashed lines are linear fits of the MSD at finite $d$.}
\label{fig:cutoff}
\end{figure}

\begin{figure}[t]
\includegraphics[width=.45\textwidth]{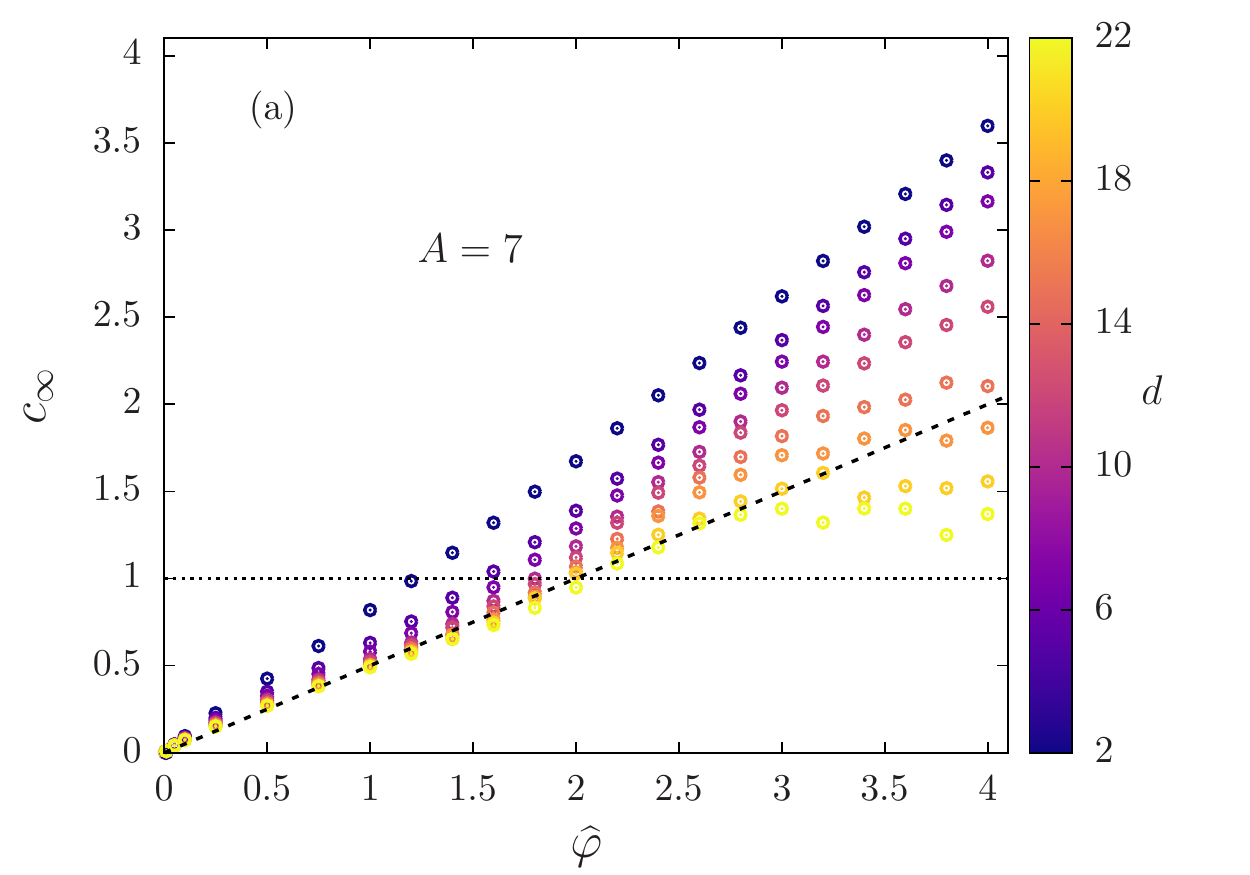}
\includegraphics[width=.45\textwidth]{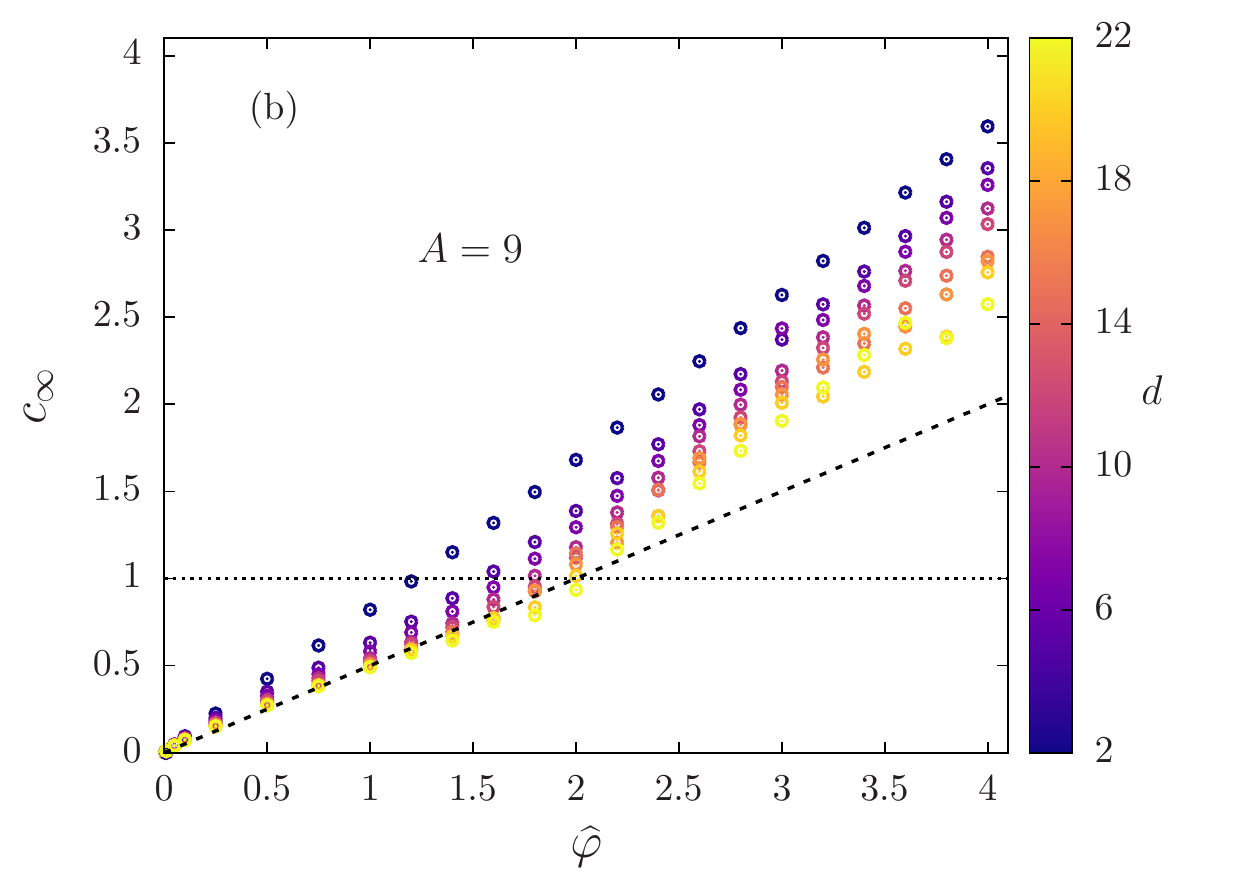}
\includegraphics[width=.45\textwidth]{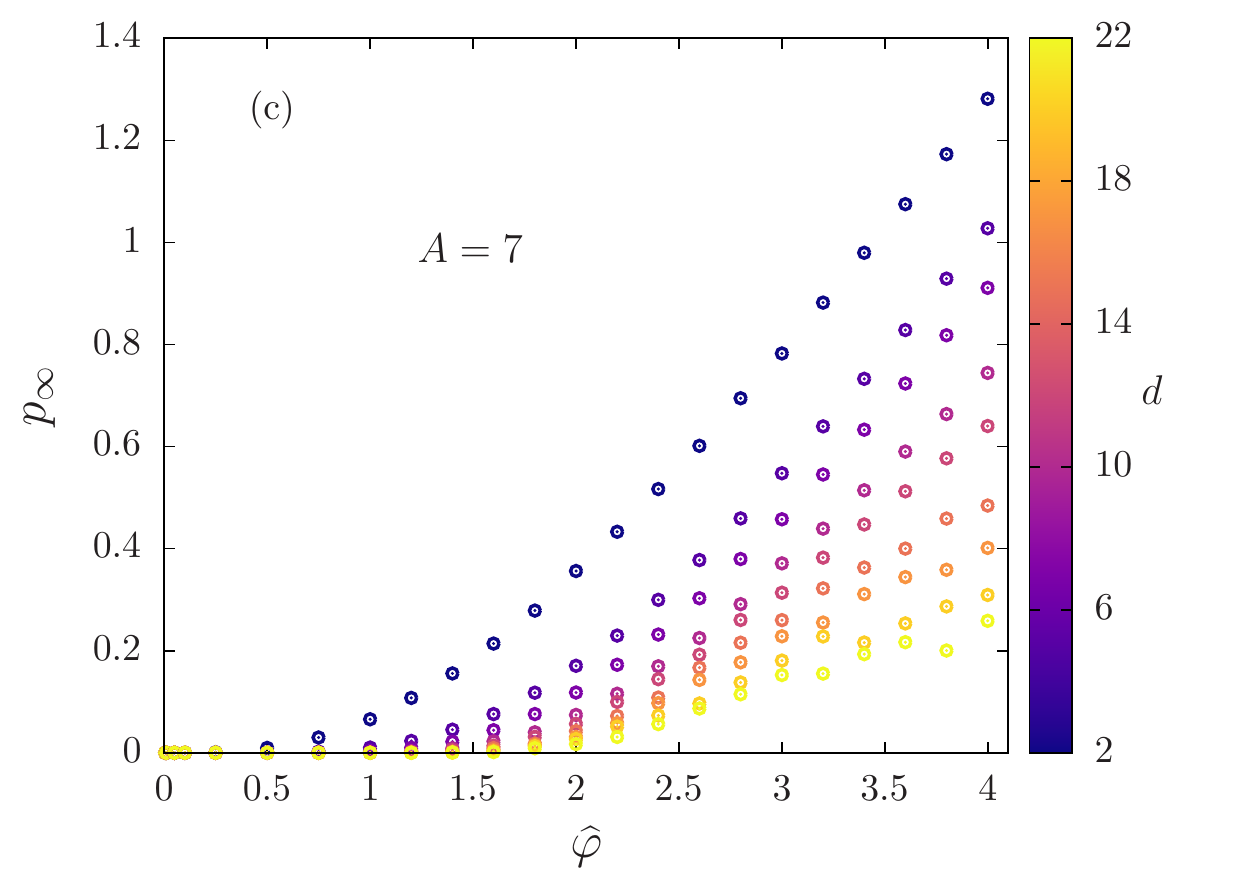}
\includegraphics[width=.45\textwidth]{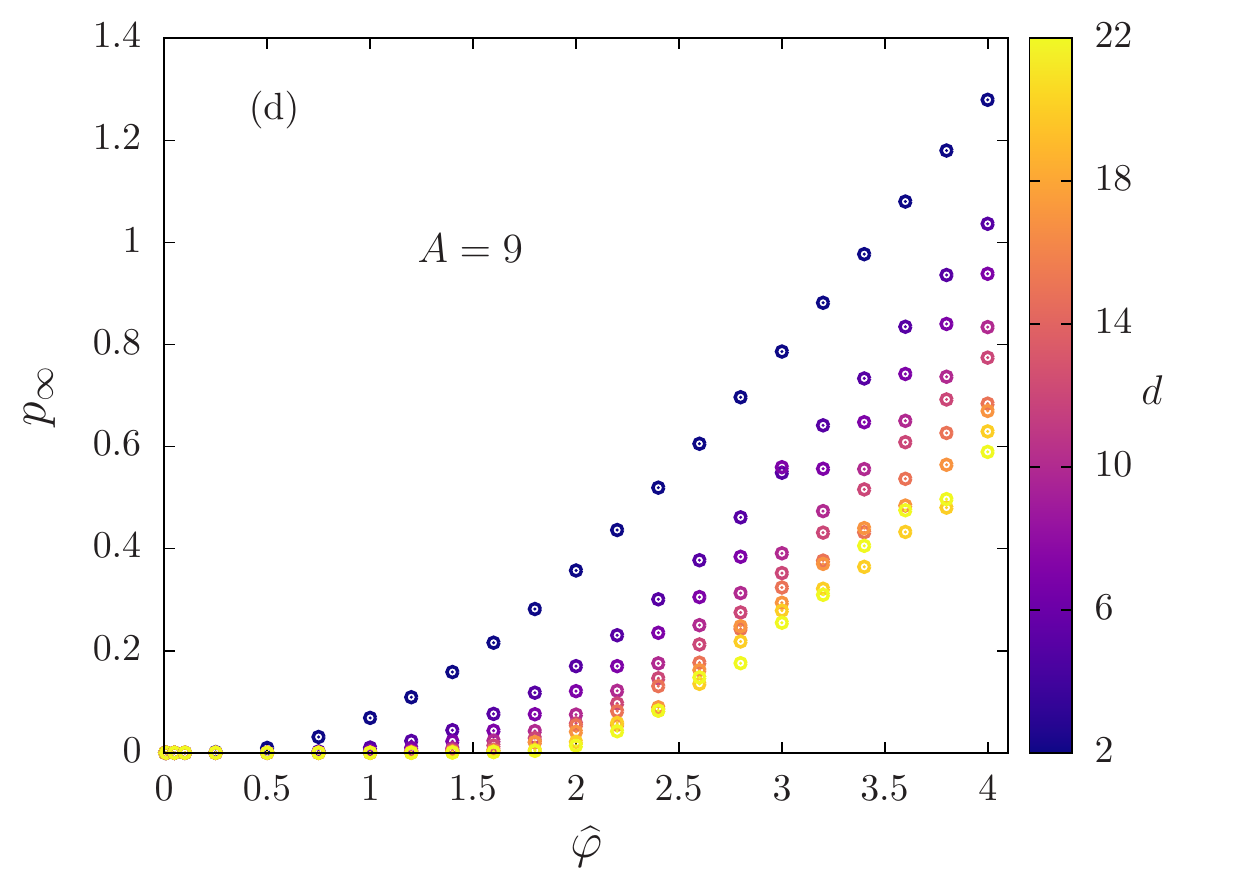}
\includegraphics[width=.45\textwidth]{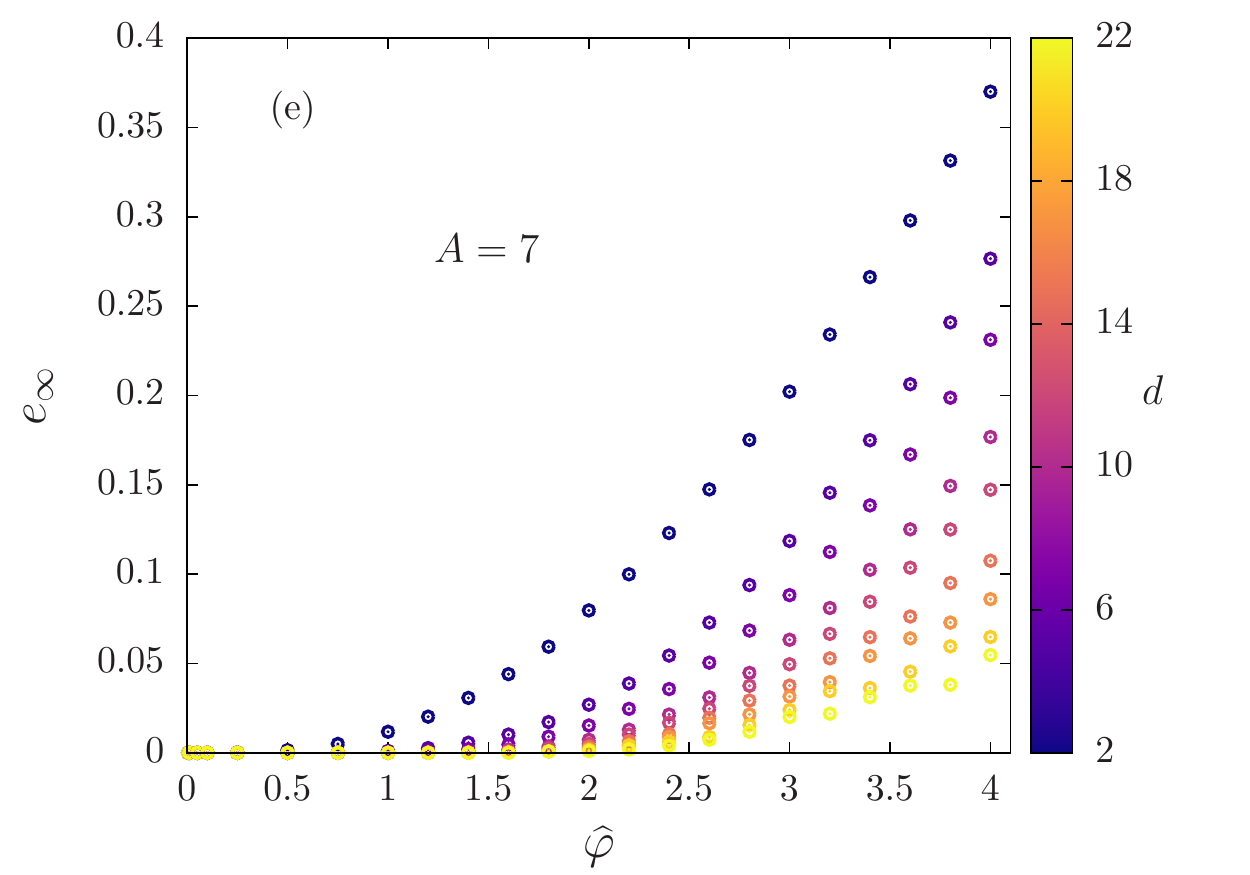}
\includegraphics[width=.45\textwidth]{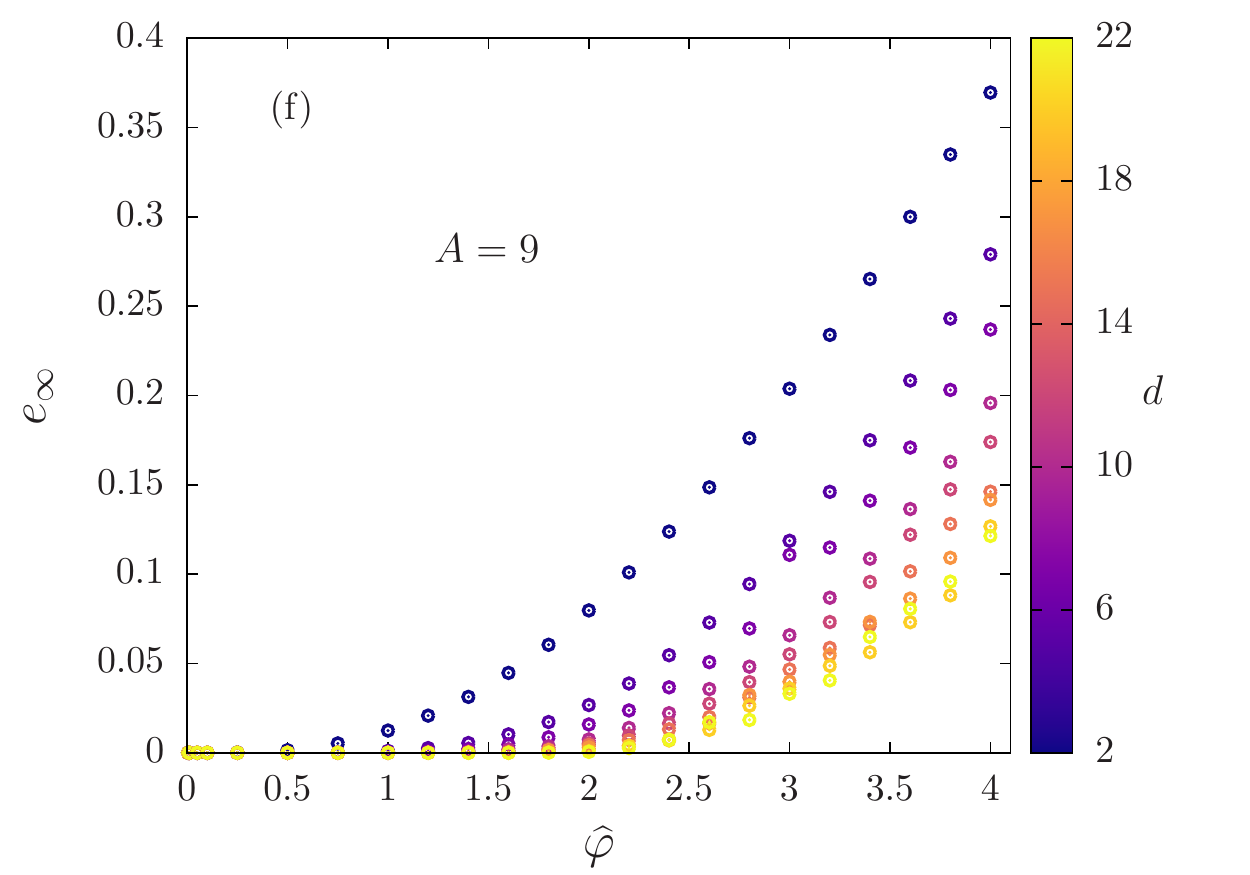}
\includegraphics[width=.45\textwidth]{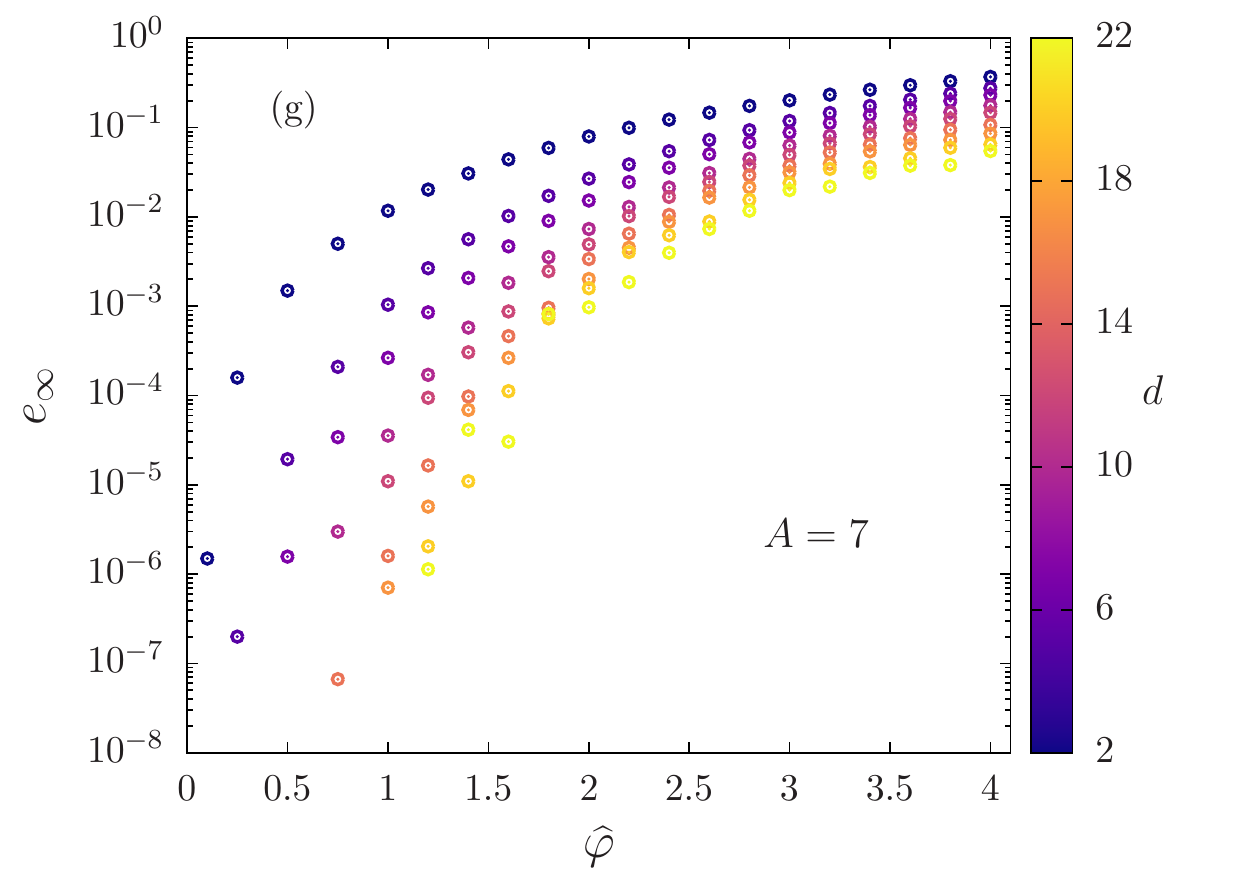}
\includegraphics[width=.45\textwidth]{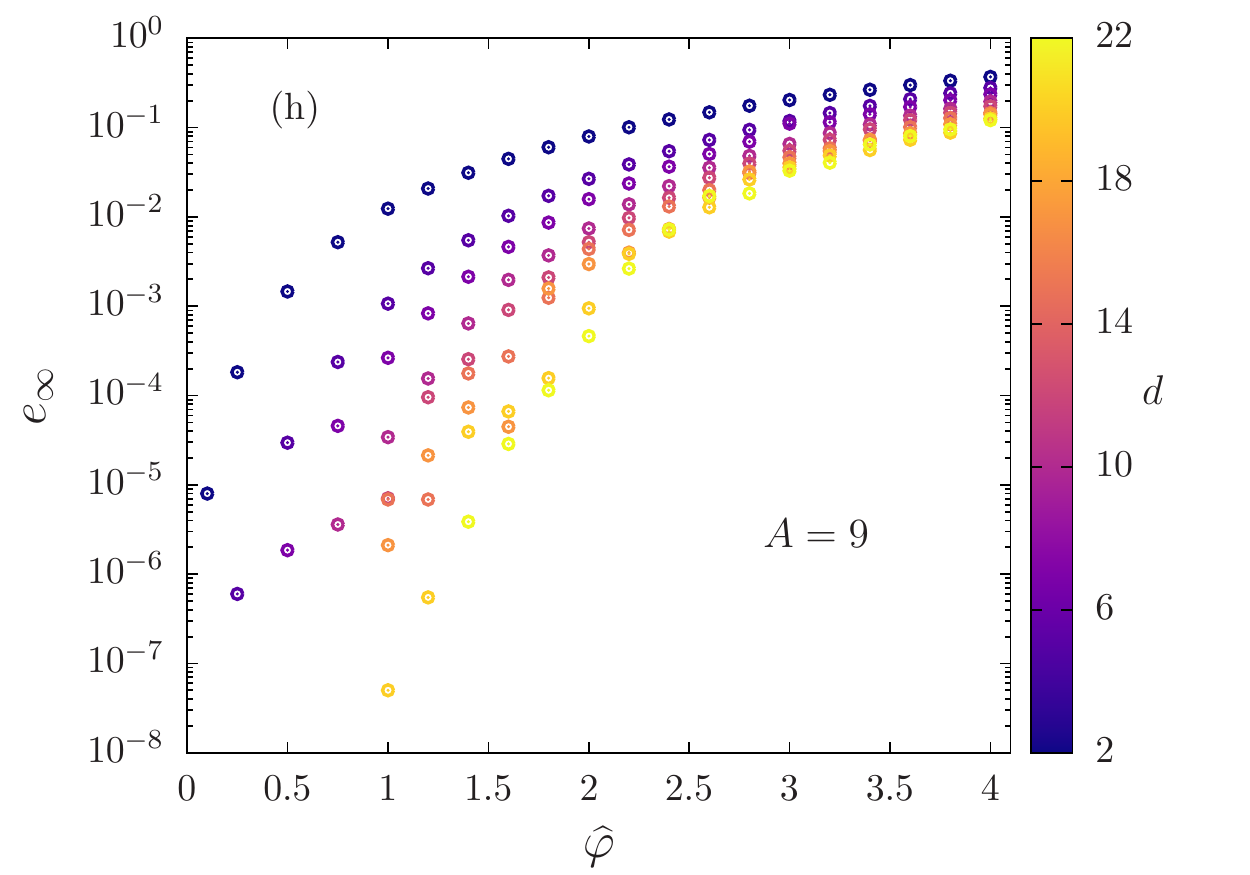}
\caption{
Average asymptotic values of
the isostatiticy index $c_\io=\lim_{t\to\io} z(t)/d$, where $z(t)$ is the number of obstacles in contact with the RLG tracer at time $t$,
of the pressure $p_\io$, and of the energy $e_\io$ (shown both in linear and log scales), as a function of density $\wh\f$, for several values of $d$ and two values of $A=7$
(panels a, c, e, g) and $A=9$ (panels b, d, f, h).
It is observed that $c_\io \approx \wh\f/2$, as predicted by the low-density expansion, up to 
the jamming transition point  $\wh\f=\wh\f_J$ where $c_\io=1$. Consistently, $p_\io$ and $e_\io$ are observed to vanish for $\wh\f\lesssim \wh\f_J$ upon increasing $d$, 
while $p_\io$ grows linearly and $e_\io$ grows quadratically in $\wh\f - \wh\f_J$ for $\wh\f \gtrsim \wh\f_J$.
}
\label{fig:Z}
\end{figure}

\subsection{Asymptotic long-time results and cutoff dependence}
\label{sec:cutoff}

We now focus on the asymptotic results at long times. For each trajectory, we compute the final energy $e_\io$, final pressure $p_\io$, final isostatic index $c_\io$, and final MSD $\D_r^\io$.
 The cutoff dependence of the probability distribution of $\D_r^\io$ over realizations of the disorder is shown in Fig.~\ref{fig:cutoff}(a,b).
 At low densities and high dimensions, it converges with increasing $A$; conversely, at high 
density it becomes double-peaked upon increasing the cutoff and it does not 
converge within the accessible range of $A$. This is because the average value of $\D_r^\io$ is larger (possibly divergent with $d\to\io$, as discussed in section~\ref{sec:timedep}) in this regime; at low $A$, the peak is due to the cutoff,
\ie all trajectories escape the sphere of radius $A/d$ and are thus unphysical, see Fig.~\ref{fig:traj}(c). 
Upon increasing the cutoff, we start to observe a cutoff-independent 
``physical'' peak, but the peak due to escaping trajectories remains visible, although it decreases with the cutoff. Clearly, the distribution would converge for larger $A$, but
the number of obstacles then becomes exceedingly large for our available computational resources.

We also compare  in Fig.~\ref{fig:cutoff}(c) the rescaled MSD computed at $d=5,\ldots,22$ in the RLG with the prediction for
$\D_r^\io$ computed from the DMFT. At low densities, the MSD converges at 
$A\approx5$ and a linear extrapolation at $d=\io$ corresponds with the DMFT result. At higher 
densities the MSD does not converge with the cutoff nor with $d\to\io$, and its value is also far from the DMFT prediction. This
suggests once again that the DMFT numerical solution is not reliable, as discussed in  section~\ref{sec:timedep}.

Even if the available computational resources do not allow us to reach convergence in the cutoff at high density, we can still
elucidate a somehow counterintuitive behavior of the dynamics, \ie the \textit{increase} of the MSD 
with density at fixed $d$. The situation is indeed inverted with respect to systems in thermal equilibrium, 
where a density increase yields the transition from diffusive to arrested dynamics and the asymptotic MSD decreases upon increasing density~\cite{manacorda2020numerical}; in this 
athermal case, in the low-density unjammed phase the tracer reaches the boundary of a zero-energy lake,
and stops its motion there, leading to a finite displacement from the origin. Upon increasing the density, the tracer starts its 
dynamics from a higher energy level and it surfes over several energy valleys before reaching the zero-energy region.
The final MSD from the origin thus increases with density. 
Whether the MSD diverges when jamming is reached remains an open problem. Under the weak ergodicity breaking scheme mentioned above~\cite{bouchaud1992weak,CK93,FFR20},
one should expect the MSD to diverge, because just
above jamming, the tracer keeps surfing and surfing without ever finding a local minimum, so its MSD keeps increasing with time. 
While the data of Fig.~\ref{fig:cutoff}(c)
for the MSD seem to increase with $d$ when $\wh\f \gtrsim \wh\f_J$ and to saturate when $\wh\f \lesssim \wh\f_J$,
they are, unfortunately, not conclusive, because our
limited computational resources do not allow us to describe this regime in detail.

More precise information is obtained by plotting the asymptotic value of the average isostaticity index as a function of density, dimension, and cutoff, see Fig.~\ref{fig:Z}.
We observe that the data converge to a stable result when $A\approx 7$ and $d\approx 20$ in the regime $\wh\f \lesssim 2.5$, in which $c_\io\approx \wh\f/2$ as predicted
by the dilute limit of section~\ref{sec:dilute}. As a consequence,
we obtain $c_\io \sim 1$ for $\wh\f_J \approx 2$, which provides
a rough estimate (to be refined below) of the jamming transition in the infinite-dimensional RLG, 
and therefore also suggests $\wh\f_J \approx 4$ for the infinite-dimensional many-body problem, a value lower than that estimated in Ref.~\cite{parisi2020theory} by a less reliable procedure. 
The origin of this discrepancy is left for future investigation; in particular the simulations of the Mari-Kurchan model quoted in Ref.~\cite{parisi2020theory} should probably be extended to higher dimension.
Consistently with this estimate, we observe that both $p_\io$ and $e_\io$ vanish upon increasing $d$ for $\wh\f \lesssim 2$, suggesting that the tracer is reaching a zero-energy region
without overlaps. On the contrary, for $\wh\f \gtrsim 2$, we observe that $p_\io$ grows linearly and $e_\io$ grows quadratically in the distance from jamming, as observed in the MB problem~\cite{OLLN02,OSLN03}.

We obtain a more precise estimate of the jamming density $\wh\f_J$ by 
extrapolating its value from the limit of $d,A \to \io$. Therefore, we fit the pressure
and energy functions as
\beq\label{eq:fit-pe}
\begin{split}
p(\wh\f) &= ab\log\left( 1 + e^{(\wh\f-\wh\f_J)/b} \right) \underset{ b \approx 0^+}{\approx} a (\wh\f-\wh\f_J) \th(\wh\f-\wh\f_J) \ , \\
e(\wh\f) &= a\left[b\log\left( 1 + e^{(\wh\f-\wh\f_J)/b} \right) \right]^2 \underset{ b \approx 0^+}{\approx} a (\wh\f-\wh\f_J)^2 \th(\wh\f-\wh\f_J) \ ,
\end{split}\eeq
using $a$, $b$ and $\wh\f_J$ as fitting parameters (distinct for the two observables).

\begin{figure}[t]
\centering
\includegraphics[width=.45\textwidth]{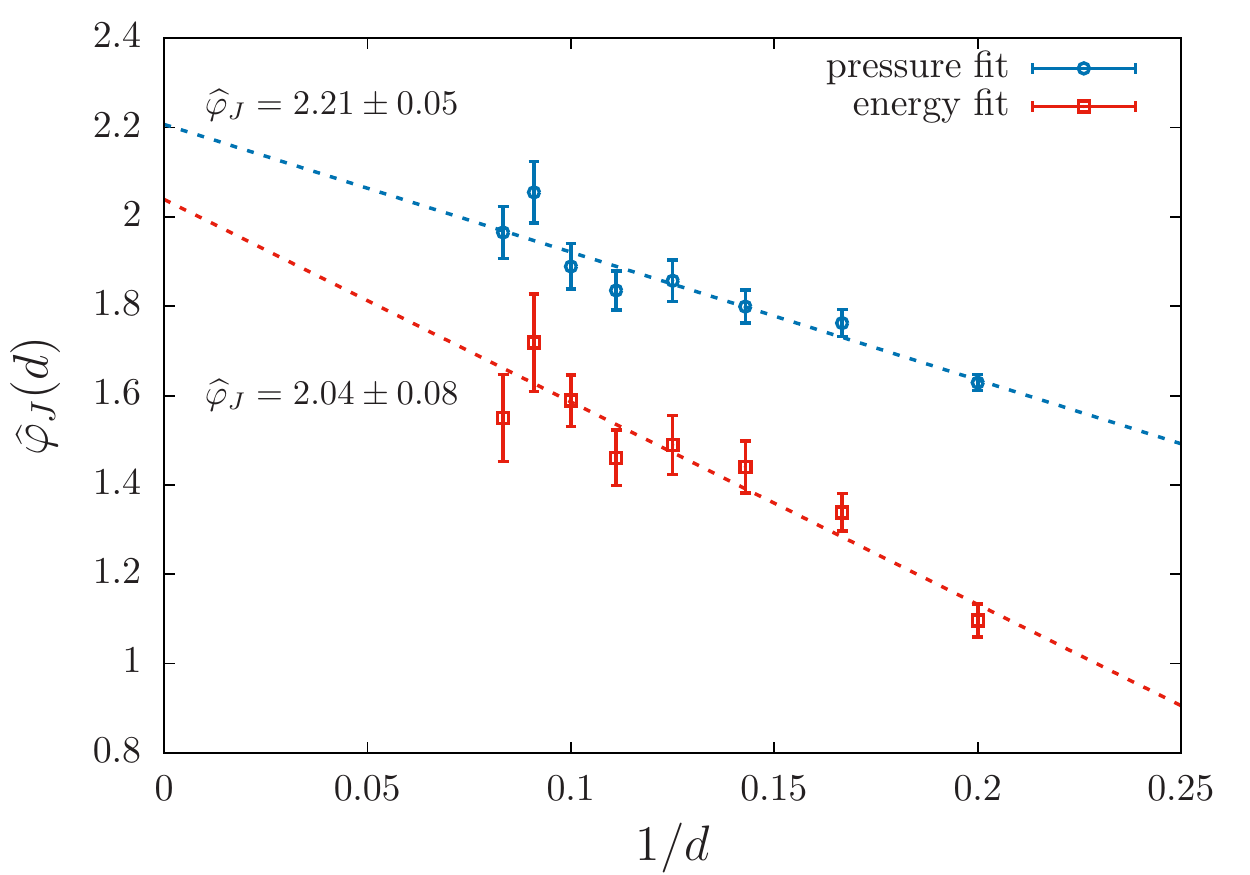}
~
\includegraphics[width=.45\textwidth]{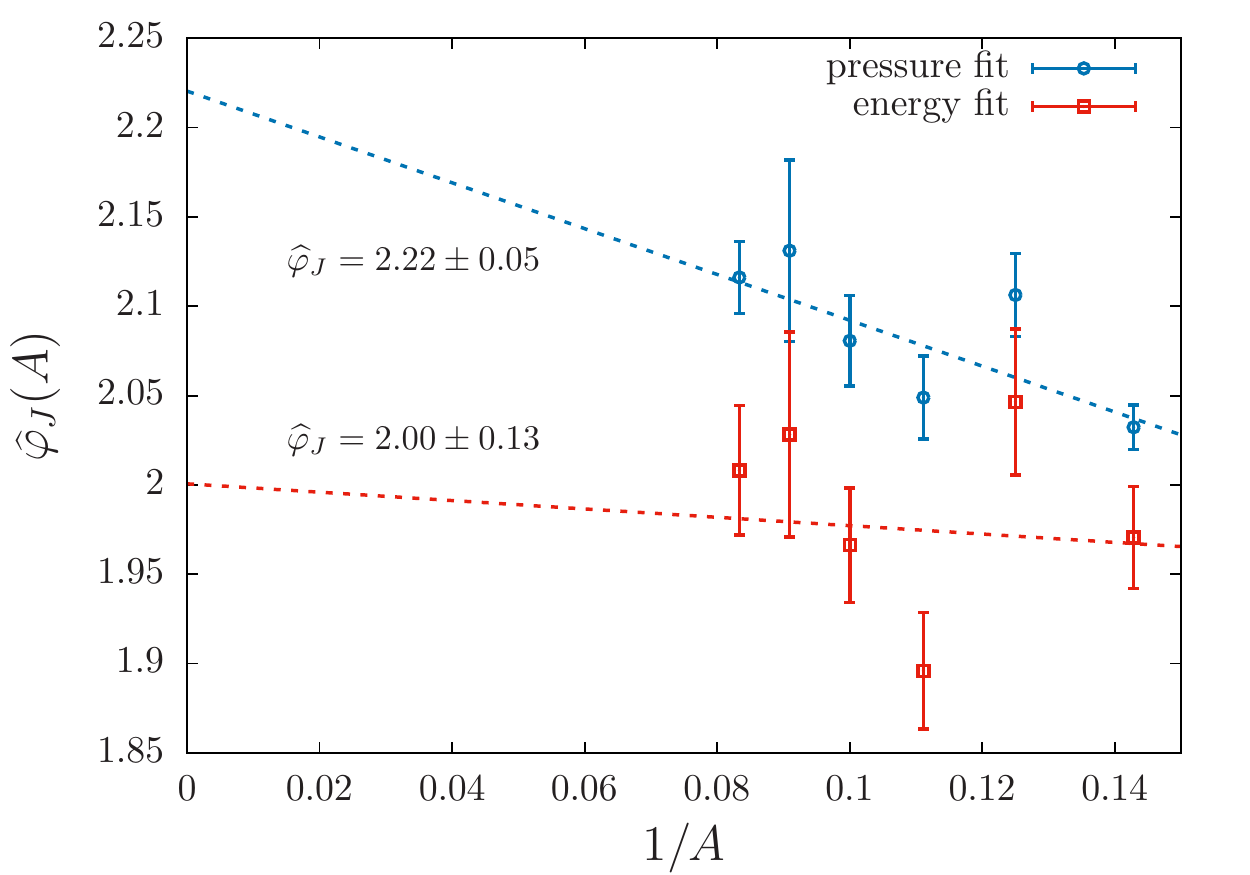}
\caption{Jamming density $\wh\f_J(d)$ vs. $1/d$ (left) or  $\wh\f(A)$ vs. $1/A$ (right). These values are obtained by first fitting the pressure and energy curves in Fig.~\ref{fig:Z}
by means of Eq.~\eqref{eq:fit-pe}, to obtain $\wh\f_J(d,A)$, which is then fitted as $\wh\f_J(d) + \a/A$ (left) or $\wh\f_J(A) +  \b/d$ (right).
The points represents the intercept of these extrapolations. Finally, the values $\wh\f_J$ represent the intercepts of a second linear fit with 
respect to the other parameter. The final extrapolation is weakly dependent on the order of the two limits; conversely, the critical values obtained fitting
the energy curves slightly differ from those obtained fitting the pressure curves.}
\label{fig:phiJ}
\end{figure}

We thus get a jamming point $\wh\f_J(d,A)$ for every $d$ and $A$; 
keeping one of the two parameters constant, one can extrapolate an
asymptotic value $\wh\f_J(d)$ taking the 
intercept of a linear fit in $1/A$. Repeating the same extrapolation in $1/d$,
one finally obtains the values for 
$\wh\f_J = \lim_{d\to\io}\lim_{A\to\io}\wh\f_J(d,A)$. The same procedure can be repeated by extrapolating first in $1/d$
and then $1/A$. Both are reported in 
Fig.~\ref{fig:phiJ}. A conservative estimate of the jamming density therefore
gives $\wh\f_J = 2.1 \pm 0.2$.

\subsection{Memory kernel and response function}
\label{sec:mem-corr}

\begin{figure}[!ht]
\includegraphics[width=.45\textwidth]{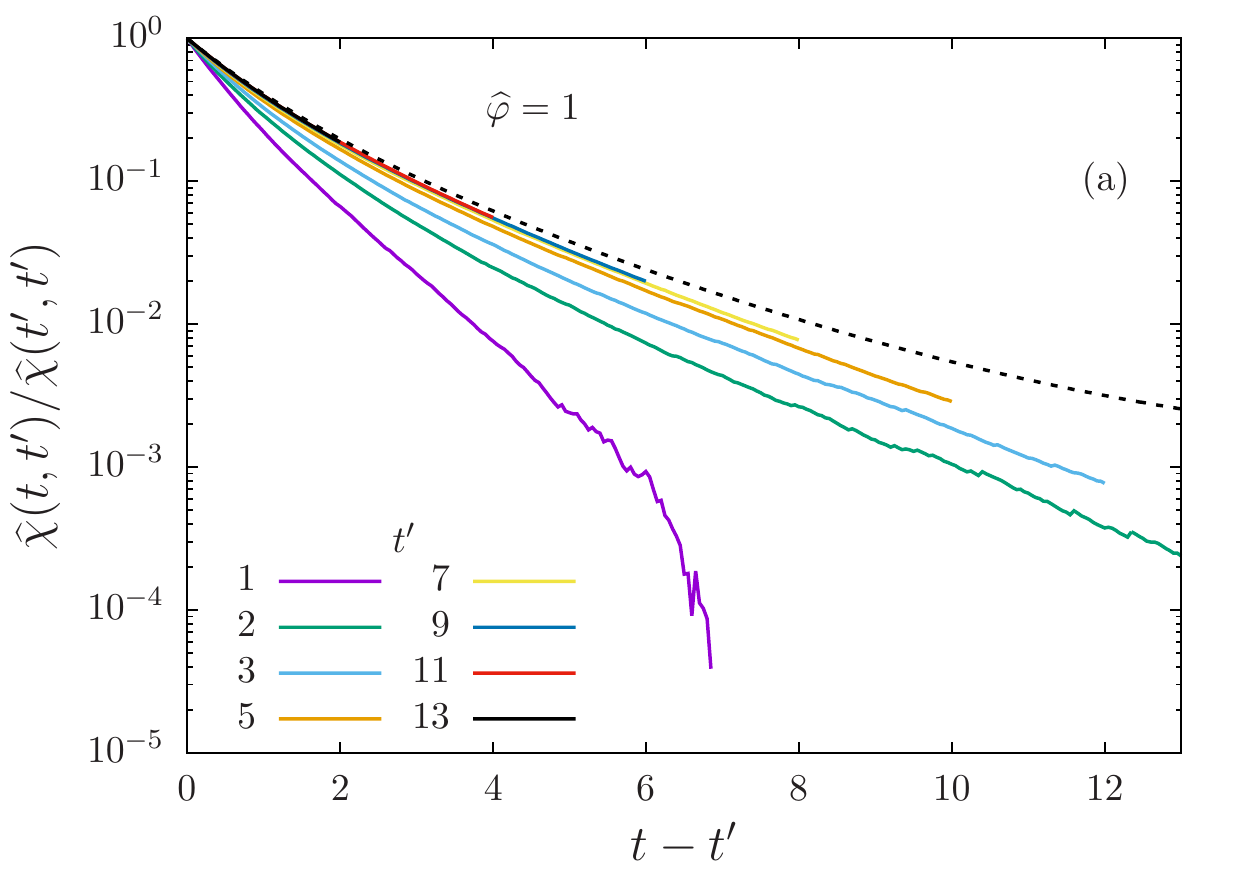}
~
\includegraphics[width=.45\textwidth]{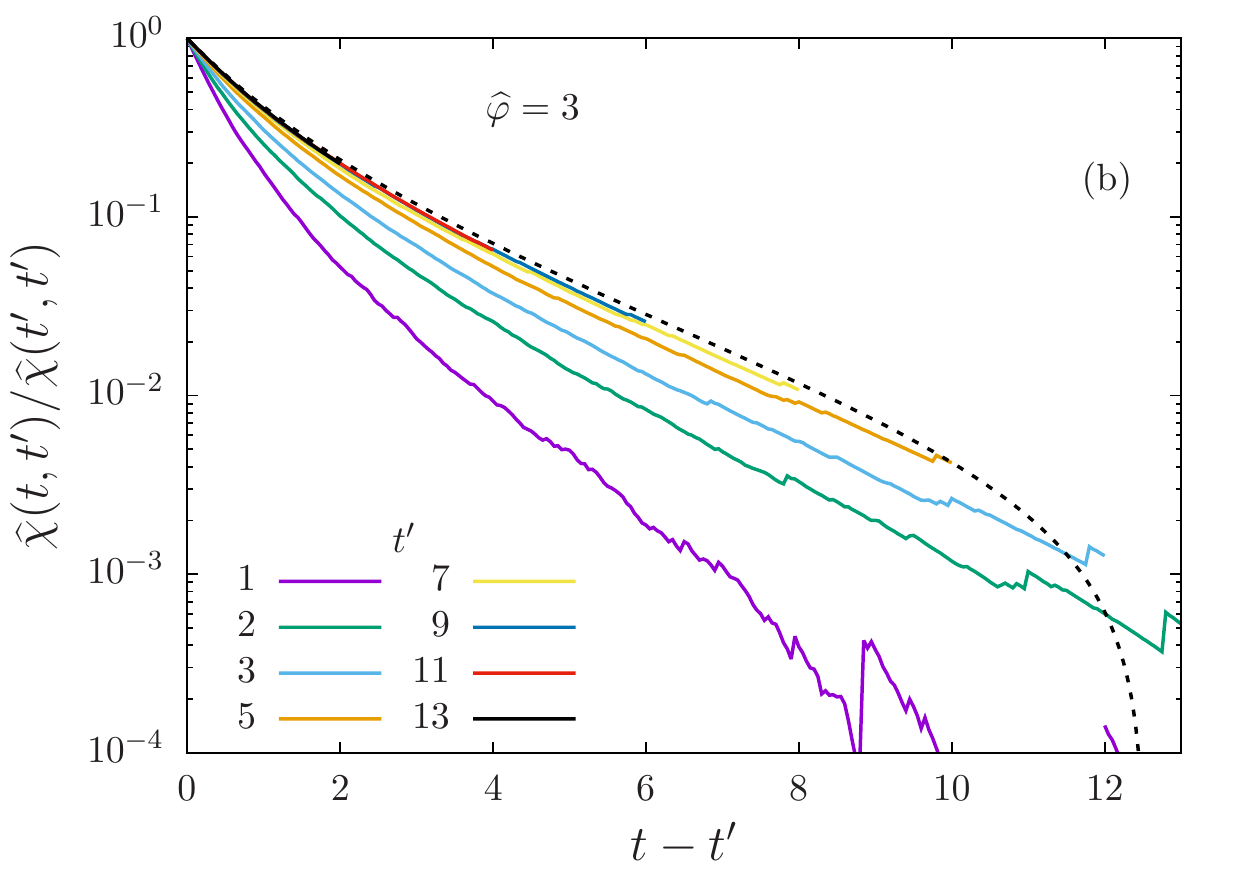}
~
\includegraphics[width=.45\textwidth]{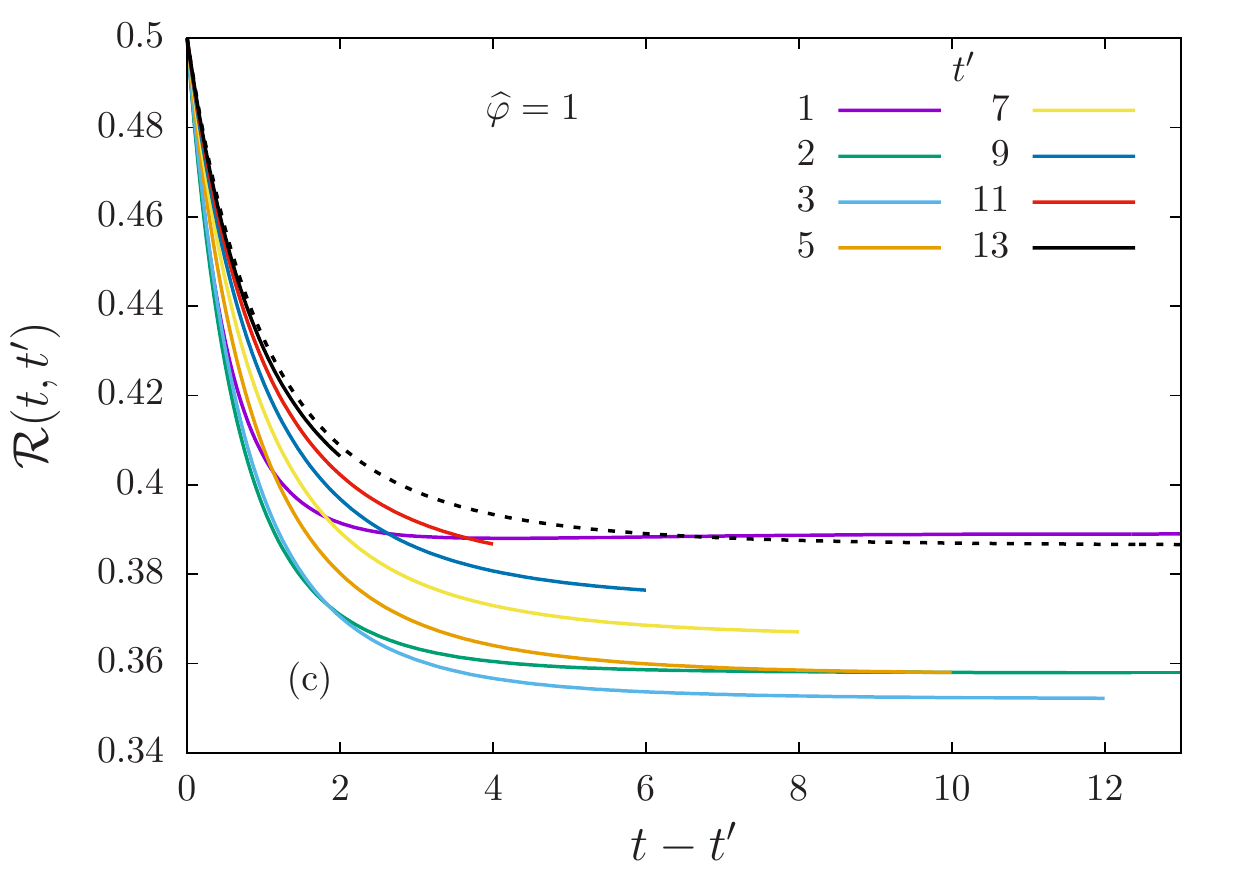}
~
\includegraphics[width=.45\textwidth]{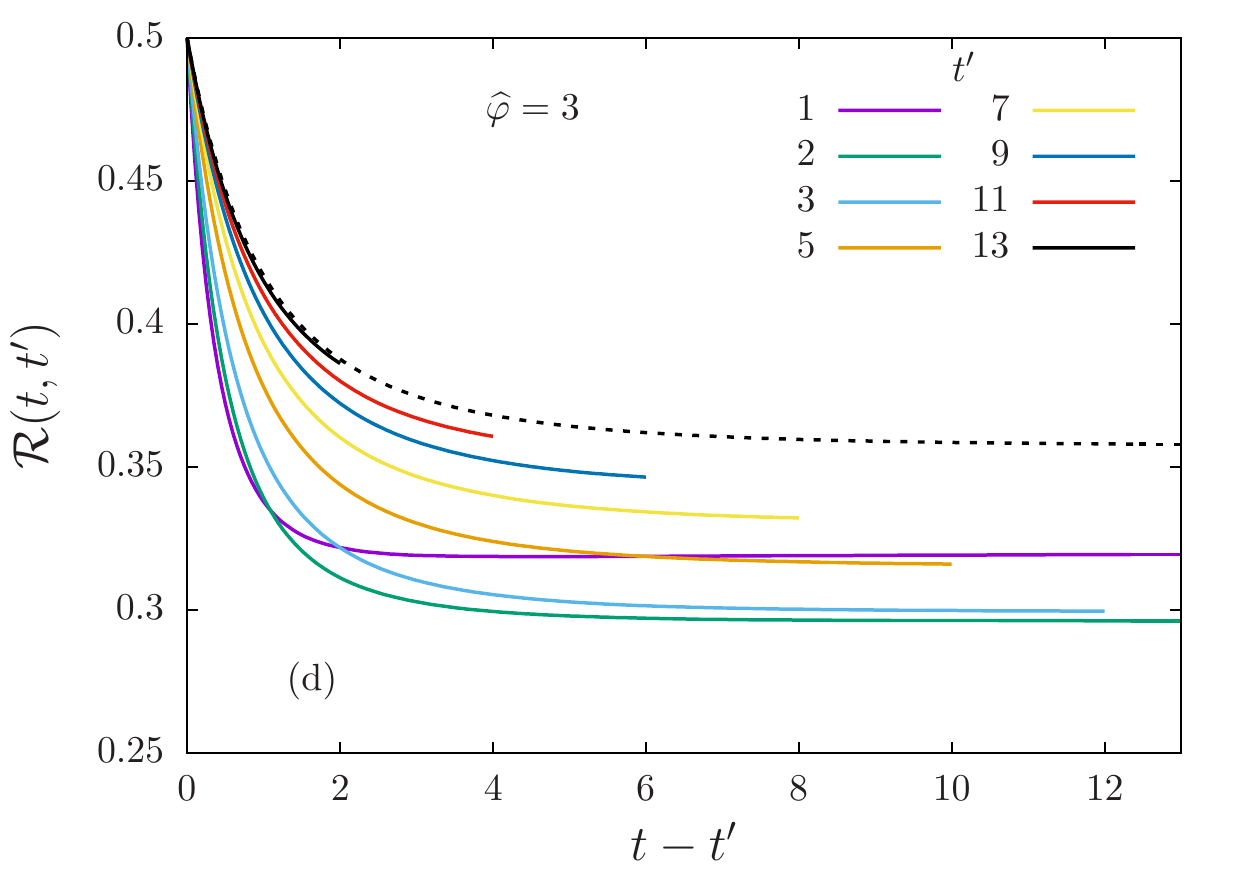}
\caption{Plot of the $\hat\chi(t,t')/\hat\chi(t',t')$ and $\cR(t,t')$ vs $t-t'$ for various $t'$ from DMFT (colored lines), to
be compared with the stationary state predictions for $\hat\chi(\tau)/c^\io$ and $\cR^\io(\tau)$ (black dashed lines).   \\
}
\label{fig:MR-R}
\end{figure}

Finally, we check the validity of the analytical computation of the long-time asymptotic memory kernel 
$\MM^\io_R(\t)$ and response function $\cR^\io(\t)$, given in Eqs.~\eqref{eqdio:MRio_gen} 
and~\eqref{eqdio:RioLap}, respectively. We plot in Fig.~\ref{fig:MR-R}(a,b) the integrated response 
kernel $\c(t,t')$ defined in Eq.~\eqref{eqdio:interesp}, normalized by its value at equal times, as function of the time difference $t-t'$.
We compare the result obtained from the numerical solution of DMFT with
the asymptotic prediction from Eq.~\eqref{eqdio:RioLap},
 at the same rescaled densities discussed 
above, setting $c_\io$ as the long-time limit of $c(t)$ obtained from the DMFT solution.
The value of $\cR^\io(t)$ are obtained by 
the numerical inversion of the solution obtained in Laplace space. The curves in Fig.~\ref{fig:MR-R}(c,d) show the convergence 
of the numerical solution to the TTI analytical solution in the unjammed phase, confirming the 
validity of the analysis of section~\ref{sec:TTI}. Surprisingly, the analytical result matches the long-time behavior also in the 
jammed phase; this last feature may be a signal of a more robust validity of the theoretical prediction derived in 
Sec.~\ref{sec:TTI}.

\section{Conclusions and perspectives}
\label{sec:conclusions}

In this paper, we have investigated the gradient descent dynamics of soft repulsive spheres in the limit $d\to\io$, focusing in 
particular on the jamming dynamical transition.

In the unjammed phase, in which GD dynamics reaches zero energy,
we have derived some analytical predictions for the long-time asymptotic dynamics, by analyzing the dynamical mean field theory equations.
We have shown that the asymptotic solution of the DMFT equations can be written in closed form in terms of a single quantity, namely the asymptotic
value $c_\io$ of the contact number (section~\ref{sec:TTI}). 
Most notably, we derived an expression for the response functions that implies a Marcenko-Pastur vibrational spectrum in the final state (section~\ref{sec:DOS}),
 as conjectured in~\cite{IS21} (see also~\cite{FPUZ15}).
Unfortunately, the value of $c_\io$ remains undetermined by the asymptotic analysis, and has to be derived from the full solution of the dynamics, as shown for example by the 
dilute limit (section~\ref{sec:dilute}). The jamming transition point, which corresponds to $c_\io=1$, thus remains undetermined.
Another important problem that remains open is the derivation of the dynamical critical exponents that characterize the divergence of the relaxation time upon approaching
the jamming transition from below~\cite{OT07,VOT11,OT20,NIB21}.

We then presented a (rather unsuccessful) attempt to solve numerically the DMFT equations. Unfortunately, our solution algorithms display
convergence problems even at rather short times, thus preventing us to obtain reliable numerical predictions for the jamming transition from DMFT.
Similar problems where encountered in Ref.~\cite{mignacco2021stochasticity}.
To obtain more insight into the problem, we then simulated the Random Lorentz Gas in several dimensions ranging from $d=2$ to $d=22$. This study
allowed us to confirm some predictions of DMFT, and more importantly to obtain an estimate of the jamming transition in $d\to\io$, \ie $\wh\f_J\sim2.1$
for the RLG and $\wh\f_J\sim 4.2$ for the many-body problem. Note that this value is smaller than the dynamical Mode-Coupling transition, which is
$\wh\f_{\rm d}=4.8$ in $d\to\io$~\cite{parisi2020theory}. This result suggests that the so-called J-line~\cite{MKK08,MK11,PZ10,OBC17} extends to densities 
below the equilibrium glass transition in large enough dimensions, contrarily to what is found in physical dimensions $d=2,3$. This results confirms the idea that the equilibrium
dynamical Mode-Coupling arrest is not relevant for the gradient descent dynamics~\cite{FFR20}, thus suggesting a very complex energy landscape in which zero-temperature
jammed states can exist even in a region that is fully ergodic in presence of thermal noise~\cite{sclocchi2021high}.

The problem that remains strikingly open, in our opinion, is that of computing the asymptotic final energy $e_\io$ reached by GD dynamics in the jammed (SAT) phase. We considered here the simplest case, in which GD starts from
a fully random infinite-temperature initial condition, see~\cite{FFR20} for the case where GD starts in equilibrium at finite temperature. 
Our numerical results for $e_\io$ are reported in Fig.~\ref{fig:Z}. Clearly, if one could compute this quantity analytically, one could
then derive the jamming density at which $e_\io=0$. Let us then review some approaches that have been proposed for the computation of $e_\io$.
\begin{itemize}
\item First of all, one might wonder whether the system simply asymptotically reaches the ground state energy, \ie $e_\io = e_{\rm GS}$, which could then be computed by a standard thermodynamic
zero-temperature calculation. However, such calculation can be done~\cite{BJZ11}, and in the inifinite-dimensional limit~\cite{SBZ18} it predicts that the ground state energy is zero up to values of $\wh\f$ that diverge when $d\to\io$. Because our numerical results  strongly suggest that $e_\io$ remains finite for finite values of $\wh\f\geq \wh\f_J$ in the limit $d\to\io$, this hypothesis can be ruled out, and the system must remain out of equilibrium in the asymptotic long-time limit.
\item The first consistent out-of-equilibrium solution for the gradient descent dynamics in the long-time limit was obtained in the pioneering work of Cugliandolo and Kurchan~\cite{CK93}. 
Their asymptotic solution is based on the following assumptions: (i) memory of the initial condition is lost, \ie $\D_r(t\to\io)\to\io$ in our formalism, 
which is deemed {\it weak ergodicity breaking}~\cite{bouchaud1992weak}; (ii) for large $t'$ and $t>t'$, there is a sharp separation between 
a short-time regime where the dynamics is time-translational invariant, $\D(t,t')=\D(t-t')$, 
and reaches equilibrium at temperature $T$ (here $T=0$ and $\D(t-t')=0$) and a long-time regime where persistent aging is observed,
with a scaling $\D(t,t') = \overline\D[\HH(t)/\HH(t')]$ with scaling functions $\overline\D[\l]$ and $\HH(t)$; (iii) in the long-time regime correlations and responses are related by a modified effective temperature
$T_{\rm eff} = T/m$; and (iv) dynamics is asymptotically {\it marginal}, \ie the eigenvalue spectrum $\rho(\lambda)$ in the final state touches zero and as a consequence the response
function and the energy have a power-law decay at long times (see section~\ref{sec:DOS}).
Based on these assumptions, Cugliandolo and Kurchan derived a consistent and exact asymptotic solution that also provides the value of $e_\io$ in the spherical pure $p$-spin model~\cite{CK93}.
It can be shown in full generality, see \eg~\cite{altieri2020dynamical} for a pedagogical derivation, that this asymptotic solution coincides with the zero-temperature version of the Monasson
replica calculation~\cite{Mo95}, with the additional condition that the replicon eigenvalue vanishes~\cite{CK93,BFP97,altieri2020dynamical}. Although the replica derivation is not needed, because the result
can be fully derived from the DMFT equations~\cite{CK93,altieri2020dynamical}, it provides a simple recipe, and the corresponding formulae have been already derived in the literature~\cite{BJZ11,SBZ18,parisi2020theory}. In appendix~\ref{app:CuKu}, we show that this asymptotic solution is inconsistent with our numerical data for the RLG, which suggests that
some of its underlying assumptions are not verified in this system. 
This is also suggested by the fact that jammed packings display non-trivial critical exponents in their structural properties~\cite{OLLN02,OSLN03,LN10,LNSW10,Wy12,MW15} that are not captured by this asymptotic solution~\cite{PZ10,BJZ11,parisi2020theory}. Hence, this solution can also be ruled out, at least in the vicinity of jamming.

\item 
An attempt to go beyond this ansatz was presented by
Montanari and Ricci-Tersenghi~\cite{MR03,MR04} and then Rizzo~\cite{Ri13}. Putting aside its dynamical justification, 
they extended the Monasson replica calculation mentioned above to take into account replica symmetry breaking.
While, unfortunately, this scheme gives inconclusive results in what concerns the calculation of $e_\io$~\cite{MR03,MR04,Ri13},
it was shown in~\cite{CKPUZ13,CKPUZ14} that it
provides the correct critical exponents of jamming, see~\cite{parisi2020theory} for a detailed discussion. Yet, the value of $e_\io$, and thus $\wh\f_J$, still remain undetermined. 
An estimate of $\wh\f_J$ within this approach (again, without a dynamical justification) was given in~\cite{CKPUZ13},  but it overestimates 
our numerical results, and it moves from $\wh\f_J = 6.26$ ($\wh\f_J=3.13$ in RLG units) to an even worse
$\wh\f_J=6.87$ ($\wh\f_J=3.43$ in RLG units) upon increasing the steps of replica symmetry breaking.
Hence, around the values of $\wh\f_J$ we find numerically, no consistent solution of the full replica symmetry breaking equations in the Monasson scheme can be found.
In summary, not only the dynamical meaning of this proposal remains obscure (except in the Cugliandolo-Kurchan scheme mentioned above), but it also cannot provide a consistent solution in the relevant regime of jamming densities.
Yet, the fact that the jamming transition displayed by the asymptotic dynamics has the same critical properties
of that obtained within this approach clearly calls for a first-principle justification, beyond the marginal stability argument put forward in~\cite{MW15}.
Other replica approaches have been attempted, see~\cite{parisi2020theory} for a review, with similar drawbacks.

\item 
Assumptions (ii) and (iii) in the Cugliandolo-Kurchan approach can be generalized to introduce a hierarchy of well separated time scales, each associated to an effective temperature~\cite{CK94,Cu02,altieri2020dynamical,kurchan2021time}. Such an approach provides a dynamical justification to full
replica symmetry breaking, because it leads to the same set of equations.
However, it is not clear to us how to fix the asymptotic dynamical energy within this approach, see~\cite{altieri2020dynamical} for a discussion.

\item 
The other assumptions of the Cugliandolo-Kurchan solution should also be checked carefully. Results obtained on the Ising $p$-spin model seem to suggest that memory of the initial
condition is not completely lost, even when the gradient descent dynamics is initialized in a fully random initial configurations~\cite{Ri13}, hence indicating strong ergodicity breaking.
Similar results have been obtained for the Sherrington-Kirkpatrick model~\cite{bernaschi2020strong} and for spherical mixed $p$-spin models~\cite{folenaprivate}.
Our own results are inconclusive on this point. The validity of the weak ergodicity breaking is then not certain, at least in the vicinity of jamming.
Concerning assumption (iv), it has also been suggested~\cite{Ri13,bernaschi2020strong} that the relaxation could be slower than a power-law (\eg logarithmic). Our numerics is consistent
with power-law relaxation, however, which suggests that marginality is a robust assumption.

\end{itemize}

Overall, we believe that despite much progress, the problem of investigating the long-time aging of GD dynamics, which was fully solved by Cugliandolo and Kurchan for the pure $p$-spin model~\cite{CK93}, remains
still open for more general models, including those investigated here. 
Our observations suggest that
strong ergodicity breaking~\cite{Ri13,FFR20,bernaschi2020strong} and
a multithermalization structure with an infinite hierarchy of time scales~\cite{CK94,altieri2020dynamical,kurchan2021time} should both be present in the
proper asymptotic solution. Preliminary (and unsuccessful) attempts at constructing such kind of solutions have been presented in Refs.~\cite{BFP97,FFR20}.
Future work will hopefully provide better numerical algorithms to solve the DMFT equations, and thus provide insight on how to make further progress.

\acknowledgments

We warmly thank P. Urbani for a series of very important exchanges during the early stages of this work, and in particular for sharing with us his closure of the long
time limit of the DMFT equations in the replica symmetric UNSAT phase of the perceptron (later published in Ref.~\cite{sclocchi2021high} in a slightly different model), 
here reported for completeness in appendix~\ref{sec:perc-UNSAT}, and his (still unpublished)
numerical solution results of the DMFT for the gradient descent dynamics of the perceptron model.
We also warmly thank L. Berthier, P.Charbonneau, P.Morse, and Y. Nishikawa for discussing and sharing with us their unpublished numerical data on GD in many-body 
dynamics, and A.Altieri, G.Biroli, L.Cugliandolo, G.Folena, S.Franz, J.Giannini, Y.Hu, S.Hwang, H.Ikeda, J.Kurchan, F.Ricci-Tersenghi, T.Rizzo, S.Sastry and G.Szamel, for several useful discussions 
related to this project.
This project has received funding from the European Research Council (ERC) under the European Union's Horizon 2020 research and innovation programme (grant agreement n. 723955 - GlassUniversality).

\section*{Data availability statement}

The codes used to solve the DMFT equations and to simulate the RLG gradient descent are available at the GitHub
repositories {\tt https://github.com/amanacorda/dmft\_jamming.git} and 
{\tt https://github.com/amanacorda/rlg\_gd.git}, respectively.

\bibliography{HSmerge,SGmerge}

\clearpage
\appendix

\section{The perceptron}

In this appendix, we show how the results of section~\ref{sec:DMFT_dinf} can be adapted to the perceptron, the simplest supervised data classifier.
We refer the reader to Refs.~\cite{FPSUZ17,ABUZ18} (and references therein) for an introduction to the model, and for recent work on its statics and dynamics.

\subsection{Dynamical mean field equations for the perceptron model}

We consider the harmonic perceptron Hamiltonian given by
\beq
H=\sum_{\mu=1}^M v(h_\mu)\ , \qquad h_\mu=\frac{1}{\sqrt N} \underline \xi^\mu \cdot \underline x - \s\ ,  \qquad v(h)=\frac12 h^2\th(-h) \ ,
\eeq
and the gradient descent (GD) dynamics:
\beq
\dot x_i(t) = -\hat \nu(t) x_i(t) - \sum_\mu \frac{\xi^\mu_i}{\sqrt{N}} v'(h_\mu(t))  \ ,
\eeq
where $\hat \nu(t)$ is a Lagrange multiplier needed to enforce the spherical constraint $|\underline x(t)|^2=N$, which can be expressed as
\beq\label{eq:nu}
\hat\nu(t) = - \frac1N \sum_\mu (h_\m(t)+\s) v'(h_\mu(t)) = -\a \la (h(t)+\s) v'(h(t)) \ra \ .
\eeq
The DMFT equations corresponding to GD dynamics 
have been derived in Ref.~\cite{ABUZ18}, and are expressed in terms of an
effective variable $h(t)$. If the initial condition is extracted at infinite temperature, \ie uniformly on the sphere $|\underline x(t)|^2=N$,
they read\footnote{Note that a minus sign is missing in Eqs.~(34) and (64) of Ref.~\cite{ABUZ18}.}
\beq\label{eq:DMFE}
\begin{split}
\dot h(t) &= -\tilde \nu(t) [h(t)+\s] - v'(h(t) - P(t)) +\int_0^t \de s M_R(t,t')[h(t')+\s] + \eta(t) \ , \qquad h(0)=h_0 \ ,\\
P(h_0) &= e^{-\frac12(h_0+\s)^2}/\sqrt{2\pi} \ , \\
M_C(t,t') &= \langle \eta(t)\eta(t')\rangle = \alpha \langle v'(h(t)) v'(h(t'))\rangle \ ,\\
M_R(t,t') &= \left.\alpha \frac{\delta \langle v'(h(t))\rangle}{\delta P(t')}\right|_{P=0} \ ,\\
\tilde \nu(t)&= \hat \nu(t) + \a \langle v''(h(t))\rangle = 
\a \langle v''(h(t)) - (h(t)+\s) v'(h(t)) \rangle = 
\int_0^t \de u\left[ M_R(t,u)C(t,u)+M_C(t,u)R(t,u) \right] \ ,\\
\partial _t C(t,t') &= -\tilde \nu(t) C(t,t') +\int_0^t \de u M_R(t,u) C(u,t') + \int_0^{t'}\de u M_C(t,u)R(t',u) \ ,\\
\partial _t R(t,t') &= \delta(t-t') -\tilde \nu(t) R(t,t') +\int_{t'}^t \de u M_R(t,u) R(u,t') \ ,
\end{split}
\eeq
where the field $P(t)$ is only used to compute $M_R(t,t')$ (with the derivative taken by keeping the kernels constant) and otherwise is set to $P=0$.
Note that using the explicit expression for $\tilde\nu(t)$ in terms of an average over the process $h(t)$,
the first five equations above provide closed equations for
$h(t)$, $M_C(t,t')$, $M_R(t,t')$ and $\tilde\nu(t)$, without the need of solving for $C(t,t')$ and $R(t,t')$.

As in section~\ref{sec:DMFT_def},
it is convenient to derive an expression of the response function that does not require an explicit computation of the functional derivative.
For this we write
\beq\label{eq:H1}
M_R(t,t') = \alpha \la\frac{\delta  v'(h(t))}{\delta P(t')} \ra_{P=0} = \alpha \la v''(h(t))\frac{\delta h(t) }{\delta P(t')}_{P=0} \ra
= \alpha \la v''(h(t)) H(t,t') \ra_{P=0} \ , \qquad H(t,t') = \frac{\delta h(t) }{\delta P(t')}_{P=0} \ .
\eeq
Taking the functional derivative with respect to $P(t')$ of the equation for $h(t)$, we obtain
\beq\label{eq:H2}
\partial_t H(t,t') = -\tilde \nu(t) H(t,t') - v''(h(t))[ H(t,t')-\d(t-t')] +\int_{t'}^t \de u M_R(t,u) H(u,t')  \ , 
\eeq
with $H(t,t')=0$ for $t<t'$. This linear equation allows one to compute the function $H(t,t')$ associated to each trajectory of $h(t)$.
Note that 
another possible approach is to use the Novikov theorem~\cite{roy2019numerical}, which gives
\beq
M_R(t,t') = \int \de u M_C^{-1}(t',u) \la v'(h(t)) v''(h(t')) \h(u) \ra \ ,
\eeq
but this expression is not very practical for our purposes, because of the need to invert $M_C(t,t')$ and of the presence of a three-point correlation.

As in section~\ref{sec:DMFT_def},
we define the integrated response:
\beq\label{eq:chidef}
\chi(t,t') = \int_{t'}^t \de u R(t,u) \ ,
\eeq
and we note that an equation for $\chi(t,t')$ can be obtained from the dynamical equation for $R(t,t')$,
\beq\label{eq:chit}
\dot \chi(t,t') = 1 - \tilde \nu(t) \chi(t,t') +\int_0^t \de u M_R(t,u)\chi(u,t') \ .
\eeq
We also define the integrated kernel of response, for $t>t'$, as
\beq\label{eq:whcdef}
\wh\c(t,t') = \tilde\nu(t) - \int^t_{t'} \de u \, M_R(t,u) \ ,
\eeq
which encodes the kernels as
\beq
\tilde\nu(t) = \wh\c(t,t) \ , 
\qquad M_R(t,t') = \th(t-t') \, \partial_{t'} \wh\c(t,t') \ .
\eeq

\subsection{Some general properties of the long-time GD dynamics}

We now discuss the general behavior of the GD dynamics in the long time limit.

\subsubsection{Long-time TTI regime}
\label{app:TTI}

Because the GD must converge to a unique final configuration, we must have
\beq
\lim_{t\to\io}C(t+\tau,t)=C^\io(\tau) =1 \ ,\qquad  \forall \t \ .
\eeq
At long times, the effective gap $h(t)$ must also converge to a constant value, $h_\infty$, which is itself a random variable, and therefore
\beq\label{eq:TTIlong}
\begin{split}
\tilde\nu_\io &= \lim_{t\to\io} \tilde\nu(t) = \a \langle v''(h_\io)-(h_\io+\s) v'(h_\io)\rangle \ ,  \\
M_C^\io(\tau) &= \lim_{t\to\io} M_C(t+\t,t) = \alpha \langle v'(h_\io)^2 \rangle \ .
\end{split}\eeq
Furthermore, in Eq.~\eqref{eq:H2}, when $t,t'\to\io$ we do not have any explicit time dependence, and as a result
\beq\label{eq:Hlong1}
\lim_{t\to\io}H(t+\tau,t) = H^\io(\t) \qquad \Rightarrow \qquad \lim_{t\to\io}M_R(t+\tau,t) = M_R^\io(\t)=\alpha \la v''(h_\io) H^\io(\t) \ra \ ,
\eeq
with
\beq\label{eq:Hlong2}
\partial_\t H^\io(\t) = -[ \tilde \nu_\io+ v''(h_\io)] H^\io(\t) + v''(h_\io)\d(\t) +\int_0^\t \de u M^\io_R(\t-u) H^\io(u)  \ .
\eeq
Note that if $h_\io > 0$, then $v''(h_\io)=0$, which implies $H^\io(\t)=0$.
Furthermore, from Eq.~\eqref{eq:whcdef} we have
\beq\label{eq:MRasy}
\wh\c^\io(\t) = \lim_{t\to\io}\wh\c(t+\tau,t) = \tilde\nu_\io - \int_0^\t \de u M_R^\io(u) \ , 
\qquad
\wh\c =  \lim_{\t\to\io} \wh\c^\io(\t) = \tilde\nu_\io - \int_0^\io \de u M_R^\io(u) \ .
\eeq
Note that the response function also satisfies
\beq\label{eq:Rio}
\lim_{t\to\io}R(t+\tau,t) = R^\io(\t) \ , \qquad
\partial _t R^\io(\t) = \delta(\t) -\tilde \nu_\io R(\t) +\int_{0}^\t \de u M^\io_R(\t-u) R^\io(u) \ ,
\eeq
and recalling Eq.~\eqref{eq:chidef} we can define
\beq
\c=\lim_{\t\to\io} \lim_{t\to\io} \chi(t+\t,t) = \int_0^\io \de u R^\io(u) \ .
\eeq

\subsubsection{Exact solution for quadratic potential}

Following section~\ref{sec:TTI},
in the special case $v(h) = h^2\th(-h)/2$ we have $v''(h)=\th(-h)$ and we can thus obtain a closed equation for $M^\io_R(\t)$. 
Recalling Eq.~\eqref{eq:Hlong1}, multiplying Eq.~\eqref{eq:Hlong2}
by $v''(h_\io)$ and taking the average, we obtain
\beq\begin{split}
\partial_\t M_R^\io(\t) &= -(1+\tilde \nu_\io) M_R^\io(\t) 
+ c_\io \d(\t) +\int_0^\t \de u M^\io_R(\t-u) M_R^\io(u)  \ ,
\qquad 
c_\io = \a \la \th(-h_\io) \ra \ ,
\end{split}\eeq
where $c_\io$ is the isostaticity index for the perceptron~\cite{FPSUZ17}.
In Laplace space, this gives
\beq\label{eq:MRLap}
s M^\io_R(s) = - (1+\tilde \nu_\io) M_R^\io(s) + c_\io + [M^\io_R(s)]^2 \qquad \Rightarrow \qquad
M^\io_R(s) = \frac{(1+ \tilde\nu_\io+s) - \sqrt{(1+ \tilde\nu_\io+s)^2 - 4 c_\io}}2 \ .
\eeq
Note that at short times we must have
\beq
M^\io_R(\t\to 0) \sim c_\io + \OO(t) \qquad\Leftrightarrow \qquad
M^\io_R(s\to\io) = \frac{c_\io}{s} + \OO(1/s^2) \ ,
\eeq
which fixes the choice of sign in Eq.~\eqref{eq:MRLap}, because the other solution has an unphysical behavior
$M^\io_R(s) \sim s$ at large~$s$.
Note also that in order for Eq.~\eqref{eq:MRLap} to be well defined we need the condition
\beq
(1+\tilde\nu_\io)^2 \geq 4 c_\io \ ,
\eeq
whose meaning will be discussed below.
The Laplace transform can be inverted and we get
\beq\label{eq:MRio_gen}
M^\io_R(\t) = \frac{\sqrt{c_\io}}{\t} e^{-(1+\tilde \nu_\io)\t} I_1(2\sqrt{ c_\io} \t) \ ,
\eeq
where $I_1(x)$ is the modified Bessel function of first kind.
Injecting the expression of $M^\io_R(\t)$ in Eq.~\eqref{eq:Rio}
in Laplace space, we get
\beq\label{eq:RioLap}
R^\io(s) = \frac1{s+\tilde \nu_\io-M_R^\io(s)}
= \frac{2}{( \tilde\nu_\io+s-1) + \sqrt{(1+ \tilde\nu_\io+s)^2 - 4 c_\io}} \ ,
\eeq
hence
\beq\label{eq:chiLap}
\chi  = R^\io(s=0) =\frac{2}{( \tilde\nu_\io-1) + \sqrt{(1+ \tilde\nu_\io)^2 - 4 c_\io}} \ .
\eeq

\subsection{Memory-less solution}

We are now going make a stronger assumption as in Eq.~\eqref{eq:fastcond}, \ie we assume that the convergence to the TTI regime described in appendix~\ref{app:TTI} is fast enough,
and the memory kernels decay fast enough, such that,
for a function $f(t)$ that tends to a constant, $f(t\to\io) = f_\io$,
one can write
\beq
\lim_{t\to \infty}\int_0^t \de t' M_R(t,t')f(t') = \int_0^\io \de \t M_R^\io(\t) f_\io = (\tilde\nu_\io - \wh \chi) f_\infty \ ,
\eeq
where we used Eq.~\eqref{eq:MRasy}.
In other words, all the previous dynamical history is lost and only the TTI regime contributes at long times. 

Because we know that the effective gap satisfies $h(t\to\io) = h_\io$, applying this assumption to the equation for $h(t)$ in Eq.~\eqref{eq:DMFE}, we obtain
\beq
0 = -\wh\c (h_\infty+\s) - v'(h_\infty)+\eta_\infty \ , \qquad \text{where} \qquad \langle \eta_\infty^2\rangle = \a \langle v'(h_\infty)^2\rangle = M_C^\infty \ .
\label{self_infinity}
\eeq
Applying the same assumption to the equation for $\tilde \nu(t)$ in Eq.~\eqref{eq:DMFE} we obtain
\beq\label{eq:nuchiM}
\wh\c =  M_C^\infty \chi \ .
\eeq
At this point, it is worth to note that $\chi$, which is the integral of the response function, can be finite or formally infinite. We analyze the two cases separately below.

\subsubsection{UNSAT phase}
\label{sec:perc-UNSAT}

The content of this appendix was communicated to us by P.Urbani at the beginning of this project, 
and is reported here for completeness; it has been published in Ref.~\cite{sclocchi2021high} in a slightly different model.
Let us suppose that $\chi$ is finite. This corresponds to a response function $R(t,t')$ that decays sufficiently fast to be integrable. In other words, the final state reached by 
GD is such that, if the system is infinitesimally displaced away from it, it returns to the same state fast enough that the perturbation is integrable in time.
Physically, this suggest that GD reaches a minimum that admits a harmonic expansion with well-defined and strictly positive frequencies; we come back to
this point below.
Using Eq.~\eqref{eq:nuchiM}, Eq.~\eqref{self_infinity} becomes
\beq
0 = -M_C^\infty \chi  (h_\infty+\s) - v'(h_\infty)+\eta_\infty \ .
\eeq 
We then need to eliminate $M_C^\infty$. 
Under the assumption that the GD converges sufficiently fast, and that the response function $R(t,t')$ is integrable, we have 
$\lim_{t\to\io} \chi(t,0) =\lim_{t\to\io} \chi(t,t')$ for all $t'$, because the memory of the initial time is lost. 
By taking the limit $t\to\io$ of Eq.~\eqref{eq:chit},
we then obtain
\beq\label{eq:Mchi2}
\wh \chi = \frac1\c \qquad \Rightarrow \qquad M_C^\infty = \chi^{-2} \ ,
\eeq
which leads to a closed equation for $\chi$ by noting that
\beq
\eta_\infty = (h_\infty+\s)/\chi +v'(h_\infty) \ ,\qquad \langle \eta_\infty^2\rangle =\chi^{-2} = \a \langle v'(h_\infty)^2\rangle \ .
\eeq
More explicitly, we have 
\beq\label{eq:Pio}
P(h_\io) = P(\h_\io) \frac{\de \h_\io}{\de h_\io} = \frac{e^{-\frac{1}2 \left( h_\infty+\s + \c v'(h_\infty)\right)^2} }{\sqrt{2\pi}} \left( 1 +\c v''(h_\infty)\right) \ ,
\qquad \chi^{-2} = \a \int \de h_\io P(h_\io) v'(h_\infty)^2 \ ,
\eeq
which is a closed equation for $\c$ from which all the other quantities can be derived.
Specializing to $v(h) = h^2 \th(-h)/2$, 
we get
\beq
\frac 1{\chi^2} = \frac \a{(1+\chi)^2} \int_{-\infty}^0 \de h h^2 \g_1(h+\s) \ ,
\eeq
which is the same equation one gets from replicas \cite{FPSUZ17} in the replica symmetric UNSAT phase of the perceptron.
In fact, Eq.~\eqref{eq:Pio} gives a finite probability to negative gaps, $h_\io<0$, which indicates that the system is in an UNSAT phase.
Furthermore, the equation for $C(t,0)$ in the limit $t\to\io$ gives
\beq\label{eq:Cio}
0 = \wh \c C(\io,0) \qquad \Rightarrow \qquad C(\io,0)=0 \ ,
\eeq
which indicates a complete loss of correlation with the initial condition. This indicates that the ground state is unique, which is consistent
with the replica symmetric ansatz and also with our initial assumption, that the response function at long times is integrable.

Given the expression of $P(h_\io)$, one can easily derive the expressions of $\tilde\nu_\io$ and $c_\io$, see Ref.~\cite{FPSUZ17}. Inserting these
expressions into the exact result in Eq.~\eqref{eq:MRio_gen}, one can show that:
\begin{itemize}
\item $M^\io_R(\t)$ decays exponentially for $\s>0$ and all values of $\a > \a_J(\s)$ such that the system is in the UNSAT phase~\cite{FPSUZ17}.
\item When $\s=0$ and $\a>2$, one has $M^\io_R(\t)\sim \t^{-3/2}$. This is the critical line on which replica symmetry is marginally broken~\cite{FPSUZ17}.
\item For $\s<0$ one obtains an inconsistency. In fact, the value of $\wh\c$ obtained from Eq.~\eqref{eq:nuchiM} corresponds to the wrong choice of sign
in Eq.~\eqref{eq:MRLap}. This indicates that the assumption of loss of memory is incorrect in this case, and we know indeed that the system displays replica
symmetry breaking and a complex landscape with multiple minima in the UNSAT phase for $\s<0$~\cite{FPSUZ17}.
\end{itemize}

\subsubsection{SAT phase}

Upon approaching the SAT phase, it is known from the replica analysis that $\c\to\io$~\cite{FPSUZ17}. Indeed, this is the only way to suppress negative values of $h_\io$, corresponding to unsatisfied
gaps, in Eq.~\eqref{eq:Pio}.
We then expect that in the SAT phase one formally has $\c=\io$ and $M_C^\io=0$, but with $\c M_C^\io=0$, as suggested by Eq.~\eqref{eq:Mchi2}.
This is also consistent with the observation that in the SAT phase, because $h_\io\geq 0$ and $v'(h_\io)=0$, one has
$\tilde \nu_\io = c_\io$, see Eq.~\eqref{eq:TTIlong}, which, once inserted into Eq.~\eqref{eq:RioLap}, gives for the response function
\beq\label{eq:RioLapSAT}
R^\io(s) 
= \frac{2}{c_\io-1+s + \sqrt{(1+ c_\io+s)^2 - 4  c_\io}} 
= \frac{1-c_\io-s + \sqrt{(1+ c_\io+s)^2 - 4 c_\io}}{2 s } \ ,
\eeq
hence
\beq
\chi  = R^\io(s=0) =\io \ ,
\eeq
\ie $\chi$ diverges in the whole unjammed phase, which indicates that the response function reaches a plateau at long times.
When $s\to 0$ we have indeed
\beq
R^\io(s) \sim \frac{1-c_\io}{s}  \qquad \Rightarrow \qquad R^\io(\t\to\io) \to 1-c_\io \ .
\eeq
Physically, this corresponds to the fact that in the SAT phase the GD dynamics stops on the boundary of the finite volume of phase space corresponding to solutions, \ie zero energy
states. A random perturbation applied to this state has a finite probability to bring the system inside this volume, and as a consequence the system will not relax back to its initial state;
the response function does not decay at long times and $\chi$ thus diverges.

Furthermore, setting $\tilde \nu_\io = c_\io$ in Eq.~\eqref{eq:MRio_gen} and using $I_1(x)\sim e^x/\sqrt{2 \pi x}$ at large $x$, 
we obtain that $\MM^\io_R(\t) \sim e^{-(1-\sqrt{c_\io})^2 \t}$ at large $\t$, provided $c_\io < 1$,
but when $c_\io=1$ (the isostatic point) we obtain $\MM^\io_R(\t) \sim \t^{-3/2}$, consistently with the results obtained in the UNSAT phase.

Because then $\wh\c=1/\c=0$
and $\la \h_\io^2 \ra=1/\c^2=0$, hence $\h_\io=0$, Eq.~\eqref{self_infinity} becomes a trivial equation for $h_\io$,
\beq
v'(h_\io)=0 \qquad \Leftrightarrow \qquad h_\io \geq 0 \ ,
\eeq
which expresses the SAT condition but leaves $h_\io$ indeterminate. Also, Eq.~\eqref{eq:Cio} leaves $C(\io,0)$ indeterminate,
which indicates that the memory of the initial state of the dynamics is not lost.
The distribution of $h_\io$, and all the observables that derive from it, such as $c_\io$, can then only be determined by the solution of the dynamics.

To obtain some insight we can consider first the free case $\alpha=0$, which implies $\tilde\nu(t)=0$, $M_C(t,t')=0$, $M_R(t,t')=0$,
and $\h(t)=0$, and we specialize to the quadratic potential.
Then, the evolution equation for $h(t)$ is simply
\beq
\dot h(t) = -v'(h(t)) = - h(t) \th(-h(t)) \qquad \Rightarrow \qquad
h(t)=
\begin{cases}
h_0 & h_0 \geq 0 \ , \\
h_0 e^{-t} & h_0 < 0 \ .
\end{cases}
\eeq
and the distribution of initial states is $P(h_0) = e^{-\frac12(h_0+\s)^2}/\sqrt{2\pi} =\g_1(h_0+\s)$. 
Introducing
\beq
\Th_n(\s) = \int_{-\io}^0 \de h_0 \, h_0^n \g_1(h_0+\s) \ ,
\eeq
we get
\beq
P(h,t) =\th(-h) e^t \g_1 \left(e^t h+\s\right) + \th(h) \g_1(h+\s) \qquad \Rightarrow \qquad P_\io(h)=\Th_0(\s) \d(h) + \th(h) \g_1(h+\s) \ .
\eeq
We also obtain
\beq\begin{split}
M_C(t,t') &= \alpha \langle v'(h(t)) v'(h(t'))\rangle = \a\int_{-\io}^0 \de h_0 \g_1(h_0+\s) h_0^2 e^{-(t+t')} = \a \Th_2(\s)e^{-(t+t')} \ ,  \\
\tilde \nu(t)&= \a \langle v''(h(t)) - (h(t)+\s) v'(h(t)) \rangle = \a \left[ \Th_0(\s) - \s \Th_1(\s) e^{-t} - \Th_2(\s) e^{-2t} \right]  \ .
\end{split}\eeq
To compute $M_R(t,t')$ we need to compute $H(t,t')$. For $h_0\geq 0$, we have $H(t,t')=0$. For $h_0<0$, we obtain
\beq
\partial_t H(t,t') = -[1+ \tilde \nu(t)] H(t,t')  +  \d(t-t') \qquad \Rightarrow \qquad 
H(t,t') = \th(t-t') e^{ - \int_{t'}^t \de u [1 + \tilde\n(u) ] } \ ,
\eeq
so that $H(t,t')$ is independent of $h_0$ and
\beq
M_R(t,t') = \a \la \th(-h(t)) H(t,t') \ra = \a \Th_0(\s) H(t,t')
= \a \Th_0(\s) \th(t-t') e^{ - \int_{t'}^t \de u [1 + \tilde\n(u) ] }
 \ .
\eeq
Note that at large times $\tilde\n(t) \to \tilde\n_\io = c_\io= \a \Th_0(\s)$ and then
\beq
M^\io_R(\t) = c_\io \th(\t) e^{ - (1 + c_\io) \t } \ , \qquad c_\io= \a \Th_0(\s) \ ,
\eeq
which indeed coincides with the small $\a$ expansion of Eq.~\eqref{eq:MRio_gen}.

\subsection{Vibrational density of states}

At long times, the system reaches a unique configuration $\ul X^* = \{x_i^*\}$. We can then linearize the dynamics around this configuration, 
with $y_i(t) = x_i(t) - x_i^*$,
and
the GD equation become
\beq
 \dot y_i(t)
 	=  -\hat \nu_\io y_i(t) -\sum_j \frac{\partial H(\ul X^*)}{\partial x_i \partial x_j}\cdot  y_j(t) +  \l_i(t) 
	=  -\sum_j \HH_{ij} \cdot  y_j(t) +  \l_i(t) \ ,
\eeq		
where $\HH$ is the Hessian in the minimum, which includes a diagonal term equal to $\hat\nu_\io$ due to the spherical constraint~\cite{FPUZ15}
and $\ul \L(t) = \{  \l_i(t)\} $ is the source term used to compute the response~\cite{ABUZ18}.
This is solved by
\beq
\ul Y(t) = \int_0^t \de t' e^{- \HH (t-t') } \ul \L(t') \ ,
\eeq
which gives
\beq\begin{split}
R^\io(t) &=\left. \frac{1}{N}\sum_{i} \frac{\d y_{i}(t)}{\d \l_{i}(0)} \right|_{ \l=0} = 
\frac{1}{N} \Tr \, e^{- \HH t }
=\frac{1}{N} \sum_{\a=1}^{N} e^{- \l_\a t }
= \int \de \l \r(\l) e^{- \l t }
\ , \qquad \r(\l) = \frac{1}{N} \sum_{\a=1}^{N} \d\left(\l - \l_\a \right) \ ,
\end{split}\eeq
where $\r(\l)$ is the vibrational density of states.
In Laplace space
\beq\label{eq:Rs2}
R^\io(s) = \int \de\l \frac{\r(\l)}{ \l + s} \ .
\eeq
We focus for simplicity to the SAT phase.
Combining Eqs.~\eqref{eq:Rs2} and \eqref{eq:RioLapSAT} we obtain the Cauchy transform of $\rho(\l)$ in the form
\beq
g(z) = \int \de\l \frac{\r(\l)}{z - \l} = - R^\io(s=- z)
= \frac{1-c_\io + z + \sqrt{(1+ c_\io- z)^2 - 4 c_\io}}{2z } \ ,
\eeq
which shows that $\r(\l)$ is the Marcenko-Pastur distribution with parameter $c_\io<1$, \ie
\beq
\r(\l) = (1-c_\io) \d(\l) + \frac{\sqrt{(\l_+-\l)(\l-\l_-)}}{2 \pi \l} \ , \qquad \l_{\pm} = (1 \pm \sqrt{c_\io})^2 \ ,
\eeq
which is consistent with the result of Ref.~\cite{FPUZ15} for the vibrational spectrum of the perceptron.
Note that
the present calculation is unable to detect the isolated eigenvalue that is responsible for the
critical slowing down upon approaching the jamming transition~\cite{HI20}.

\section{An alternative formulation of the DMFT equations}
\label{app:B}

In this appendix, we derive an alternative but equivalent formulation of the DMFT equations, 
that can be useful for both analytical and numerical purposes.

\subsection{Perceptron}

Using an integration by parts, we can then rewrite all the response contributions as\footnote{
The last relation must be carefully applied to a function $f(t,t')$ containing an Heaviside theta contribution. Indeed, in that case one has
\beq
f(t,t') = f_r (t,t') \, \th(t-t')  \ , \nonumber
\eeq
being $f_r(t,t')$ a regular function; therefore
\beq\nonumber
\begin{split}
-\tilde\nu(t) \, f(t,t') + \int^t_{t'} \de u \, M_R(t,u) \, f(u,t') &= - \wh\c(t,t') f_r(t',t') \th(0) - \int^t_ {t'} \de u \, \wh\c(t,u) \left[ \dot f_r(u,t') \th(u-t') + f_r(u,t') \d(u-t') \right] \\
&= - \wh\c(t,t') f_r(t',t') - \int^t_ {t'} \de u \, \wh\c(t,u) \dot f_r (u,t')  \ ,
\end{split}
\eeq
counting only once (or two halves) the theta-delta contribution.
}
\beq\label{eq:inteparts}
-\tilde\nu(t) \, f(t) + \int^t_{t'} \de u \, M_R(t,u) \, f(u) = - \wh\c(t,t') f(t') - \int^t_ {t'} \de u \, \wh\c(t,u) \, \dot f(u) \ ,
\eeq
having assumed $t>t'$.
We can then write the equation for $h(t)$ as 
\beq
\dot h(t) = - v'(h(t)) - \wh\c(t,0)[h(0)+\s] - \int_0^t \de u \, \wh\c(t,u) \dot h(u) + \eta(t) \ ,
\eeq
and 
the dynamics can be determined self-consistently in terms of the two kernels $\wh\c(t,t')$ and $M_C(t,t')$. 
From Eqs.~\eqref{eq:DMFE}, 
the latter
is given by $M_C(t,t') = \alpha \langle v'(h(t)) v'(h(t'))\rangle$ while
 the former
is given, using Eqs.~\eqref{eq:H1} and \eqref{eq:H2}, by
\beq
\wh\c(t,t') =  \a\moy{ v''(h(t)) F(t,t') -[h(t)+\s]  v'(h(t)) } \ ,\qquad
F(t,t') = 1 - \int^t_{t'} \de u \, H(t,u) \ .
\eeq
Given the structure of Eq.~\eqref{eq:H2}, 
we can write $H(t,t') = H_r(t,t') \theta(t-t')$, with $H_r(t,t) = v''(h(t))$, and using Eq.~\eqref{eq:inteparts} we obtain
for $t>t'$:
 \beq\label{eq:HH2}
\partial_t H_r(t,t') =  - v''(h(t)) H_r(t,t') -\wh\c(t,t') H_r(t',t') - \int^t_{t'} \de u \, 
 \wh\c(t,u)\partial_u H_r(u,t') \ .
 \eeq 
We can then write
\beq\begin{split}\label{eq:deF}
\partial_t F (t,t') &= -H_r(t,t) - \int^t_{t'} \de u \, \partial_t H_r(t,u)  \\
&= - v''(h(t)) + \int^t_{t'} \de u \, v''(h(t)) H_r(t,u)
+ \int^t_{t'} \de u \wh\c(t,u) H_r(u,u)
+ \int^t_{t'} \de u\int^t_{u} \de w \, 
 \wh\c(t,w)\partial_w H_r(w,u) \ .
\end{split}\eeq
Now we use the relation
\beq\label{eq:B6}
\int^t_{t'} \de u \int^t_u \de w \, \wh\c(t,w) \partial_w H_r(w,u) = \int^t_{t'} \de w \, \wh\c(t,w) \int^w_{t'} \de u \, \partial_w H_r(w,u)
\eeq
to rewrite the last two terms in Eq. \eqref{eq:deF} as
\beq\label{eq:B7}
\int^t_{t'}  \de u \, \wh\c(t,u) H_r(u,u)
+ \int^t_{t'} \de u\int^t_{u} \de w \,  \wh\c(t,w)\partial_w H_r(w,u) = - \int^t_{t'} \de w \, \wh\c(t,w) \partial_w F(w,t') \ ,
\eeq
and we obtain the final form of the equation for $F(t,t')$:
\beq
\partial_t F (t,t') = - v''(h(t)) F(t,t') - \int^t_{t'} \de u \, \wh\c(t,u) \partial_u F(u,t')  \ .
\eeq
Note that $F(t,t')$ is only defined for $t\geq t'$ and $F(t,t)=1$.
To summarize, it is useful to collect the final closed set of equations, equivalent to Eqs.~\eqref{eq:DMFE}, as follows:
\beq
\begin{split}
\dot h(t) &= - v'(h(t)) - \wh\c(t,0)[h(0)+\s] - \int_0^t \de u \, \wh\c(t,u) \dot h(u) + \eta(t) \ , \qquad h(0)=h_0 \ ,  \\
P(h_0) &= e^{-\frac12(h_0+\s)^2}/\sqrt{2\pi} \ , \\
M_C(t,t') &= \langle \eta(t)\eta(t')\rangle = \alpha \langle v'(h(t)) v'(h(t'))\rangle \ ,\\
\wh\c(t,t') &=  \a\moy{ v''(h(t)) F(t,t') -[h(t)+\s]  v'(h(t)) } \ , \qquad \text{for } t>t'  \ ,\\
\partial_t F (t,t') &=  - v''(h(t)) F(t,t') - \int^t_{t'} \de u \, \wh\c(t,u) \partial_u F(u,t') \ ,\qquad \text{for } t>t' \text{ with } F(t,t)=1 \ .
\end{split}
\eeq

\subsection{Infinite-dimensional particles}

The definition of integrated response given in Eq.~\eqref{eqdio:interesp} has two main advantages:
\begin{itemize}
\item In the equilibrium case we have
\beq 
\wh\c(t,t') = \b \MM_C(t-t') \ , \qquad
\wh\c (t,t) = \b \MM_C(0) = \k  \ .
\eeq
\item It allows one to rewrite all the response contributions as
\beq
-\k(t) \, f(t) + \int^t_{t'} \de u \, \MM_R(t,u) \, f(u) = - \wh\c(t,t') f(t') - \int^t_ {t'} \de u \, \wh\c(t,u) \, \dot f(u) \ ,
\eeq
having assumed $t'<t$ as usual, thus getting rid of the spring term attached to $f(t)$.
\end{itemize}
The response equation then takes the form:
\beq
\begin{split}
\cR(t,t') &= \cR_r(t,t') \, \th(t-t') \ , \qquad \cR_r(t',t') = \frac1{2\wh\z}  \ , \\
\wh\z \frac{\partial}{\partial t} \cR_r(t,t') &= -\frac1{2\wh\z} \, \wh\c(t,t') - \int^t_{t'} \de u \, \wh\c(t,u) \frac{\partial}{\partial u} \cR_r(u,t')  \ ,
\end{split}
\eeq
while the correlation equation is, recalling that $\CC(t,0)=0$,
\beq
\wh\z \frac{\partial}{\partial t} \CC (t,t') = - \int^t_0 \de u \, \wh\c(t,u) \frac{\partial}{\partial u} \CC(u,t') + \int^{t'}_0 \de u \, \MM_C (t,u) \cR(t',u) \ .
\eeq
We can write similarly the equation for $y(t)$, recalling that $y(0)=0$,
\beq
\wh\z \, \dot y(t) = - \int^t_0 \de u \, \wh\c(t,u) \dot y(u) - \redv'(h(t)) + \X(t) \ , \qquad
h(t) = h_0 + y(t) + \D_r(t) \ ,
\eeq
The dynamics can be thus determined self-consistently if one knows the two kernels $\wh\c(t,t')$ and $\MM_C(t,t')$; the former
is defined as
\beq
\begin{split}
\wh\c(t,t') &= \frac{\wh\f}2 \int \de h_0 \, e^{h_0} \moy{ \redv''(h(t)) + \redv'(h(t)) - \redv''(h(t)) \int^t_{t'} \de u \, H(t,u)  }_{h_0} \\
&= \frac{\wh\f}2 \int \de h_0 \, e^{h_0} \moy{ \redv''(h(t)) F(t,t') + \redv'(h(t)) }_{h_0} \ ,
\end{split}
\eeq
having now defined
\beq
F(t,t') = 1 - \int^t_{t'} \de u \, H(t,u) \ .
\eeq
The equation for $H(t,t') = H_r(t,t') \theta(t-t')$, with $\wh\z H_r(t,t) = v''(h(t))$,  becomes
 \beq\label{eqdio:H2}
 \wh\z \frac{\partial}{\partial t} H_r(t,t') =  - \redv''(h(t)) H_r(t,t') -\wh\c(t,t') H_r(t',t') - \int^t_{t'} \de u \, 
 \wh\c(t,u)\frac{\partial}{\partial u} H_r(u,t') \ .
 \eeq 
We can write
\beq\begin{split}\label{eq:deF_spheres}
\wh\z \frac{\partial}{\partial t} F (t,t') &= -\wh\z H_r(t,t) - \int^t_{t'} \de u \, \wh\z\frac{\partial}{\partial t}H_r(t,u)  \\
&= - v''(h(t)) + \int^t_{t'} \de u \redv''(h(t)) H_r(t,u)
+ \int^t_{t'} \de u \wh\c(t,u) H_r(u,u)
+ \int^t_{t'} \de u\int^t_{u} \de w \, 
 \wh\c(t,w)\frac{\partial}{\partial w} H_r(w,u) \ ,
\end{split}\eeq
%Now we use the relations
%\beq
%\int^t_{t'} \de s \int^t_s \de u \, \wh\c(t,u) \partial_u H_r(u,s) = \int^t_{t'} \de u \, \wh\c(t,u) \int^u_{t'} \de s \, \partial_u H_r(u,s)
%\eeq
%to rewrite the last two terms in Eq. \eqref{eq:deF_spheres} as
%\beq
%\int^t_{t'}  \de u \, \wh\c(t,u) H_r(u,u)
%+ \int^t_{t'} \de s\int^t_{s} \de u \,  \wh\c(t,u)\partial_u H_r(u,s) = - \int^t_{t'} \de u \, \wh\c(t,u) \partial_u F(u,t') \ ,
%\eeq
which can be simplified using relations analogous to Eqs.~\eqref{eq:B6} and \eqref{eq:B7},
to obtain the equation for $F(t,t')$:
\beq
\wh\z \frac{\partial}{\partial t} F (t,t') = - \redv''(h(t)) F(t,t') - \int^t_{t'} \de u \, \wh\c(t,u) \frac{\partial}{\partial u} F(u,t')  \ .
\eeq

\section{Equivalence between the many-particle problem and the Random Lorentz Gas in infinite dimensions}
\label{app:equivalence}

The equivalence between a many-body (MB) system of spherical particles and the Random Lorentz Gas (RLG) in high dimension has been 
exploited throughout section~\ref{sec:num_dinf}. This equivalence is based on a mapping of the mean field equations for the one-particle and two-particle
dynamics for the MB~\cite{AMZ18} and RLG~\cite{biroli2021mean} problems.
We restrict here to the equilibrium case for simplicity. 
The equations 
for the one-particle dynamics, namely \cite[Eq.~(24)]{AMZ18} and \cite[Eqs.~(49,50)]{biroli2021mean}, are equivalent 
and lead to the same evolution equation for the MSD $D(t)$, namely
\beq
\frac{\z}2 \dot D(t) = T - \frac{\b}2 \int_0^t \de t' \, M(t-t') \, \dot D(t') \ .
\eeq
In the arrested phase, this equation leads to a plateau 
$D_\io = 2 T^2/ M_\io$. We have to understand if the coefficients are also the same or 
need to be rescaled by a proper factor. We then compare the two-particle processes, \ie \cite[Eq.~(123)]{AMZ18} 
and \cite[Eqs.~(55,56)]{biroli2021mean}, getting the following self-consistent equations:
\beq
\text{MB:} \qquad 
\left\{
\begin{split}
\z \dot \rr(t) &= -\b \int_0^t \de t' \, M(t-t') \, \dot \rr(t') - 2\nabla v(\rr(t)) + \sqrt2 \, \bX(t) \\
\moy{\X_\m(t)\, \X_\n(t')} &= \d_{\m \n} \left[ 2\z T \, \d(t-t') + M(t-t') \right] \\
M(t) &= \frac{\r}{d} \int \de \rr_0 \, e^{-\b v(\rr_0)} \moy{ \nabla v(\rr_0) \cdot \nabla v(\rr(t))}_{\rr_0}
\end{split}
\right. 
\ ,
\eeq
and
\beq
 \text{RLG:} 
  \qquad
 \left\{
\begin{split}
\z \dot \rr(t) &= -\b \int_0^t \de s \, M(t-t') \, \dot \rr(t') - \nabla v(\rr(t)) + \bX(t) \\
\moy{\X_\m(t)\, \X_\n(t')} &= \d_{\m \n} \left[ 2\z T \, \d(t-t') + M(t-t') \right] \\
M(t) &= \frac{\r}{d} \int \de \rr_0 \, e^{-\b v(\rr_0)} \moy{ \nabla v(\rr_0) \cdot \nabla v(\rr(t))}_{\rr_0}
\end{split}
\right.
\ .
\eeq
The  two equations above differ only for the factors 2 in the first lines, while the others are formally equivalent. 
Introducing the rescaling
\beq\label{eq:D4}
\z_{\rm MB} = 2 \z_{\rm RLG} \ , \quad \rho_{\rm MB} = 2\r_{\rm RLG} \ , \quad M_{\rm MB}(\r_{\rm MB},t) = 2 M_{\rm RLG}(\r_{\rm RLG},t) \ ,
\eeq
the systems of equations coincide. For the plateau value of the MSD one finds
\beq
D^\io_{\rm RLG} (\r_{\rm RLG}) = \frac{2 T^2}{M^\io_{\rm RLG} (\r_{\rm RLG})}  = 2 \frac{2T^2}{M^\io_{\rm MB}(\r_{\rm MB})}
= 2 D^\io_{\rm MB}(\r_{\rm MB}) \ .
\eeq
In summary, the MB dynamics is equivalent to that of the RLG,
but with a value of density twice as smaller,
a reference time scale
$\t = \z^{-1}$ twice as smaller, and a MSD twice as larger.
The same results on density and MSD have been derived in Ref.~\cite{biroli2021mean}.

To conclude the discussion, it is useful to examine the one-time quantities given in Eq.~\eqref{eq:one-time-q}, which are all sums of pair interactions.
Consider for example the energy. In the MB case, what remains finite in the limit $d\to\io$ is 
the average energy {\it per degree of freedom}, \ie
\beq
e_{\rm MB}(t) = \frac1{Nd }  \sum_{i<j} \la v(\xx_i-\xx_j) \ra = \frac1{2Nd} \sum_{i\neq j} \la v(\xx_i-\xx_j) \ra =  
 \frac1{2d} \sum_{j (\neq i)} \la v(\xx_i-\xx_j) \ra \ .
 \eeq
In the RLG, the average energy per degree of freedom is
 \beq
e_{\rm RLG}(t) =  \frac1{d} \sum_{i=1}^N \la v(\xx-\XXX_i) \ra \ .
 \eeq
 Hence, the RLG energy does not have the factor 2 in front.
 Because each term in the sum contributes a factor $\int^{\infty}_{-\infty} \, \de h_0 \, e^{h_0}  \moy{\redv(h(t)}_{h_0}$ in the $d\to\io$ limit, the factor $1/2$ in front of the energy in equation Eq.~\eqref{eq:one-time-q} is not present for the RLG.
Taking into account Eq.~\eqref{eq:D4}, we obtain
\beq
e_{\rm MB}(\rho_{\rm MB}, t) = e_{\rm RLG}(\rho_{\rm RLG}, t) \ ,
\eeq
\ie the two models have the same energy per degree of freedom when the state points are properly mapped as
in Eq.~\eqref{eq:D4}. Similar considerations apply to any observables that is a sum of pair contributions, and in particular
to the number of contacts per degree of freedom, \ie the isostaticity index defined in Eq.~\eqref{eq:one-time-q}. Hence, the jamming density at which $c=1$ is twice as smaller in the RLG than in the MB problem.
Note that for the RLG the number of degrees of freedom is $d$, hence the isostaticity index is $c=z/d$, and at jamming we have $z=d$, \ie the tracer has $d$ obstacles in contact.

To conclude, let us note that the value of the memory at equal times is a static quantity, \ie in the equilibrium case
${M(t=0) = \frac{\r}{d} \int \de \rr_0 \, e^{-\b v(\rr_0)} |\nabla v(\rr_0)|^2}$. However, the previous reasoning about static quantities
does not apply here, because the memory is not a sum of pair interactions, but rather the square of such a sum~\cite{AMZ18}. This is why an additional factor
of two appears in Eq.~\eqref{eq:D4}.

\section{Cugliandolo-Kurchan asymptotic solution in the jammed phase}
\label{app:CuKu}

We discuss here the predictions of the asymptotic solution first proposed in~\cite{CK93}, with a single time scale.
For convenience, we exploit the fact that this solution is formally analogous to the replica scheme of Monasson~\cite{Mo95}.
We thus start from 
the
replica-symmetric Monasson free energy at density $\wh\f$, temperature $T$, and effective temperature $T/m$~\cite[Eq.(7.34)]{parisi2020theory},
\beq\label{eqC7:PhimRS}
\begin{split}
-\b \Phi(m ; \wh\f, T, \D)= & \frac{d }2 \log\left(\frac{2\pi e}{d}\right)   + \frac{d \, (m-1)}2 \log\left(\frac{\pi e \D}{d^2}\right)  + \frac{d}2 \log m 
 +\frac{ d \wh \f}{2} \int_{-\io}^\io \de h \, e^h \left[  q(\D,\b;h)^m  - 1\right]  \ ,
\end{split}\eeq
with~\cite[Eqs.(4.69,4.74)]{parisi2020theory}
\beq\label{eqC4:qdef}
q(\D,\b;h) = \g_\D \star e^{-\b \redv(h+\D/2)} = \int_{-\infty}^\infty \frac{\de z}{\sqrt{2\pi \D}}e^{-\frac{z^2}{2\D}}e^{-\b \redv(h-z+\D/2)}   \ .
\eeq
The caging order parameter $\D$ must be determined by optimization of the free energy,
and the replicon~\cite[Eq.(7.54)]{parisi2020theory}
\beq
\begin{split}
\l_R &= 1  - \frac{\wh\f}2 \D^2\int_{-\io}^\io \de h e^h q(\D,\b;h)^m  \left( \frac{\de^2}{\de h^2} \log q(\D,\b;h) \right)^2 
\end{split}\eeq
must be positive to ensure the consistency of the replica symmetric calculation.

We then 
consider soft harmonic spheres with $\redv(h) = h^2 \th(-h)/2$, setting $\ee=1$ for simplicity, and
we take the limit $T\to 0$ with $\beta m = y$ and $\beta \D = \chi$~\cite{BJZ11,parisi2020theory}.
We define~\cite[Eq.(9.73)]{parisi2020theory}
\beq\label{eqC9:finitSHS}
f(\Delta,\b,h) = \log \g_{\D}\star e^{-\b \frac{1}2 h^2 \th(-h) } = \log q(\D,\b;h-\D/2)  \ , 
\eeq
with~\cite[Eq.(9.76)]{parisi2020theory}
\beq\label{eqC9:f0h}
\begin{split}
 \lim_{T\to 0, \, \D = \chi T} \D f(\Delta ,\b,h)  = - \frac{\c}{1+\c} \frac{h^2 \th(-h)}2 \ .
\end{split}\eeq
In this zero temperature limit, within the Monasson approach, 
the free energy reduces to the Legendre transform of the complexity~\cite{BJZ11}, \ie the logarithm of the number of minima of energy $e$, \ie we have $-\b \Phi \to \SS = \max_e [ \Sigma(e) - y e]$
with
\beq
\begin{split}
\SS(y ; \wh\f, \chi)= & \frac{d }2 \log\left(\frac{2\pi e}{d}\right)   - \frac{d }2 \log\left(\frac{\pi e \chi}{d^2 y}\right)   
 +\frac{ d \wh \f}{2} \left[ \FF\left( \frac{1+\c}{y} \right) -1 \right] \ , \\
\FF(x) &=   1+ \int_{-\io}^\io \de h \, e^{h } \left\{  e^{-  \frac{h^2 \th(-h)}{2x}   }  - 1\right\} =
\int_{-\io}^0 \de h \, e^{h -  \frac{h^2}{2x}   }    \ ,
\end{split}\eeq
from which the equation for $\chi$ is
\beq\label{eq:xeq}
\frac1{\wh \f} = \frac{\c}y \FF'\left(\frac{1+\c}{y} \right) = \left(x  -\frac{1}y\right) \FF'(x) \ , \qquad x = \frac{1+\c}{y}  \ ,
\eeq
 the energy is
\beq\label{eq:Monen}
e(y ; \wh\f, \chi) = - \frac{\partial \SS}{\partial y} =- \frac{d}2 \left\{\frac1y-  \wh \f \FF'\left[\frac{1+\c}{y} \right]\frac{1+\c}{y^2}  \right\}
=- \frac{d}2 \frac{1-  \wh \f x \FF'(x) }{y}  = \frac{d}2\frac{1}{y^2 (x-1/y)} \ ,
\eeq
where in the last step we used Eq.~\eqref{eq:xeq}, and the replicon is
\beq\label{eq:Monrep}
\l_R = 1  - \frac{\wh\f}2 \left( \frac{\c}{1+\c}  \right)^2 \FF\left[\frac{1+\c}{y} \right] = 1  - \frac{\wh\f}2 \left( \frac{x-1/y}{x}  \right)^2 \FF(x)    \ .
\eeq

\begin{figure}[t]
\includegraphics[width=.49\textwidth]{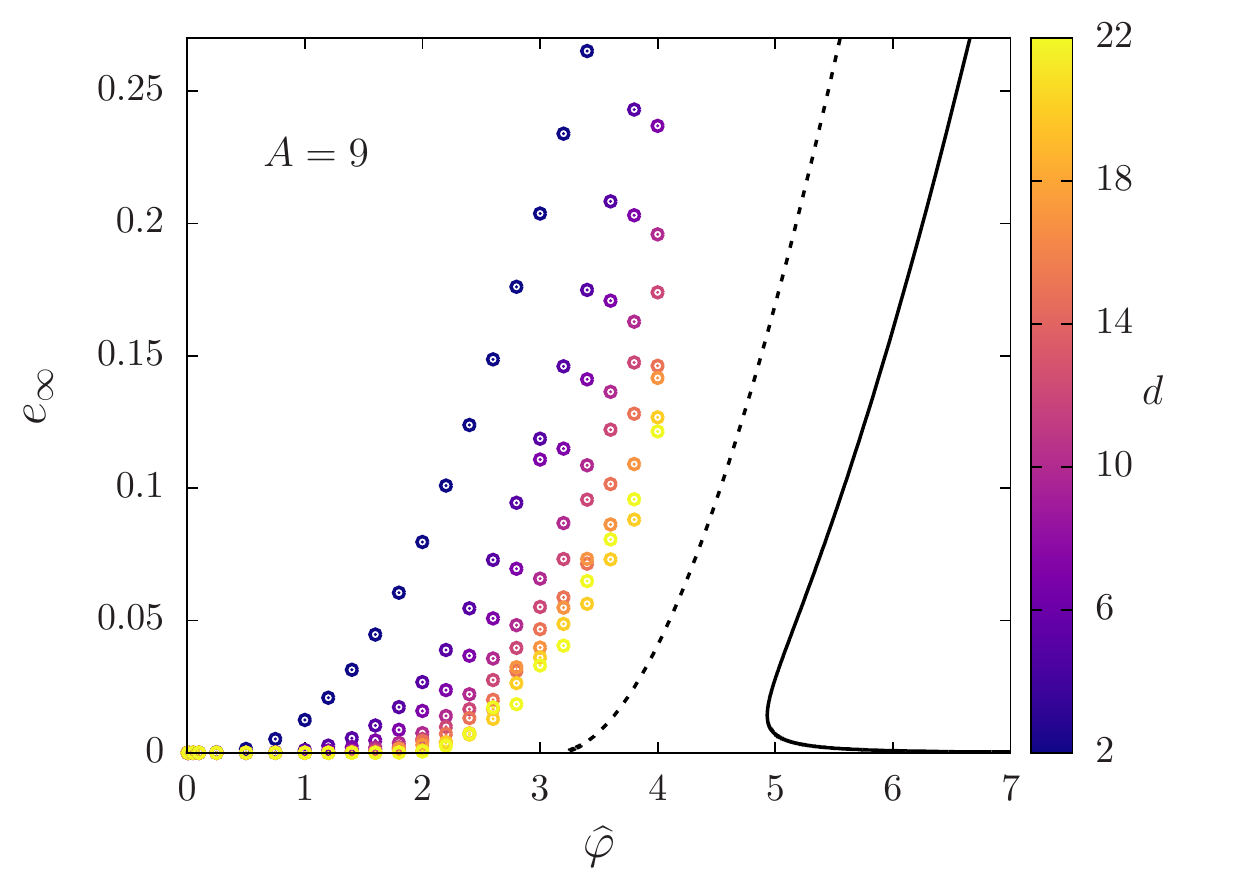}
\includegraphics[width=.49\textwidth]{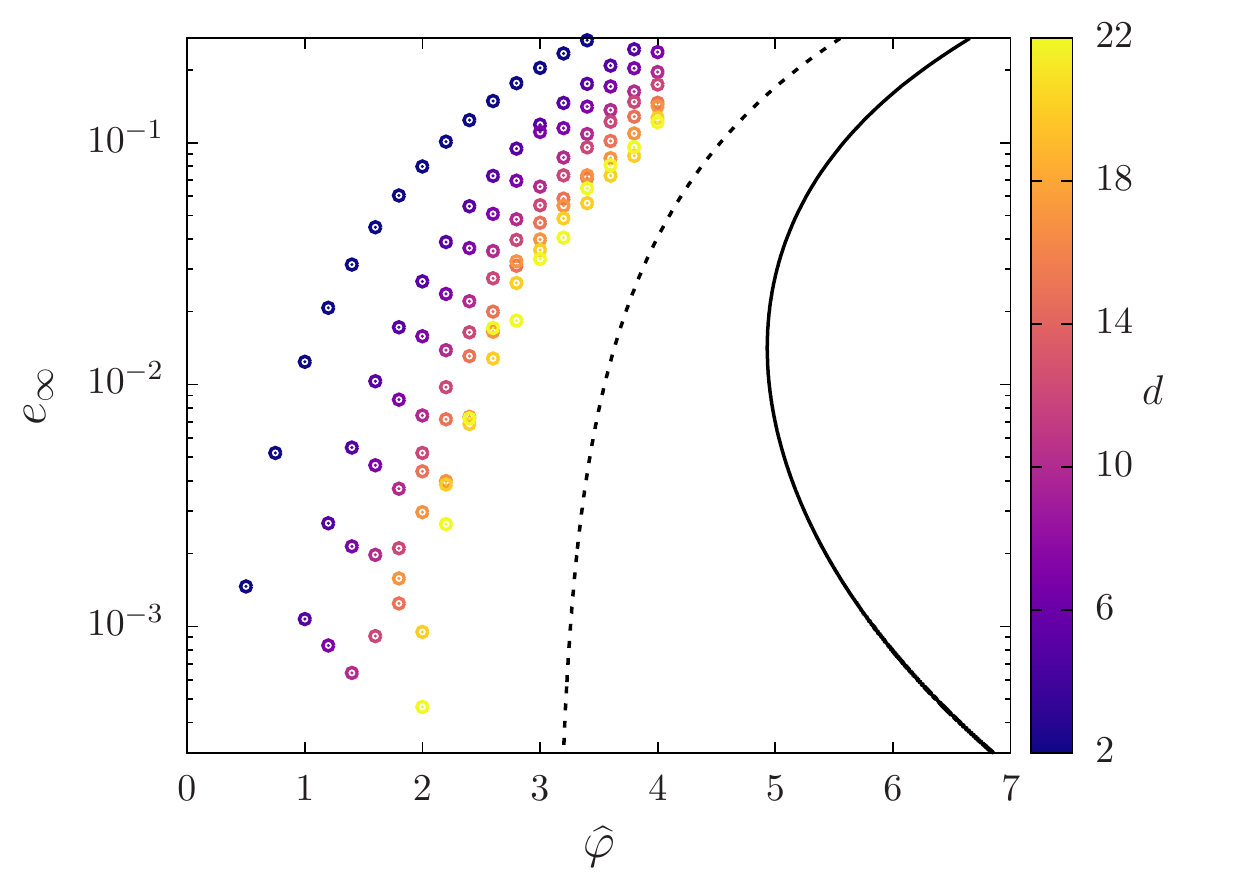}
\caption{
Numerical results for $e_\io$ versus $\wh\f$ in the RLG (same data as in Fig.~\ref{fig:Z}(f), with cutoff $A=9$) 
compared with
the Cugliandolo-Kurchan asymptotic prediction with one single time scale, using the marginal stability condition (full line), or the energy at which the solution for $\D$ disappears (dashed line).
The left panel is in linear scale, while the right panel is in log-linear scale.
Both predictions are inconsistent with the numerical results.
}
\label{fig:Monasson}
\end{figure}

Some results are shown in Fig.~\ref{fig:Monasson}. The full line is
the Cugliandolo-Kurchan asymptotic prediction with one single time scale, \ie the Monasson calculation at zero temperature given by Eqs.~\eqref{eq:xeq} and \eqref{eq:Monen}, together with the marginal stability condition $\l_R=0$ from Eq.~\eqref{eq:Monrep}. 
The curve is clearly not in agreement with the data, and is unphysically re-entrant at low energies. Replacing the condition of marginal stability with the energy at which the solution for $x$ in Eq.~\eqref{eq:xeq} disappears~\cite{BJZ11,parisi2020theory} (dashed line) provides a qualitatively better result, but still inconsistent with the numerical results, and is 
dynamically unjustified (the line ends at $e_\io$ at $\wh\f_J=6.26$ as given in~\cite{PZ10,BJZ11}).

\end{document}